\documentclass[12pt]{iopart}

\expandafter\let\csname equation*\endcsname\relax
\expandafter\let\csname endequation*\endcsname\relax
\usepackage{amsmath}

\usepackage{iopams}

\usepackage{soul}
\usepackage{graphicx,color}  % needed for figures
\usepackage{amsfonts,amssymb,amsthm}
\usepackage{mathtools}
\usepackage{tabularx}
\usepackage{comment}
\usepackage{cancel}
\usepackage{bigints}
\usepackage{soul}
\usepackage{ulem}
\usepackage{float}

\theoremstyle{definition}

\newcommand\reallywidehat[1]{%
\savestack{\tmpbox}{\stretchto{%
  \scaleto{%
    \scalerel*[\widthof{\ensuremath{#1}}]{\kern.1pt\mathchar"0362\kern.1pt}%
    {\rule{0ex}{\textheight}}%WIDTH-LIMITED CIRCUMFLEX
  }{\textheight}% 
}{2.4ex}}%
\stackon[-6.9pt]{#1}{\tmpbox}%
}
\newcommand{\pb}[3]{\left\lbrace #1 ,#2\right\rbrace_{#3}}

\usepackage[colorlinks,citecolor=red,urlcolor=black,bookmarks=false,hypertexnames=true]{hyperref} 
\begin{document}

\setlength{\abovedisplayskip}{8pt}
\setlength{\belowdisplayskip}{8pt}

\setlength{\abovedisplayshortskip}{8pt}
\setlength{\belowdisplayshortskip}{8pt}

\normalem

%\widetext  

\title[Hybrid Geometrodynamics.]{%
\Large Hybrid Geometrodynamics:\\
\large A Hamiltonian description of classical gravity coupled to quantum matter.}
\date{\today}

\author{J L Alonso$^{1,2,3,4}$, C Bouthelier-Madre$^{1,2,3,4}$, J Clemente-Gallardo $^{1,2,3,4}$and D Martínez-Crespo $^{1,3,4}$ }
\address{$^1$ Departamento de F\'{\i}sica Te\'orica, Universidad de Zaragoza,  Campus San Francisco, 50009 Zaragoza (Spain)}
\address{$^2$ Instituto de Biocomputaci{\'{o}}n y F{\'{\i}}sica de Sistemas Complejos (BIFI), Universidad de Zaragoza,  Edificio I+D, Mariano Esquillor s/n, 50018 Zaragoza (Spain)}
\address{$^3$ Centro de Astropart{\'{\i}}culas y F{\'{\i}}sica de Altas Energías (CAPA), Universidad de Zaragoza, Zaragoza 50009, (Spain)}
\address{$^4$ Instituto de Biocomputaci{\'{o}}n y F{\'{\i}}sica de Sistemas Complejos (BIFI), Universidad de Zaragoza,  Edificio I+D, Mariano Esquillor s/n, 50018 Zaragoza (Spain)}

\begin{abstract} 
We generalize the Hamiltonian picture of General Relativity coupled to classical matter, known as geometrodynamics, to the case where such  matter is described by a Quantum Field Theory in Curved Spacetime, but gravity is still described by a classical metric tensor field over a spatial hypersurface and its associated momentum. Thus, in our approach there is no non-dynamic background structure, apart from the manifold of events, and the gravitational and quantum degrees of freedom  have their dynamics inextricably coupled. Given the Hamiltonian nature of the framework,  we work with the generators of hypersurface deformations over the manifold of quantum states. The construction relies heavily on the differential geometry of a fibration of the set of quantum states over the set of gravitational variables.  {\color{black}An important mathematical feature of this work is the use of Minlos's theorem to characterize Gaussian measures over the space of matter fields and of Hida distributions to define a common superspace to all possible Hilbert spaces with different measures, to properly characterize the Schr\"odinger wave functional picture of QFT in curved spacetime. }This allows us to relate states within different Hilbert spaces in the case of vacuum states or measures that depend on the gravitational degrees of freedom, as the ones associated to Ashtekar's complex structure. This is achieved through {\color{black}the inclusion of a quantum Hermitian connection} for the fibration, which will have profound physical implications. The most remarkable physical features of the construction are norm conservation of the quantum state (even if the total dynamics are non-unitary), the clear identification of the hybrid conserved quantities and the description of a dynamical backreaction of quantum matter on geometry and \textit{vice versa}, which shall modify the physical properties the gravitational field would have in the absence of backreaction.
\end{abstract}
\vspace{2pc}
\maketitle

\section{Introduction}
In the pursuit of understanding the fundamental nature of our universe, the need of a holistic description unifying quantum mechanics and general relativity remains one of the greatest challenges in modern theoretical physics. Within this context, the concept of Quantum Gravity (QG) has emerged, aiming to reconcile the quantum behavior of matter with the curvature of spacetime by making quantum the geometry itself. One of the main approaches, inspired by the similarities with quantum mechanics, is Canonical Quantization of Gravity \cite{Kie}. Such quantization program was the main motivation for the construction of geometrodynamics; the starting point of the quantization process was precisely such Hamiltonian formulation of gravity.  In  fact, the central result of such canonical quantization is Wheeler–DeWitt equation, resulting from the quantization of the constraints present in geometrodynamics. Because of this, the study of geometrodynamics is still of interest for the community, summarized in \cite{HKT76} from ADM's original approach to the language of Hamiltonian generators of hypersurface (hereafter, \textit{hsf.}) deformations over the space of 3-metrics and their momenta.

Nevertheless, most of the phenomena associated to QG are far from falsifiable in any foreseeable future. A less ambitious approach relies on Semiclassical Gravity (SG) theories, which treat matter as quantum states while describing spacetime using classical variables. One example of them is the result of the semiclassical limit to the Wheeler-deWitt equation \cite{Kie2} where, even if the gravity is technically still considered quantum, its dynamics follows classical paths over which one may consider quantum corrections. Another common approach is based on the Moller-Rosenfeld equation, $G_{\mu\nu}=\langle\Psi\mid\hat T_{\mu\nu}\mid\Psi\rangle$, where the expectation value of the quantum energy-momentum tensor acts as matter source for Einstein's equations.

The key question in these approaches  is how quantum matter (or in general, quantum degrees of freedom) influences and shapes the classical variables describing the geometry of spacetime. This  problem becomes even more complex when one realizes the absence of a unique vacuum state and the ambiguity surrounding the notion of  particle in curved spacetime. Thus, finding a mathematically consistent description of the backreaction of quantum matter on classical spacetime has become a formidable quest with fundamental phenomenological implications \cite{wald1977}.  In turn, the backreaction of the matter fields on gravity also affects the propagation of the fields themselves, which forces us to find a joint dynamical description to portray this intertwining, {\color{black} even without considering QG or its semiclassical limit, in the spirit of \cite{Til1,Til2}.
}

In this context we situate our work. We are  interested in unraveling to which extent quantum matter fields modify classical general relativity. Given the greater similitude of geometrodynamical language to quantum mechanics, instead of modifying Einstein's equations, we seek a description where both quantum matter and classical gravity are described in a Hamiltonian way. One of the key differences with M\"oller-Rosenfeld equation from this perspective is the dynamical arising, leaf by leaf, of the geometry together with the quantum matter Cauchy data from their coupled evolution along a foliation of spacetime. Thus, we do not quantize covariant solutions to field equations, but canonical Cauchy data on a spatial hypersurface. Consequently, our fields and operators do not depend explicitly on time, as does the quantum energy-tensor in the covariant quantization approach. The price to pay, of course, is the loss of general covariance of the framework. The associated boons is the  clarity of the definitions of the conserved quantities and their coupling to gravitational degrees of freedom. 
%; there is no need to construct a consistent quantum stress-energy tensor (as in \cite{wald1977}) in this framework.
Nevertheless, as happened in classical geometrodynamics, covariance is recovered for the solution curves to the dynamics, once the so called Hamiltonian and momenta constraints are enforced.

As it happens in the case of classical geometrodynamics, we think that the proposal of this hybrid geometrodynamics can be seen as a previous step to the quantization program, and we think it is still interesting to build the whole formalism of QFT in curved spacetime and gravity under a unified Hamiltonian language. Perhaps, given the difficult falsifiability of quantum gravity, most of the interesting phenomenology may already arise at this level. In the same sense as in molecular dynamics, where Ehrenfest's dynamics (describing molecules with classical nuclei and quantum electrons under a coupled Hamiltonian evolution) reproduce with reasonable exactitude the full quantum dynamics thanks to the hierarchical masses and confinement widths between both subsystems, we may think of hybrid geometrodynamics as an effective theory whose validity will be related to the hierarchy between the gravitational and quantum scales.
Of course, this idea is not new, and a top-down approach based on a suitable WKB truncation from quantum canonical gravity to QFT coupled to classical geometry was recently developed in \cite{maniccia2023qft}, recovering a rigourous picture of QFT in Curved Spacetime. 

In our work, instead of top-down we will go bottom-up and we will  consider firstly classical geometrodynamics; secondly, we develop a quantum field theory over curved spacetime in a Hamiltonian fashion and, lastly, we couple them in a suitable way to build consistent hybrid geometrodynamics. These ingredients will define a Cauchy problem where the initial data must be defined fulfilling certain constraints. In geometrodynamics, such constraints are derived from physical principles, namely the path independence criterium. This criterium can be easily understood: given that the field data contained in a spatial hsf. should be equal to the one induced by the spacetime field solutions through the foliation map, if some initial field data is evolved from an initial hsf. through two different ``paths" (sequences of intermediate hsfs.), until a same final spatial hsf. is reached, the field content in such final hsf. should be the same. To fulfill this integral criterium, it must satisfy its differential form, which is the need of commutativity of the differential generators of different deformations at each hsf. in each path. Given that the generators of hsf. deformations form a non-trivial Poisson algebra, they will only commute if the superhamiltonian  and supermomenta are zero on shell. These are the so called Hamiltonian and momenta constraints. See \cite{HKT76} for the canonical construction of geometrodynamics.  From a Lagrangian perspective, lapse ($N_s$) and shift ($N_s^i$) functions are Lagrange multipliers for these constraints. We will consider spacetime a globally hyperbolic Lorentzian manifold in order to define a proper Cauchy problem. Notice that, while we stick to the metric description of General Relativity, the general framework also applies to any alternative description of gravity.

On the other hand, in \cite{HKT76}, even though most of the discussion is kept general, the matter sources explicitly considered are classical fields. It is our aim to substitute such classical fields by a quantum theory rewritting QFT in Curved Spacetime in a Hamiltonian formalism where the spacetime itself is not a given background as is the case in \cite{Holl}, but is dynamically coupled to QFT. In order to do so, we firsly consider the Schrödinger wave functional picture for quantum fields in curved spacetime \cite{Long1996TheSW}, as it is easily adaptable to the non-covariant Hamiltonian framework as in \cite{Heslot,Kibble79}. Then we promote the geometric variables from a given background to be part of the kinematics. The dynamics will be encoded in a set of coupled \textit{hybrid equations of motion} of gravitational variables and QFT states, posing the hybrid Cauchy problem.

To rigorously build a Hamiltonian picture of QFT in curved spacetime, we secure our mathematical grounds with tools from infinite dimensional calculus and differential geometry in such framework. As we need to build differential tensors on our manifold of fields (classical and quantum), we will ask that manifold to be defined in such a way that derivatives can be rigorously proved to exist.  In this paper we will reduce such  mathematical technicalities to the minimum, addressing the interested reader to \cite{David,David2} for technical details. Thus, the manifold of classical fields will be assumed  to be the dual (see \cite{David,kriegl1997}) to some nuclear-Frechèt  space, denoted by $\mathcal{N}$, of functions defined over a spatial hsf. $\Sigma$ (for example, differentiable functions of compact support  $\mathcal{N}=C^\infty_c(\Sigma)$ when $\Sigma$ is compact). This choice ensures the existence of a well-defined differential calculus on the fields phase space. For analogous reasons, quantum states can be modelled using the set of Hida functions  $(\mathcal{N})$ that will densely populate any Hilbert space whose domain is the manifold of classical fields. Being the set of Hida functions a nuclear-Fréchet space, it allows us to perform Fréchet differential calculus and ultimately, differential geometry. This set of functions will be completed under the choice of a scalar product, which is given by a Gaussian measure $D\mu$ over the space of fields, to form a Rigged Hilbert space $(\mathcal{N})\subset L^2(\mathcal{N}',D\mu)\subset (\mathcal{N})'$. Such a measure can be characterized through Minlos' theorem \cite{minlos1959generalized} and it presents interesting properties for the Schr\"odinger functional picture of QFT (see e.g. \cite{David}).  

 Moreover, such measure is closely related to the quantization procedure (in Corichi et. al.  \cite{CCQ04}, given by the choice of complex structure  $J_C$ for the classical space of fields), or to  what in the literature is usually referred as the choice of vacuum state. The main difficulty arises from the fact that the quantization procedure or the vacuum state are time dependent, as \textit{a priori} they depend  on the leaf 3-metric over the spatial hsf., lapse and shift (as $J_C$ 
 does in the static or stationary cases, see \cite{ashtekar75}). This is a widely known fact in the literature, associated to phenomena such as norm loss of quantum states and quantum completeness \cite{Schneider}.
In this line, we consider that this  dependence  introduces a parametric family of quantizations and a family of non-unitarily-equivalent Hilbert spaces. Given that the dynamics along the foliation will necessarily  change such geometric parameters, we need a way to relate states (and operators) within these Hilbert spaces. To do so we will introduce, analogously to the case of time dependence Hilbert spaces in \cite{mostafazadeh2018energy}, the parallel transport for quantum states along curves over the space of pure geometrodynamical variables. In fact, the quantum states will become covariant sections over a fibration with base the geometrodynamical variables and fiber the space of wave functionals, and an (almost) Hermitian quantum connection will allow us to relate inequivalent Hilbert spaces. A different notion of quantum connection over a different bundle, but with certain similarities, has been introduced in \cite{Canarutto2004QuantumCA}. In the realm of algebraic quantum field theory, some works, such as \cite{torre1999functional}, have treated the functional picture from the point of view of algebraic QFT, addressing some problems, such as the lack of unitarity of time evolution in the functional picture. Our approach, however, is based on a submanifold of field Cauchy data (not on covariant solutions of field equations), which constitutes the domain of Schr\"odinger wave functionals. To our knowledge, the mathematical construction we present constitutes a novelty for the Schr\"odinger functional picture of QFT on a Cauchy hypersurface. The novelty relies in the proper definition of leaf-dependent Gaussian measures (because $D\mu$ depends on $J_C$) over distributions that provide leaf-dependent scalar products for the wave functionals. In turn, it is also a novelty that the subsequent parametric family of Hilbert spaces are considered to be inside the common superset of Hida distributions and related by the Hermitian connection, which is precisely designed to compensate for the change of the scalar product under changes of gravitational variables. 

The QFT under consideration will be the result of a leaf-dependent quantization procedure of a classical field theory. Such classical field theory will be built in a Hamiltonian formalism together with gravity and, in the spirit of geometrodynamics, the generating functions of hypersurface deformations over such classical phase space will already fulfill Dirac's closing relations for a Poisson bracket (hereafter, PB)  combining geometry and classical matter. Once we provide QFT with a quantum PB, we will also have Hamiltonian generating functions of hypersurface deformations at the quantum level that reproduce the same algebraic relations. We will assume that our classical field theory is regularized when written in terms of distributions, to avoid the presence of certain anomalies \cite{tsamis1987factor} (usually solved with spatial point-splitting) when we try to quantize it. On the other hand, regarding renormalization, a similar analysis to the one in  \cite{BERNARD1977201,Ebo2} could be carried out in this Hamiltonian framework, although in this case the geometry of spacetime is not known \textit{ab initio}, but constructed dynamically and coupled to the quantum states. In this work, our focus will be on the formal interplay between the geometric structures of Hamiltonian gravity and QFT.

On the other hand, we maintain the geometrodynamical principle of equivalence as stated in \cite{HKT76} and we consider for quantization classical theories whose classical matter supermagnitudes only depend locally (not derivatively) on the 3-dim metric tensor $h$  and do not depend on  its momentum $\pi_h$, nor on the lapse or shift functions. This property is dubbed \textit{ultralocality} in  \cite{teitelboim1973hamiltonian}. This excludes non-minimally coupled theories from this scheme, even though a similar analysis could be carried out in that case.

In conclusion, the main result of our work will be the Hamiltonian representation of hsf. deformations for a hybrid theory of quantum field matter and classical gravitation and the definition of the hybrid constraints that enforce \textit{path independence} as in \cite{HKT76}. In order to do so, given that the quantization procedure is dependent on the gravitational variables, the quantum connection  will need to fulfil some constraints, which we show that can be satisfied, at least for the case of the scalar field.

Note that a Hamiltonian picture characterizing formally the backreaction of QFT on gravity for a Friedman-Roberson-Walker universe and a single foliation of spacetime was developed by Husain and Singh in \cite{Husa} and \cite{husain2021quantum}. Our formalism diverges from theirs both in the mathematical formalism (we add the quantum Hermitian connection and the geometric-dependent measure and Hilbert spaces, the use of the superset of Hida distributions and the symplectic geometry and Poisson structures for this space of quantum states) and physical implications (its application to generic spacetimes, norm conservation, relating non-unitarily equivalent Hilbert spaces, potential effects of the quantum connection in relation with particle creation), but the spirit is similar.  
%Given the amount of technicalities devoted to the foundational construction of hybrid geometrodynamics and the remarkable length of this paper we will not solve any particular physical example, but illustrative examples (such as those considered in \cite{husain2021quantum, husain2022dynamics}) will be considered in future works. 

The structure of the paper is as follows. 
In section \ref{sec:2}, we introduce the essential mathematical ingredients and notation from geometrodynamics, namely: the notation for the geometry of hsfs. and foliations, the closing relations of hsf. deformations, its representation in phase space and the relations that any matter theory must fulfill as long as it is coupled in a non-derivative way with the 3-metric. For a more detailed presentation of these tools, we address the reader to \cite{Dirac50,Dirac51, HKT76}.

Section \ref{sec:3} is devoted to the Hamiltonian representation of Schrödinger wave functional picture for a leaf dependent quantization procedure,  where the quantum states are defined as covariant sections for the quantum connection. This construction is partially based on the mathematical tools presented in \cite{David,David2}. 

In section \ref{sec:4} we combine the ingredients from the previous sections in a Hamiltonian description for both the classical 3-geometry and the QFT for matter, to build Hybrid Geometrodynamics. We identify the hybrid generating functions that, under a hybrid PB, reproduce the algebraic relations of the generators of hsf. deformations. Then, we impose the Hamiltonian and momenta constraints at the hybrid level and present the hybrid equations of motion. As happened in the hybrid Hamiltonian model for finite dimensional systems defined by Ehrenfest's dynamics \cite{Alonso2012}, the resulting dynamics cannot be unitary because of the non-linearity arising from the backreaction. Nevertheless, the quantum connection provides with norm conservation for the quantum states along the dynamics, even if both the states and the scalar product (or vacuum state) are dependent on the gravitational degrees of freedom. Lastly, we discuss some physical implications of these hybrid dynamics regarding their phenomenology, in particular in relation with the phenomenon of quantum completeness \cite{Schneider,Schneider2}. In section \ref{example}, we provide an example of the derivation of the quantum connection compatible with geometrodynamics for the holomorphic quantization in curved spacetime of the scalar field. In section \ref{sec:6}, we summarize our main results and discuss possible future research lines.

\section{A brief summary of classical geometrodynamics\protect\footnote{This section is used to fix notation and to provide a shallow summary of geometrodynamics. For a deeper introduction check the notes in the supp. material and, e.g. \cite{HKT76} and references therein.}}
\label{sec:2}

Geometrodynamics subtitutes spacetime by a foliation in terms of spatial hsfs. (all isomorphic to $\Sigma\sim\mathbb{R}^3$) and the covariant formalism for fields by a Hamiltonian representation, which evolves the spatial field content (both geometrical and matter) from the initial hsf. to the final one, defining a Cauchy problem. The group of deformations of spatial hsfs. in Lorentzian spacetime is lifted to the set of fields on $\Sigma$ through a PB and a representation of its generators as generating functions. The generating function of local normal deformations, $\mathcal{H}(x)$, is called \textit{superhamiltonian} (sH) while for local tangential deformations, $\mathcal{H}_i(x)$, they are called \textit{supermomenta} (sM).

For any combination of matter and geometric fields, in order to reproduce Dirac's algebraic relations of the generators of hsf. deformations, the generating functions must fulfil the following closing relations (hereafter, CR):
\begin{equation}
\label{closeHH}
\lbrace \mathcal{H}_x,\mathcal{H}_{x^\prime}\rbrace=\mathcal{H}^i_x\delta_{,i}(x,x^\prime)-\mathcal{H}^i_{x^\prime}\delta_{,i}(x^\prime,x)
\end{equation}
\begin{equation}
\label{closeHM}
\lbrace \mathcal{H}_{ix},\mathcal{H}_{x^\prime}\rbrace=\mathcal{H}_x\delta_{,i}(x,x^\prime)
\end{equation}
\begin{equation}
\label{closeMM}
\lbrace \mathcal{H}_{ix},\mathcal{H}_{jx^\prime}\rbrace=\mathcal{H}_{ix^\prime}\delta_{,j}(x,x^\prime)+\mathcal{H}_{jx}\delta_{,i}(x,x^\prime)
\end{equation}
where both the generating functions and the PB is for the whole theory (containing matter and geometry). These relations are a requirement to reproduce the kinematics of hsf. deformations \cite{HKT76} and to define a physical evolution ~\cite{Glow20}.

Although the framework is general, we will follow the 3+1 description of metric based Einstenian gravity, as in \cite{HKT76}. Thus, the canonical degrees of freedom for gravitational variables are given by a Riemannian 3-metric $h_{ij}(x)$ (induced on the hsf. by the spacetime metric $g_{\mu\nu}$) and its associated conjugate momenta $\pi^{ij}(x)$ (related with the extrinsic curvature of the leaf). The gravitational phase space, with $(h_{ij},\pi^{ij}_h)\in\mathcal{M}_G$, is therefore constructed as a cotangent bundle:
\begin{equation}
\label{MG}
    \mathcal{M}_G=T^\star \text{Riem}(\Sigma)\sim \text{Riem}(\Sigma)\times \text{Riem}^\prime(\Sigma)\;
\end{equation}
locally isomorphic to the Cartesian product of the Riemannian 3-metrics and symmetric (0,2)-tensor densities of weight 1 (which we have denoted by $\text{Riem}^\prime(\Sigma)$). 

We can now define a PB over the smooth functions over $\mathcal{M}_G$ given by:
\begin{equation}
\label{eq:classicalPB}
\lbrace f,g\rbrace_G=\int_\Sigma d^3x \delta_{[h_{ij}(x)}f\delta_{{\pi}^{ij}_h(x)]}g \quad \forall f,g\in C^\infty(\mathcal{M}_G)\;.
\end{equation}
 For further analysis, check the notes on functional calculus in the supp. material. At an operational level, it is enough to consider that such derivatives are local (here  $\delta(x,x^\prime)$ is associated to the Lebesgue measure of $\Sigma$) and independent from each other:
\begin{equation}
\frac{\delta}{\delta {\pi_h^{ij}(x)}}{\pi_h^{kl}(x^\prime)}=\frac{\delta}{\delta {h_{kl}(x)}}{h_{ij}(x^\prime)}=\delta^{kl}_{ij}\delta(x^\prime,x)\quad
\text{and}\quad  \frac{\delta}{\delta {\pi_h^{ij}(x)}}{h_{kl}(x^\prime)}=\frac{\delta}{\delta {h_{ij}(x)}}{\pi_h^{kl}(x^\prime)}=0\, .
\end{equation}
For metric-based gravity in this pure geometrodynamical case (i.e. without matter), the generating functions of hsf. deformations, sM and sH, are given by:
\begin{equation}
\label{SH}
\mathcal{H}_{i}^G(x)=-2h_{i\alpha}(x)D_j\pi^{\alpha j}_h(x)\quad\text{and}\quad \mathcal{H}^G(x)=\dfrac{1}{2}\dfrac{(2\kappa)}{\sqrt{h}}(G_{ijkl}\pi^{ij}_h\pi^{kl}_h)(x)-(2\kappa)^{-1}\sqrt{h(x)}R(x)
\end{equation}
where $D_j$ are $h$-covariant spatial derivatives, $R$ is the scalar curvature for $h_{ij}(x)$ and $G_{ijkl}:=\frac{1}{2 \sqrt{h}}\left(h_{i k} h_{j l}+h_{i l} h_{j k}-h_{i j} h_{k l}\right)$ is De Witt's metric.

The phase space $\mathcal{M}_G$ can be extended to contain matter fields, $\varphi_i$, and their momenta, $\pi^i$, for any tensor field \footnote{As in the gravitational case, functions and distributions must be related through a Gel'fand triple and a dense subset of the cotangent space  must be used to perform Fréchet calculus and densely define  differential structures, as in \cite{David}. This is briefly summarized in the supp. material.}. We denote the set of all matter fields by  $\mathcal{M}_F$:
\begin{equation}(\varphi_1,\cdots,\varphi_n,\pi^1,\cdots,\pi^n)\in\mathcal{M}_F\;,
\end{equation}
 As in the gravitational case, one can define a PB over $C^\infty(\mathcal{M}_F)$, as:
\begin{equation}
\lbrace f,g\rbrace_M:=\delta_{[\varphi_{ix}}f\delta_{\pi^{ix}]}g:=\delta_{\varphi_{ix}}f\delta_{\pi^{ix}} g -\delta_{\varphi_{ix}}g\delta_{\pi^{ix}} f\;.
\end{equation}
To clarify the notation, $i,j,k$ are discrete indices where upper and lower repeated indices follows Einstein's convention. On the other hand,  $x,y,z$ represent continuous indices, and repeated upper and lower indices implies integration over $\Sigma$.\footnote{Objects with upper index, as $\pi^x$, are densities of weight 1. With lower index, e.g. $\varphi_x$, they are test functions. To perform Fréchet derivatives one must restrict (see supp. material) the momenta $\pi^x$ to belong to a certain dense subset of  distributions $\mathcal{N}'$ which is representable under a measure, such that
$\lbrace f,g\rbrace_M=\int\limits_{\Sigma}d^3x \partial_{[\varphi_i(x)}f\partial_{{\pi}^i(x)]}g$, 
where ${\pi}^i(x)$ is the representation of $\pi^{i,x}$ under Lebesgue measure.}
\begin{comment}
  Note that this implies the presence of an auxiliar Gelfan'd triple with a Hilbert space mediating the duality given by the natural scalar product constructed from the chosen volume 1-form, as is explained in the appendix for the geometric case.   
\end{comment}
Thus, the total phase space of geometry and matter denoted as $\mathcal{M}_C$ is given by:
\begin{equation}
\label{eq:Fc}
\mathcal{M}_C:=\mathcal{M}_G\times\mathcal{M}_F\quad\text{and} \quad\mathcal{F}_C:= \mathcal{M}_N\times \mathcal{M}_G\times \mathcal{M}_F\;.
\end{equation}
For later use, we have defined another product manifold $\mathcal{F}_C$, where $\mathcal{M}_N$ contains the lapse $N(x)$ and shift $N^i(x)$ functions associated with the space-time foliation (defined e.g. in the supp. material or \cite{HKT76}). $\mathcal{F}_C$ can be considered as a trivial bundle with base manifold  $\mathcal{B}:=\mathcal{M}_N\times\mathcal{M}_G$ and fiber $\mathcal{M}_F$: the base contains the geometrical information of the theory, and the fiber, the matter information.
 \begin{comment}
Notice that $\mathcal{M}_N$, containing the information about the space-time foliation, is different from the other two pieces as their dynamics are not associated to a particular kinematical structure, but will always be associated to the choice of foliation and thus act as time dependent parameters. Therefore, given the path independence principle, even if some tensors for the matter theory may depend on them (thus, their inclusion in the base of the fibration), we anticipate that physical magnitudes, as well as the Poisson tensors over the gravitational and matter submanifolds, will be independent of $\mathcal{M}_N$. 
\end{comment}
Now, on the diff. functions $C^\infty(\mathcal{M}_C)$, one can define a total PB as the sum of the two previous PBs:
\begin{equation}
\label{totalPB}
\lbrace\cdot,\cdot\rbrace=
\lbrace\cdot,\cdot\rbrace_G+
\lbrace\cdot,\cdot\rbrace_M\;:C^\infty(\mathcal{M}_C)\times C^\infty(\mathcal{M}_C)\rightarrow C^\infty(\mathcal{M}_C)\;,
\end{equation}
 This total PB fulfills Jacobi, Leibniz, bilinearity and antisymmetry because matter and geometric PBs fulfilled them previously and there is a certain compatibility condition between them. We will delve into this condition later for the quantum case.

 Now, we can represent in a Hamiltonian way the generators of hsf. deformations for this system that includes matter fields.  Invoking the GD analog to the equivalence principle as in \cite{HKT76}, we know that the total supermagnitudes must be the sum of the pure geometrodynamical magnitudes $\mathcal{H}_{\alpha }^G$ and the matter supermagnitudes $\mathcal{H}_{\alpha}^M$:
\begin{equation}
\label{classicaltotalsH}
\mathcal{H}(x)=\mathcal{H}^G(h,\pi_h;x)+\mathcal{H}^M(h,\varphi,\pi;x)
\quad\text{and}\quad
    \mathcal{H}_{i}(x)=\mathcal{H}_{i}^G(h,\pi_h;x)+\mathcal{H}_i^M(h,\varphi,\pi;x)\;.
\end{equation}
The geometrodynamical equivalence principle of \cite{HKT76} implies that matter supermagnitudes are ultralocal (non-derivative dependence on $h$ and independent of $\pi_h$). In turn, the purely gravitational sH and sM, $\mathcal{H}_{\alpha}^G(x)$, do not depend on matter fields. The total sH and sM must fulfil the CR \eqref{closeHH},\eqref{closeHM} and \eqref{closeMM} for $\lbrace,\rbrace$ in order to reproduce Dirac's algebra. Consequently, the  gravitational sH and sM must also fulfil them for $\lbrace,\rbrace_G$, because the matter PB, $\lbrace,\rbrace_M$, is null on them. This is the case for  (\ref{SH}). The matter sH also fulfills \eqref{closeHH} for $\lbrace,\rbrace_M$, as, due to  its ultralocality and antisymmetry of the PB, we have:
\begin{equation}
   \lbrace \mathcal{H}^G(x),\mathcal{H}^M(x^\prime)\rbrace_G+\lbrace \mathcal{H}^M(x),\mathcal{H}^G(x^\prime)\rbrace_G=0\;.
\end{equation}

Regarding the CR for matter sM, given that $\mathcal{H}^M_i$ does not depend on $h,\pi_h$, it must fulfill also \eqref{closeMM} for $\lbrace,\rbrace_M$, because $
\lbrace \mathcal{H}_{i}(x),\mathcal{H}_{j}^M(x^\prime)\rbrace_G=0$. In turn, to fulfill \eqref{closeHM} we need:
\begin{equation}
\label{sMGM-sHM}
\lbrace \mathcal{H}_{i}^G(x),\mathcal{H}^M(x^\prime)\rbrace_G+\lbrace \mathcal{H}^M_{i}(x),\mathcal{H}^M(x^\prime)\rbrace_M= \mathcal{H}^M(x)\delta_{,i}(x,x^\prime)\;.
\end{equation}
The crossed gravitational-matter term yields:
\begin{equation}
\label{sMG-sHM}
\lbrace \mathcal{H}_{i}^G(x),\mathcal{H}^M(x^\prime)\rbrace_G=2h_{i\alpha}(x)D_{x^j}\dfrac{\delta\mathcal{H}^M (x^\prime)}{\delta h_{\alpha j}(x)}
\end{equation}
and thus the matter sector term should provide:
\begin{equation}
\label{sMM-sHM}
\lbrace\mathcal{H}_{i}^M (x),\mathcal{H}^M(x^\prime)\rbrace_M=-2h_{i\alpha}(x)D_{x^j}\dfrac{\delta\mathcal{H}^M (x^\prime)}{\delta h_{\alpha j}(x)}+\mathcal{H}^M(x)\partial_{x^i}\delta(x,x^\prime)\;.
\end{equation}

\begin{comment}
To do so, we will first move from the implicit general hyperplane equation where the function $\tau$ of spacetime coordinates $X^\mu$ is set to a constant $s$,  to a representation of the hyperplane in terms of four parametric equations for the spacetime coordinates $X^\mu$  of three embedding variables $x_\sigma^i$:
\begin{equation}
\tau(X^\mu)=s\Rightarrow X^\mu=X^\mu(x_\sigma^i) .
\end{equation}
This allows us to write the induced spatial metric $h_{ij}$ in terms of the derivatives of the parametric equations and the spacetime metric: 
\begin{equation}
    h_{ij}(x_\sigma)=^4g_{\mu\nu}(X^\eta(x_\sigma))X^\mu_{,i}X^\nu_{,j} .
\end{equation}
\end{comment}
As an example that will come in handy in the future, for the classical scalar field denoted by $\varphi_x,\pi^x$, the matter supermagnitudes that fulfil these CR are:
\begin{equation}
\mathcal{H}^M(x):=\dfrac{1}{2}\sqrt{h}\left(\dfrac{\pi^2_\varphi}{h}+h^{ij}\varphi_{,i}\varphi_{,j}+m^2\varphi^2+V(\varphi)\right)(x)\quad\text{and}\quad\mathcal{H}^M_i(x)=\pi(x)\varphi_{,i}(x)\;.
\end{equation}

Regarding dynamics, once a foliation of spacetime is chosen, the evolution is given by the mapping of the fields along the sheaf of hypersurfaces. In the 3+1 perspective, from a hsf. one can access the infinitesimally neighbouring one through a normal deformation (of size given by $N(x)$) and a tangential deformation (of size given by $N^i(x)$). Now that we have such deformations represented in a Hamiltonian way, the evolution along the hypersurface is the solution to the Hamiltonian dynamics given by:
\begin{equation}
\dfrac{d F}{ds}=\lbrace F, H\rbrace+\partial_s(N^{\mu x})\dfrac{\delta}{\delta{N^{\mu x}}}F+\partial_s F\quad \forall F\in C^\infty(\mathcal{F}_C\times\mathbb{R})\;,
\end{equation}
where $F$ is any functional over $h,\pi_h,\varphi,\pi,N,N^i$, with a possible differentiable dependence on $s\in\mathbb{R}$. Note that the role of time is being played by the label $s$ of the hsfs. in the foliation. The total Hamiltonian $H$ is given by the contraction of the lapse and shift functions with the corresponding generating functions $\mathcal{H}(x)$, $\mathcal{H}_i(x)$:
\begin{equation}
\label{totalHamiltonianClassical}
H(s)=N^x(s)\mathcal{H}_x+N^{ix}(s)\mathcal{H}_{ix}\;.
\end{equation}
Note that lapse and shift are $s$ dependent functions (each hsf. with label $s$ relates with the following one through different linear combination of normal and tangential deformations), and, as such, the formalism is analogue to having an explicitly time dependent Hamiltonian function. Making use of \eqref{classicaltotalsH}, $H(s)$ can be decomposed as $H(s)=H_G(h,\pi_h,N(s),N^i(s))+H_M(h,\varphi,\pi,N(s),N^i(s))$, with: 
\begin{equation}
H^G(h,\pi_h,N(s),N^i(s)):=N^x(s)\mathcal{H}_x^G+N^{ix}(s)\mathcal{H}_{ix}^G\quad\text{and}
\end{equation}
\begin{equation}
\label{classicalfieldhamiltonian}
H^M(h,\varphi,\pi,N(s),N^i(s)):=N^x(s)\mathcal{H}_{x}^M(h,\varphi,\pi)+  N^{ix}(s)\mathcal{H}_{ix}^M(h,\varphi,\pi)\;.
\end{equation}
We anticipate that the quantization of $H^M$ will yield the Hamiltonian operator governing the quantum dynamics. On a side note, it is not uncommon to consider $N(x,s)$ or $N^i(x,s)$ depending on Cauchy data such as $h_{ij}(x)$, because of geometric requirements on the definition of the foliation (e.g. section 5 of \cite{HKT76}). This is not an impediment for this Hamiltonian representation as explained in \cite{HKT76}, nor for the quantization of $H^M$, because the quantization procedure used in the next section (even though it is $h$-dependent) is performed solely on the matter degrees of freedom, $\varphi,\pi$, leaving $h,\pi_h$ completely classical. For simplicity, we exclude from the current construction any possible dependence of $N$ or $N^i$ on the matter Cauchy data.

We will now implement the\textit{ path independent principle} as defined in \cite{HKT76}. If we pose this Cauchy problem with the same initial Cauchy data, the field content on the final hsf. must be the same independently of the path followed from one initial hsf. $\Sigma_0$ to a final one $\Sigma_{s_f}$. Note that this path is given by the lapse and shift functions $\forall s\in[0,s_f)$. Naturally, independence of the path implies, at the infinitesimal level, commutativity of the generators over phase space for any pair of local hsf. deformations:
\begin{equation}
\label{constraints}
\lbrace \mathcal{H}_{\mu x}, \mathcal{H}_{\nu x^\prime}\rbrace\simeq 0\ \quad \forall\  \mu,\nu= 0,\cdots,3\quad\Longrightarrow \quad\mathcal{H}(x)\simeq 0\quad\text{and}\quad\mathcal{H}_i(x^\prime)\simeq 0
\end{equation}
where $\simeq$ means equality  \textit{on shell}, i.e. once the physical constraints are satisfied. Given that the total PBs \eqref{closeHH},\eqref{closeHM} and \eqref{closeMM} are linear on sH and sM, the left-hand side of \eqref{constraints} implies the Hamiltonian and momenta constraints, defined on the right-hand side.

In the following sections, we extend this construction to include a quantum scalar field (instead of a classical one) coupled to gravity, defining a Hamiltonian hybrid quantum-classical system. We will show that a quantum PB can be defined which turns the analog of \eqref{totalPB} into a well defined hybrid PB. In turn, we will find the generating functions of hsf. deformations inside the quantum and hybrid Poisson algebras. Both the CR (given their kinematical nature) and constraints (to preserve path independence) will have their analogue in the hybrid case. In fact, the CR derived are a must for any matter theory, as long as it has non-derivative coupling with the metric\cite{HKT76}.
\section{Quantum theory for Cauchy data in a curved spacetime.}
\label{sec:3}%13.5 paginas.
 Our objective is the representation of the kinematics of hsfs. for the quantum states in a Hamiltonian way, for which we need a PB  for functions on the space of quantum states, $M_Q$, and the quantum supermagnitudes that will fulfil the matter CR for it. Thus, in this section, we will construct the Hamiltonian formalism of QFT of matter Cauchy data on a hypersurface with a given geometric content. This construction is more complicated than in flat spacetime, since quantization mapping for matter Cauchy data depends on the gravitational degrees of freedom, lapse and shift.  Consequently, the ingredients defining quantum theory  (quantum states, operators, scalar products, ...) become different for each hsf. of the foliation, characterized by different values for $\xi=(h_{ij},\pi_h^{ij},N,N^i)$. Therefore, a careful analysis of the dependence on $\xi$ of $M_Q$ (as well as operators over it and its scalar product) is necessary before adding QFT to geometrodynamics. To our knowledge, this approach is different from the usual one of QFT in curved spacetime, as it is able to relate within non-unitarily equivalent Hilbert spaces while conserving their norm.  Besides, we do not consider the  quantization of covariant solutions of field equations, but of canonical matter fields on each leaf, so the $\xi$-dependence of the quantization is related with the parametric family of complex structures on Cauchy data considered in \cite{agullo2015unitarity}.

The procedure is very similar to some results on time-dependent Hilbert spaces $\mathcal{H}_t$ (see e.g. section 2 of \cite{mostafazadeh2018energy}), where one represents with a unique element in an original Hilbert space $\Psi_\epsilon(t)\in\mathcal{H}_{t_0}$ the infinitesimally evolved quantum state that belongs to the infinitesimally next Hilbert space, $\Psi (t+\epsilon)\in\mathcal{H}_{t_0+\epsilon}$.  Hence, $\Psi_\epsilon$ is determined by a parallel transportation from $\mathcal{H}_{t_0+\epsilon}$ to $\mathcal{H}_{t_0}$ under a certain connection. Naturally, associated to this connection one can define the covariant time derivative. The key mathematical tools are provided by the theory of vector bundles, e.g. the space of quantum states $\mathcal{M}_Q$ is related to a Hermitian vector bundle with fiber isomorphic to the Hilbert spaces and base $\mathbb{R}$. Our work will also make use of this tools, but, instead of having time-dependent Hilbert spaces, we have $(h_{ij},\pi^{ij}_h, N,N^i)$-dependent Hilbert spaces.  Besides, these Hilbert spaces in QFT, being its domain infinite dimensional, are not unitarily equivalent. Thus, the bundle we consider has, in general, a common superset of these Hilbert spaces as fiber manifold, and the space $\mathcal{M}_G\times\mathcal{M}_N$ as base manifold. In this work, we will focus on the formal geometric structures characterizing the interplay between classical Hamiltonian gravity and QFT encoded in this bundle and refer to \cite{David,David2} for most of the functional analysis and infinite dimensional calculus.

\subsection{Geometry of classical phase space and  leaf dependent quantization.}
As it is commonly done in Quantum Mechanics, the set of quantum operators $\mathcal{B}(\mathcal{H})$ is obtained from functions on the classical field theory by a quantization mapping $Q:C^\infty(\mathcal{M}_F)\rightarrow\mathcal{B}(\mathcal{H})$. On the other hand, the states are suitable representations of the dual of such $C^\ast$ algebra of operators obtained through $Q$. If $Q$ did not depend on $\xi$, states would be elements of a certain $L^2$-space, over which the operators are self-adjoint. However, as in \cite{mostafazadeh2018energy}, we need to extend our space of states, which we will denote by $\mathcal{M}_Q$. For the sake of simplicity, we will consider only pure states.

 To provide $\mathcal{M}_Q$ with a PB structure we will use the geometrical formulation of Quantum Mechanics  \cite{Kibble79,Heslot,ashtekar1999geometrical} (generalized to QFT in \cite{David,David2}), where this structure is canonical. Thus,  $\mathcal{M}_Q$ will be represented by a Kähler manifold and quantum observables will be described as real sesquilinear functions of the quantum states in the form $F_A(\Psi)=\langle \Psi, \hat A \Psi\rangle$, for $\hat A$ an Hermitian operator on the Hilbert space,  which form a Poisson algebra with respect to the natural PB associated with the symplectic tensor of the Kähler structure. In addition, the quantization procedure links  states and operators with the following geometrical structures on  the space of classical matter fields, $\mathcal{M}_F$.
 
 Firstly, $\mathcal{M}_F$ is already endowed with a canonical symplectic form:
 \begin{equation}
 \label{omegaC}
\omega_C=-d\varphi_x\wedge d\pi^x\;,
 \end{equation}
which is $\xi$-independent, as so was the PB in \cite{HKT76} and the Poisson tensor ($\omega_C$'s (weak) inverse). Considering $T^*\mathcal{F}_C$ (with $\mathcal{F}_C=\mathcal{B} \times \mathcal{M}_F$ ) this independence can be written as:
\begin{equation}
\label{LeafIndependence}
\mathcal{L}_{X_B}\pi_F^*\omega_C=0,\quad\text{where}\quad \pi^*_F\omega_C\in (\wedge^2T^\star(\mathcal{M}_F\times\mathcal{M}_G\times\mathcal{M}_N))\;,
\end{equation}
here $\pi_F$ denotes the canonical projection on the cotangent bundle $T^*\mathcal{M}_F$ and the Lie derivative is taken with respect to any vector field, $X_B\in T\mathcal{B}$. Notice that we have chosen to see $\omega_C$ as a two form over the whole manifold just to define the independence as being constant along any curve on the base.
%, even if it only acts on vertical vectors (matter directions) of $\mathcal{F}_C$ with respect to the projection $\tau:\mathcal{F}_C\to \mathcal{B}$,
This independence comes from the fact that the fields $\varphi_x$ and their momenta $\pi^x$ are fundamental kinematical variables, considered independent of $\xi$ in the Hamiltonian formalism. In the following, this property will be referred to as \textit{leaf independence} for any tensor or function fulfilling the  same condition as the symplectic form  does in eq. \eqref{LeafIndependence}. Besides, the $(h,\pi_h)$-independence of $\omega_C$ is necessary for the PB given by the sum $\lbrace,\rbrace_G+\lbrace,\rbrace_M$ to fulfil Jacobi identity. This is the compatibility condition for \eqref{totalPB} mentioned in the previous section.

We may now add a complex structure, $J_C$, to $\mathcal{M}_F$ to form a Kähler structure. Such complex structure has no meaning at the classical level, but prepares the classical field theory for quantization.  In fact, all the possible choices of inequivalent quantizations (choices of inequivalent vacua and measures)  are reduced to the choice of such a classical complex structure \cite{Wald,ashtekar75}, whether one works with geometric quantization \cite{Oeckl} or with algebraic quantization \cite{CCQ04}. While $\omega_C$ was leaf-independent, $J_C$ is usually chosen to be $\xi$-dependent for physical reasons \cite{ashtekar75,agullo2015unitarity,CCQ04}. The complex structure is in Kahler compatibility $\forall\,\xi\in\mathcal{B}$, defining a $\xi$-dependent Riemannian form $\mu_C(\cdot,\cdot)=\omega_C(\cdot,J_C\cdot)$ and a Hermitian metric $h_C(\cdot,\cdot)=(\mu_C(\cdot,\cdot)+i\omega_C(\cdot,\cdot))/2$. This, in turn, allows the construction through Minlos's 
theorem \cite{minlos1959generalized} of a Gaussian measure $D\nu_\xi$ over the space of fields, and its unique identification through its characteristic functional:
\begin{equation}
\label{eq:measurednu}
C_p(\alpha,\gamma):=
\int\limits_{\mathcal{M}_F\sim\mathcal{N}^\prime\times\mathcal{N}^\prime}
    D\nu_\xi(\varphi^{\mathbf{x}},\pi^{\mathbf{x}} )
        e^{i(\alpha_x\varphi^x+\gamma_x\pi^x)}
        =\exp\Big({-\frac{h_C^{-1}[(\alpha,\gamma),(\alpha,\gamma)]}{2}}\Big)\;,
\end{equation}
where $\alpha_x,\gamma_x\in\mathcal{N}$ and we had to redefine the space of fields to contain distributions denoted by $\mathcal{N}^\prime$, instead of test functions, as otherwise a measure cannot be defined. We will come back to this later, but, in general, we will refer to \cite{David,David2} for the mathematical details. What is relevant for this work is that this $D\nu_\xi$ measure is $\xi$-dependent (hence, the subindex), as so was the Hermitian metric $\langle,\rangle_C$ through its dependence on $J_C$.  This Gaussian measure induces (at each $\xi$) also a measure $D\mu_\xi$ on a (Lagrangian) submanifold of the space of fields (for example, the submanifold $\mathcal{N}^\prime$ containing solely $\varphi^x$, but not $\pi^x$, or the holomorphic one used in section \ref{example}). The new measure is identified by its own characteristic functional, e.g. for the submanifold of real fields $\varphi^x$, it is given by: 
\begin{equation}
\label{eq:measuredmu}
C(\alpha):=
\int\limits_{\mathcal{N}^\prime}
    D\mu_\xi(\varphi^\mathbf{x})
        e^{i\alpha_x\varphi^x}
        =\exp{\Big(-\frac{\alpha_x\Delta^{xy}\alpha_y}{4}\Big)}\quad\forall\,  \alpha_x\in\mathcal{N}\;,
\end{equation}
Where the bilinear $\Delta^{\mathbf{xy}}/2$ is called the covariance of $D\mu_\xi$ and it is extracted from $h_C^{-1}$ (and, thus, $J_C$ dependent) and the particular Lagrangian submanifold used in the description. See \cite{David2} and the example in sec. \ref{example} for further details in the construction. This defines the set of Hilbert spaces, $L^2(\mathcal{N}^\prime,D\mu_\xi)$, to which the quantum states must belong at each $\xi$, and which are not unitarily equivalent for different $\xi\in\mathcal{B}$. Therefore, each $D\mu$ in turn, defines a Hermitian metric $\langle\cdot,\cdot\rangle_Q$ for said quantum states such that, for any two vectors $\Psi_1$ and $\Psi_2$ of one of this Hilbert spaces, we have:
\begin{equation}
\langle\Psi_1,\Psi_2\rangle_\xi:=\int\limits_{\mathcal{N}^\prime} D\mu_\xi\bar\Psi_1(\bar\phi)\Psi_2(\phi)
\end{equation}
where here, $\phi$ denotes, in general, the domain of the functions of this Hilbert space. Therefore, $\langle,\rangle_\xi$ denotes a smoothly $\xi$-dependent Hermitian metric at the quantum level. As in \cite{mostafazadeh2018energy}, we can define a Hermitian bundle and provide it with an Hermitian connection (for these tools of complex differential geometry, see e.g. \cite{kobayashi2014differential}).

Note however that these $L^2$ spaces are not unitarily equivalent. Nevertheless, they all have a common dense subset given by Hida test functions $(\mathcal{N})$, and a common superset (in which they are dense) given by Hida distributions, denoted by $(\mathcal{N})^\prime$, which are common tools in white noise theory \cite{Hida}. This allows us to consider at each $\xi$ a Gelfand triple $(\mathcal{N})\subset L^2(\mathcal{N}^\prime,D\mu_\xi)\subset(\mathcal{N})^\prime$, allowing us to define on $(\mathcal{N})$ differential structures (as it is a Frechet nuclear space \cite{kriegl1997}) and densely extend them by linearity all the way up to $(\mathcal{N})^\prime$, (see the notes in the supp. material, based on \cite{David,David2}). In the current work, this is necessary for the definition of the quantum PB and to be able to jointly represent several non-unitarily equivalent Hilbert spaces inside a common superspace. Thus, instead of the Hermitian bundle with a Hilbert space in the fiber, we consider a fibration $\tau_{\mathcal{F}}:\mathcal{F}\to\mathcal{B}$, where the base manifold is defined as $\mathcal{B}:=\mathcal{M}_G\times\mathcal{M}_N$ and the fiber is given by $(\mathcal{N})^\prime$, which is big enough to contain all the possible Hilbert spaces $L^2(\mathcal{N}^\prime,D\mu_\xi)$ we must consider. This fibration is, only in a dense sense, a Hermitian bundle, as $\langle,\rangle_Q$ is defined only on a dense subset $D(\langle,\rangle_\xi)\subset (\mathcal{N})'$ different for each point of the base and representing the Hilbert space of physical quantum states at that $\xi$, but all of them containing the common dense subset $(\mathcal{N})$. This implies that $(\mathcal{N})'$ is too big to represent efficiently the quantum states.

As in the quantum mechanics case of \cite{mostafazadeh2018energy} we will bind the representation of quantum states to be covariant sections for the (local on $\xi$ representation of the) Hermitian connection, assigning an element of a (possibly different) Hilbert space to each point of the base. As a result, the quantum states will be explicitly $\xi$-dependent.
%derivative w.r.t. the leaf variables $\xi\in\mathcal{B}$ will be key for the quantum supermagnitudes to fulfill the CR of Dirac's group, eqs. \eqref{closeHH},\eqref{closeMM},\eqref{sMM-sHM}.\\
In section \ref{example}, it is discussed that, in geometric quantization, this section nature arises as early as in the definition of functions adapted to the polarization.
\subsection{Functional representation of quantum states and covariant sections.}
Returning to our fibration, $\tau_{\mathcal{F}}:\mathcal{F}\to \mathcal{B}$  which is locally trivialized into subsets of
\begin{equation}
\mathcal{B}\times(\mathcal{N})^\prime\quad\text{with base}\quad\mathcal{B}:=\mathcal{M}_G\times\mathcal{M}_N\;,
\end{equation}
we realize that the fiber is adapted to the usual  functional description of quantum states (Hida distributions were chosen for this purpose also in \cite{David2}). Thus, we have all the geometric structures necessary for quantum mechanics defined differently at each point $\xi\in\mathcal{B}$ and we need a way to relate them based on the Hermitian connection.  Let us introduce the notion of connection in our fibration. Let $\gamma^M(s)$ be a curve in $\mathcal{B}$. We ask ourselves how our representation of a quantum state $\Psi_0$ transforms along $\gamma^M$. To answer that, we must lift the curve on the base to the whole $\mathcal{F}$, such that it projects on $\gamma^M$ by the projection $\tau_{\mathcal{F}}$. This requires a definition of connection such that the \textit{horizontal} directions are defined locally on $\mathcal{F}$. With this connection, the notion of horizontal lift is well defined and it provides a locally unique curve $\gamma^{\mathcal{F}}$ on $\mathcal{F}$ such that $
\gamma^M(s=0)=\xi_0; \quad \gamma^{\mathcal{F}}(s=0)=\Psi_0; \quad \tau_{\mathcal{F}}\left (\gamma^{\mathcal{F}}(s)\right )=\gamma^M(s)$.

\begin{figure*}[h]
        \centering
        \includegraphics[height=8cm]{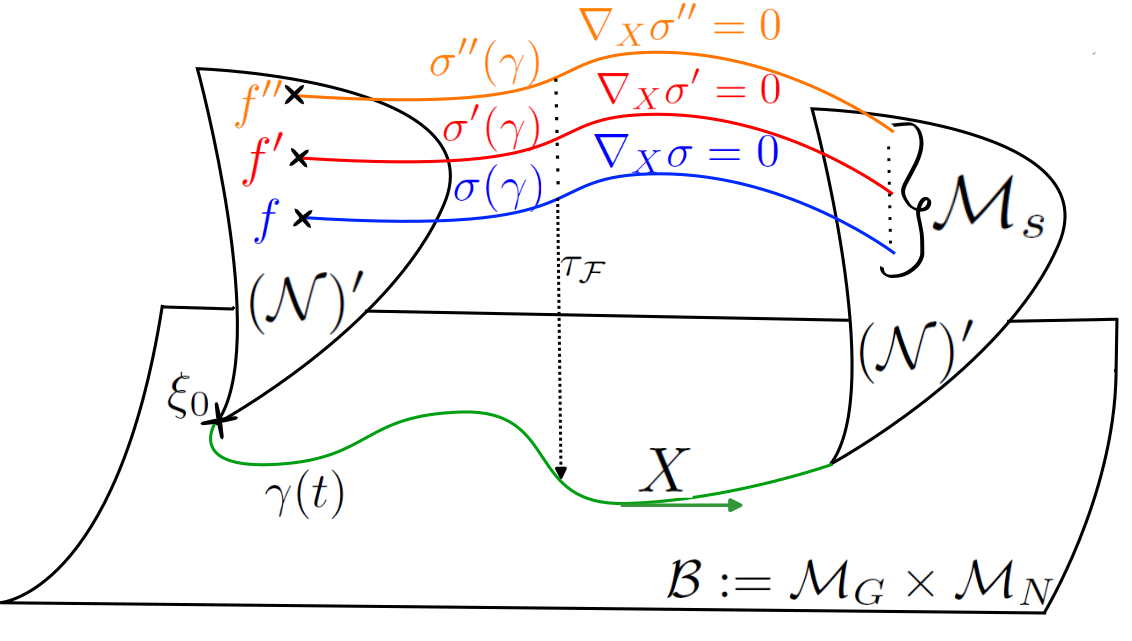}
        \caption{Fibration with base $\mathcal{B}$ and fiber $(\mathcal{N})^\prime$, and the set of parallel transported sections  over it, $\mathcal{M}_s$, associating a Hida distribution to each point of a curve $\gamma$ along $\mathcal{B}$. The sections are defined for different points of the fibration ($f,\;f^\prime,\;f^{\prime\prime},\;\cdots$) fulfilling $\tau_{\mathcal{F}}(f)=\tau_{\mathcal{F}}(f^\prime)=\tau_{\mathcal{F}}(f^{\prime\prime})=\xi_0$ and the covariance condition, $\nabla_X\sigma=0\;\forall X\in\mathfrak{X}(\mathcal{B})$. }
        \label{fig:StatesAsSections}
        \centering
        \vspace{0.5cm}
        \includegraphics[width=13cm]{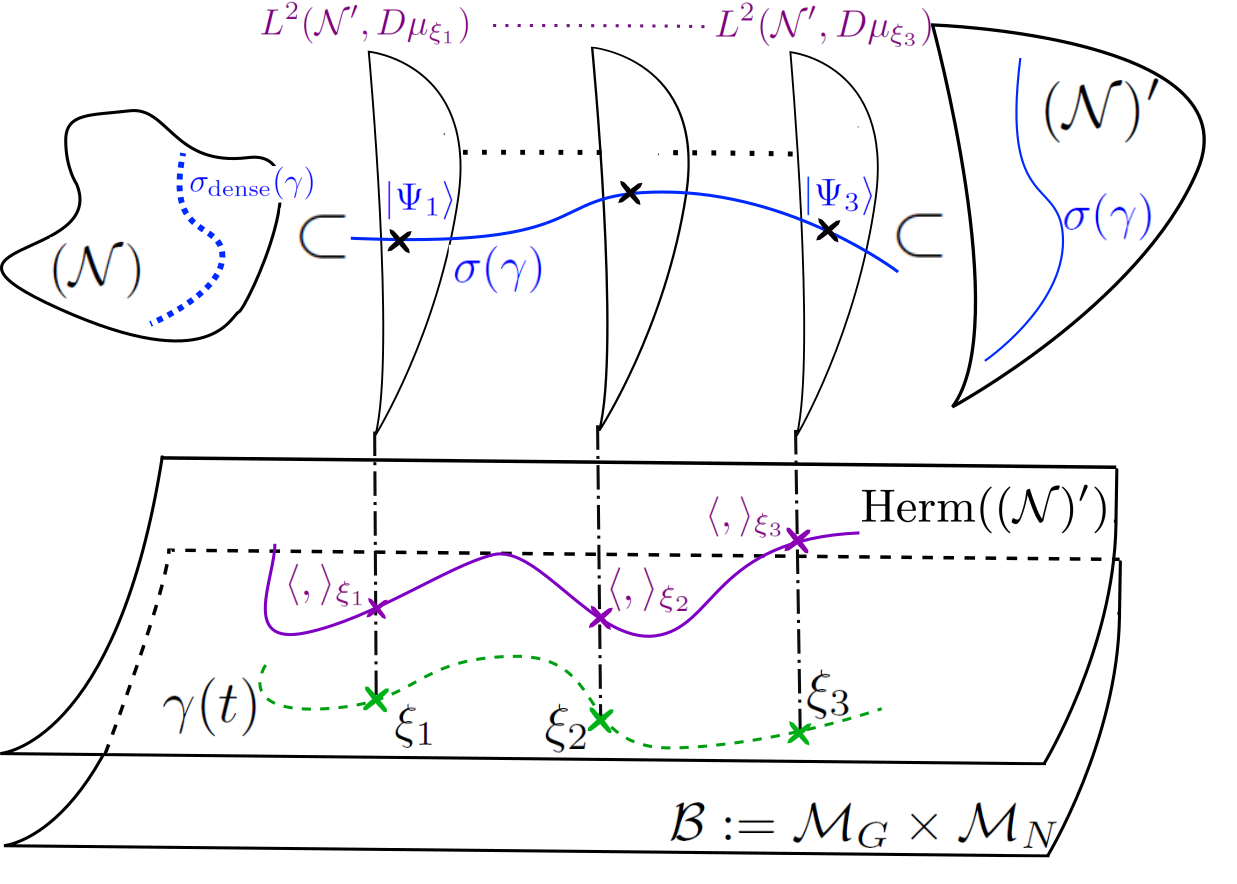}
        \caption{One of the sections of figure \ref{fig:StatesAsSections}, seen locally on $\xi_i=\gamma(t_i)$  as $\mid\Psi_i\rangle$, a vector inside a different Hilbert space at each point of the base. Such Hilbert spaces are defined depending on $\xi\in\mathcal{B}$, given the $\xi$-dependence of the Hermitian product $\langle,\rangle_\xi\in \text{Herm}((\mathcal{N})^\prime)$ (represented by an intermediate layer). Thus,  each $L^2$  forms a different Gel'fand triple with $(\mathcal{N})^\prime$ (containing the union of all the $L^2$ spaces needed) and $(\mathcal{N})$ (the dense common subspace to all of them). Thus, the curve on it represents (under the  scalar product, $\langle,\rangle_{\gamma(t)}$) a dense representation as weight-1-densities of the curve of distributions on $(\mathcal{N})^\prime$.}
        \label{fig:GelfandTriples}
\end{figure*}

In a local description, the connection is associated with a one-form $\Gamma\in \Lambda^1(\mathcal{B})\times \mathrm{Lin}((\mathcal{N}))$, (being $\mathrm{Lin}((\mathcal{N}))$ linear operators whose domain contains Hida test functions) which allows us to define the covariant derivative of a section  $\sigma:\mathcal{B}\to \mathcal{F}$, using the parallel transport defined through the horizontal lift of base curves:
$$
\nabla_X \sigma(\xi)=d\sigma(X)-\hat\Gamma(\xi, X)\sigma(\xi), \quad X\in \mathfrak{X}(\mathcal{B})
$$
In particular,  covariant sections with respect to the connection, $\nabla_X \sigma(\xi)=0,$ define a sequence of vectors describing the change suffered by the quantum state with an initial state $\sigma(\xi_0)=\Psi_0$ due solely to the change on the quantum structures determined by a change of $\xi\in\mathcal{B}$, when no quantum evolution is exerted on it. In other words, this defines the evolution of the $\xi$-dependent representation of a quantum state due to the evolving geometry, but under a vanishing (matter) Hamiltonian operator. If we consider a non-trivial Hamiltonian operator $\hat H$ (or, in particular, the Hamiltonian representation of a certain deformation of the hsf. $\Sigma$), the total evolution of the quantum state will be the sum of the image under a covariant section of the geometric transformation (thus, a horizontal change, with respect to the connection) and the effect of Schrödinger functional equation, which is vertical on the fiber  of $\mathcal{F}$. Therefore, it is relevant to define the subset $ \mathcal{M}_{\text{s}}$ of sections on the bundle $\mathcal{F}$ which are covariant for all vector fields on the base of the fibration, $X\in  \mathfrak{X}(\mathcal{B})$ :
\begin{equation}
 \mathcal{M}_{\text{s}}:=\lbrace\sigma\in\Gamma(\mathcal{F})\,\vert\, \nabla_X\sigma=0\  \forall\, X\in \mathfrak{X}(\mathcal{B}) \rbrace 
\end{equation}
 In terms of these objects (which are illustrated in Figure \ref{fig:StatesAsSections}) we may consider the horizontal distribution on $T\mathcal{F}$ representing the constraints of the transformations of quantum states to be compatible with geometry transformations, given the leaf dependence of the Hilbert space to which they belong. It is defined by the horizontal vector fields on $\mathcal{F}$, or, equivalently, by the directions which are tangent to the former covariant sections taking values at a given point of the fibration $f\in \mathcal{F}$:
\begin{comment}
ss
\begin{multline}
\mathrm{Hor}^\nabla(f)=\left \{ T\sigma_{\tau(f)} (\mathfrak{X}(\mathcal{B})) \quad  f\in \mathcal{F}\, \vert \quad\right . \\ \left .\,  \forall \, \sigma(\tau(f))=f; \, \, \nabla_X\sigma=0 \quad \forall X\in  \mathfrak{X}(\mathcal{B})  \right \},
\end{multline}
\end{comment}
\begin{equation}
\mathrm{Hor}^\nabla(f)=\left \{ T\sigma_{\tau(f)} (\mathfrak{X}(\mathcal{B})),\, \;  \forall\, \sigma\in\mathcal{M}_s\vert\;\sigma(\tau(f))=f \right \},
\end{equation}
where $T\sigma_{\tau(f)}(\mathfrak{X}(\mathcal{B}))$ represents the image of the differential of the mapping $\sigma:\mathcal{B}\to \mathcal{F}$ evaluated at the point on the base given by the projection $\tau(f)\in \mathcal{B}$ acting on all vector fields on the base manifold. Obviously,  the tangent space on a point of the fibration can be decomposed as:
\begin{equation}
\label{eq:decomposition}
T_f\mathcal{F}=V_f\mathcal{F}\oplus \mathrm{Hor}^\nabla(f); \qquad \forall f\in \mathcal{F},
\end{equation}
where $V_f\mathcal{F}$ represents the set of vertical vectors at $f$. This definition of $\mathrm{Hor}^\nabla(f)$ provides us with a way of defining  tangent vectors to curves on $\mathcal{F}$ that are only sensitive to trajectories  on the geometrodynamical data, i.e. they are horizontal.  Thus, it is also physically intuitive that tangent elements to physical states should belong to the tangent bundle $T\mathcal{F}$ that locally, $T_f\mathcal{F}$, can be decomposed as in eq \eqref{eq:decomposition}, where $V_f\mathcal{F}$ represent infinitesimal generators of inner transformations to a given Hilbert space associated to the measure $D\mu_{\tau(f)}$ while $\mathrm{Hor}^\nabla(f)$ represent generators of  isometries  between non-unitarily-equivalent Hilbert spaces characterized by measures $D\mu_\xi$, $D\mu_{\xi^\prime}$ associated to infinitesimally close base points, $\xi,\xi^\prime\in\mathcal{B}$.  In this sense, under the complete evolution,  the local quantum state $ \Psi_f \in L^2(\mathcal{N}', D\mu_f)$ will   transform into $\Psi_{f'}$ for $f'\in \mathcal{B}$, which will be outside the original Hilbert space $L^2(\mathcal{N}', D\mu_f)$, and inside a new Hilbert space $L^2(\mathcal{N}', D\mu_{f'})$. This is why the fiber of $\mathcal{F}$ must be Hida distributions, as it is a space that contains all the $L^2(\mathcal{N}^\prime,D\mu_\xi)$ through which the quantum state will evolve following the (lift of the) curve over $\mathcal{B}$. Remember, nevertheless, that $(\mathcal{N})^\prime$ it is too large to accommodate solely the physical quantum states and one should restrict (locally) the construction to a suitable subspace of it  such as $(\mathcal{N})$ (see Figure \ref{fig:GelfandTriples} for a graphical summary of this construction).

\begin{figure*}[h]
        \centering
        \includegraphics[width=17cm]{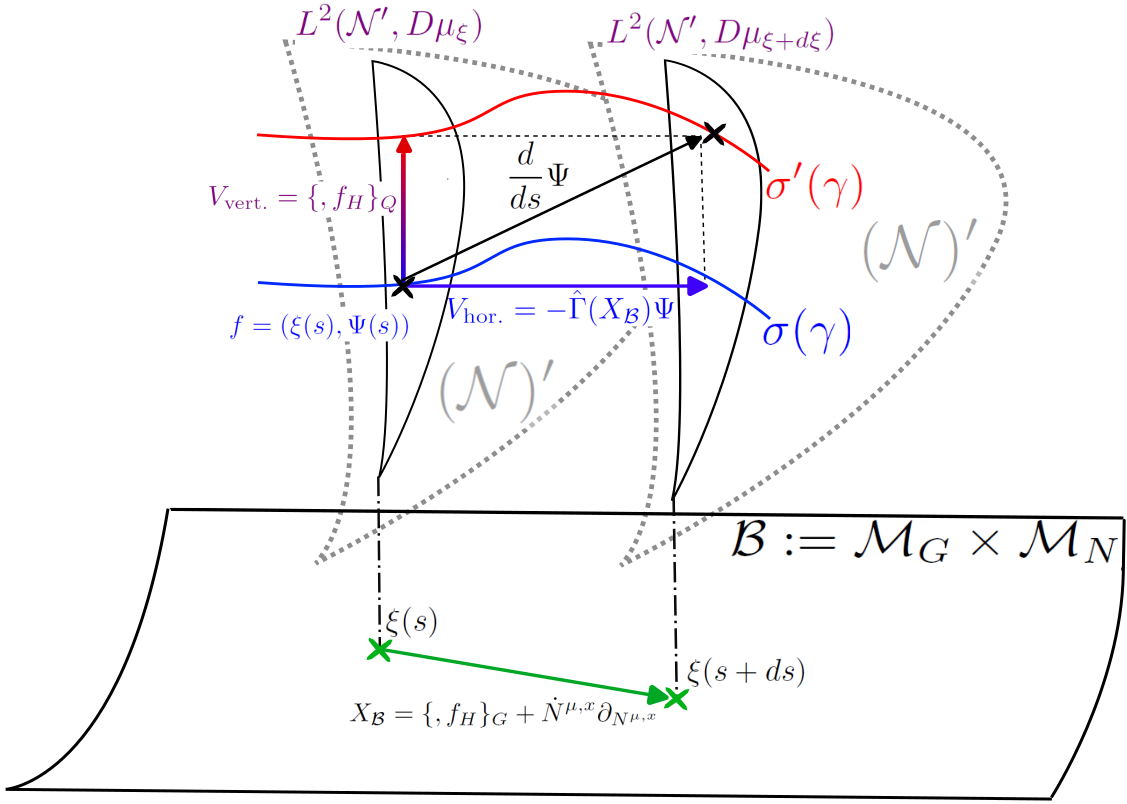}
        \caption{Illustration of the decomposition of a tangent vector to the bundle in its vertical and horizontal decomposition. In particular, we are considering the infinitesimal evolution of the hybrid states, between two infinitesimally close hsfs., $\Sigma_s\rightarrow\Sigma_{s+ds}$. Thus the vertical part is given by the ``solely quantum" evolution, i.e. $V_{\text{vert.}}(\Psi)=\lbrace \Psi,f_H\rbrace_Q=\dfrac{-i}{\hbar}\hat{H}\Psi$. This part of the dynamics is internal for the original Hilbert space, for the initial $\xi(s)\in\mathcal{B}$, and can be seen as a change between the covariant sections (from black to red) in $\mathcal{M}_s$, without changing the evaluation point. On the other hand, the horizontal one lifts the tangent field on the base through the connection. Therefore, it does not change from one covariant section to another, but it provides the change of Hilbert space due to the change of evaluation point to $\xi(s+ds)$. The sum of both of them conform $\frac{d\Psi}{ds}$, which is a generic change of quantum state, changing both covariant section \textit{and} evaluation point. Remember that the change of Hilbert spaces forces us to see the quantum state inside different (for each $\xi$) subsets (the $L^2$ spaces) of $(\mathcal{N})^\prime$, as it engulfs all the $L^2$ spaces,  which is why it is chosen to be the fiber of the given bundle (represented faintly in the background).}
        \label{fig:HorizontalVertical}
\end{figure*}

\begin{comment}

{\color{green}
¿Este parrafo es necesario?

It is important to remark that the construction is, of course, not unique. Every choice of a connection for $\mathcal{F}$, defines a different notion of covariant derivative. Choosing a different connection implies considering equivalent different quantum states for different geometric points (i.e,. different evolutions of the quantum states because of the changing geometry). Nonetheless, as the construction is geometrical, each connection would define a horizontal lift to the bundle of complex structures, scalar products, linear operators or any other tensor chosen to define a fiber on $\mathcal{B}$. \sout{A natural consistency condition is to choose the same connection for all the tensors.} When doing that, it makes sense to consider normalized quantum states for all geometrical points, since the scalar product and the states can be chosen to transform covariantly with respect to the same connection (see below).  
}
\end{comment}
With these tools, as we anticipated, the quantum states must be defined as the {\color{black} images of the parallel transported sections with respect to the connection:
\begin{comment}
\begin{multline}
\label{statesaresections}
\mathcal{M}_Q:=\left\{ \sigma_\Psi (\xi)\quad \forall \xi \in\mathcal{B},  \forall \sigma_\Psi \in \Gamma \mathcal{F}\text{ such that}, \right . \\ \left . \nabla_X \sigma_\Psi=0\quad \forall X\in \mathfrak{X}(\mathcal{B}) \right\}\;.
\end{multline}
\end{comment}
\begin{equation}
\label{statesaresections}
\mathcal{M}_Q:=\left\{\Psi:=\sigma_\Psi (\xi)\in(\mathcal{N})^\prime
\,
\vert\,
\xi \in\mathcal{B},\sigma_\Psi \in \mathcal{M}_s\right\}
\end{equation}
From Equation \eqref{eq:decomposition}, we can safely consider that $\mathcal{M}_Q$ is isomorphic to the fiber of $\mathcal{F}$, but, presented like this, the dependence of the states with the base point is more clear:
\begin{equation}
\label{statesdependonxi}
\partial_{\xi_0}\Psi=\left(\partial_{\xi}\sigma_\Psi(\xi)\right)\vert_{\xi_0}=-\hat\Gamma(\xi_0)\Psi\;.
\end{equation}
}
{\color{black}
 \footnote{Note that we work with the notation $\partial_\xi\sigma_\Psi(\xi)=-\hat\Gamma(\xi)\sigma_\Psi(\xi)$, to denote the connection associated to any local variable of the base. This is just a way to alleviate the notation for the covariant derivative along a general direction on the base given by $v_B^x:=(\dot{h}_{ij;x}\partial_{h_{ij;x}}+\dot{\pi}^{ij;x}\partial_{\pi^{ij;x}}+\dot{N}^x\partial_{N^x}+\dot{N}^{ix}\partial_{N^{ix}})\in T\mathcal{B}$, which we would denote by:
$$
v^{i;x}_B \partial_{\xi^{i;x}}\sigma_\Psi(\xi)= -(\dot{h}_{ij;x}\hat\Gamma_{h_{ij;x}}+\dot{\pi}^{ij; x}\hat\Gamma_{\pi^{ij;x}}+\dot{N}^x\hat\Gamma_{N^x}+\dot{N}^{ix}\hat\Gamma_{N^{ix}})\sigma_\Psi
$$
where, as before, the contraction of $i,j$ indices implies summation, while the contraction of  the continuous $x$-indices implies integration over $\Sigma$.}
Thus, with a covariant section $\sigma\in\mathcal{M}_s$ and a point $\xi\in\mathcal{B}$ characterizing geometry, we obtain the \textit{instantaneous} quantum state as $\Psi=\sigma(\xi)$. This will be our quantum Cauchy data and we know that it transforms as $\nabla_\xi\Psi=0$. This must not be confused with the \textit{history} of the quantum state, i.e. the solution curve to the dynamics. As part of the evolution is vertical the \textit{history} can not be covariant. Following the former isomorphism between $\mathcal{M}_Q$ and $\mathcal{B}\times\mathcal{M}_s$, the history can be represented as a curve over $\mathcal{M}_s$ (the vertical part of the evolution are the changes from one covariant section to another) together with a curve over $\mathcal{B}$ (the horizontal part are the changes of the evaluation point for the covariant sections), yielding together the evolution of $\Psi$. These properties will be relevant in the following section and are illustrated in Figure \ref{fig:HorizontalVertical}.}
\begin{comment}
\sout{
We have generically denoted $\partial_\xi\sigma_\Psi(\xi)=-\hat\Gamma(\xi)\sigma_\Psi(\xi)$, but, what we mean by this is that we have a \textit{local}  connection for each local variable in the base:}
%\begin{gather}
%\partial_{h_{ij}(x)}\sigma_\Psi(\xi)=-\hat\Gamma_{h_{ij}}(\xi;x)\sigma_\Psi(\xi)\\
%\partial_{\pi^{ij}(x)}\sigma_\Psi(\xi)=-\hat\Gamma_{\pi^{ij}}(\xi;x)\sigma_\Psi(\xi)\\
%\partial_{N(x)}\sigma_\Psi(\xi)=-\hat\Gamma_{N}(\xi;x)\sigma_\Psi(\xi)\\
%\partial_{N^i(x)}\sigma_\Psi(\xi)=-\hat\Gamma_{N^i}(\xi;x)\sigma_\Psi(\xi)
%\end{gather}
\end{comment}

%Remember, though, that even if the covariant sections are just a tool to define the horizontal distribution $\mathrm{Hor}^\nabla$ on the bundle $\mathcal{F}$, they represent the transformations of the quantum states under the effect of pure geometrodynamical (and lapse and shift) transformations. Hence, they represent the kinematics of the quantum field theory in the geometrical background where it is defined.
%Note that, by definition, the each covariant section, as defined above, once evaluated on the point of the base characterizing the geometrodynamical data $\xi_0\in\mathcal{B}$ on the current $\Sigma_s$, yield a certain element of $\mathcal{M}_Q$, isomorphic to an element of the fiber $(\mathcal{N})^\prime$ which is a valid quantum state.\\

Note that, at the level of QFT in a given Curved Spacetime, once the foliation is chosen, the evaluation point $\xi_0$ is a given time dependent parameter. This is the object of study of \cite{David2}. In  section \ref{sec:4}, in which we construct hybrid geometrodynamics, we promote such evaluation point to be part of the kinematics, extending the phase space to contain the quantum sections and the base point over which they must evaluated. The objective of the following construction is to derive the constraints on the Hermitian connection $\hat\Gamma$ that are necessary in order to have hybrid geometrodynamics, providing a physical compatibility condition for the $\xi$-parametric family of quantizations.

\textit{Physical remark.} The part of the dynamics of $\Psi$ in which the connection is involved  corresponds to the directions at $T_\Psi \mathcal{F}$ which are horizontal with respect to the connection. This is just an infinitesimal change of \textit{representation}, because of the change of  the quantization procedure with $\xi$, of the same abstract quantum state. Obtaining the integral curve for $\Psi$ for such tangent vectors defines series of quantum states which are equivalent from the probabilistic point of view (for the $\xi$--dependent measures), since they correspond to changes in their representation due to the geometry only.
{\color{black}
\subsection{Poisson algebra of quantum observables and preservation of norm.}
}
Classical functionals on $\mathcal{M}_F$ are usually mapped through quantization $Q$ to be self-adjoint operators over a Hilbert space.  Given that $Q$ depends on $J_C$, it becomes leaf dependent. As for states, we can define a bundle with base manifold $\mathcal{B}$ and fiber the space of linear operators on a different Hilbert space at each $\xi$ associated to $\langle,\rangle_\xi$. As before, we will consider the covariant sections for the Hermitian connection, providing the horizontal directions. In fact, one must realize that the notion of adjointness is referred to the leaf dependent scalar product $\langle,\rangle_\xi$. More explicitly, we denote  by $\hat A^{\dagger_\xi}$ the adjoint of an operator $\hat A$ for the scalar product $\langle,\rangle_\xi$, given by $D\mu_\xi$, so that:
\begin{equation}
\label{adjointbasedependent}
\langle\Psi_1\mid\hat A\Psi_2\rangle_\xi=\langle\hat A^{\dagger_\xi}\Psi_1\mid\Psi_2\rangle_\xi\quad\text{but}\quad\langle\Psi_1\mid\hat A\Psi_2\rangle_{\xi_2}\neq\langle\hat A^{\dagger_\xi}\Psi_1\mid\Psi_2\rangle_{\xi_2},
\end{equation}
i.e., for $\langle,\rangle_{\xi_2}$ with $\xi_2\neq\xi$ (given by  a different Gaussian measure $D\mu_{\xi_2}$), $\hat A^{\dagger_\xi}$ no longer defines the adjoint operator.
Thus, the $\dagger$ operator and the set of self-adjoint operators ($\hat A=\hat A^\dagger$) are $\xi$-dependent as so was the Hilbert space. For example, in the functional representation, $Q(\pi^x)$ is given by a covariant derivative which must be self-adjoint w.r.t. $D\mu_\xi$ and a vacuum phase \cite{CCQ04}, which adds a $\xi$-dependent imaginary multiplicative term (see section \ref{example} for the geometric quantization analogue).
\begin{comment}
The necessity for this sectional nature of the quantum states can be further justified under the prism of the $C^\star$-algebras. For the observables under consideration, we may think of the leaf dependence of the former structures as a $\xi$-parametric family of representations of such $C^\star$-algebra: both operators and the scalar product(from which the statistical notion of EV is taken) depend on $\xi$, and, as a natural consequence, if we identify a certain abstract quantum state with the set of probabilities assigned to possible results of physical measurements of observables, such object acquires a  $\xi$-parametric family of  functional representations $\sigma_\psi:\mathcal{B}\rightarrow (\mathcal{N})^\prime$.\\
\end{comment}

With this in mind, we define the quantum algebra of observables $\mathcal{A}_Q$ (as in \cite{Kibble79} and, for the hybrid case, \cite{Alonso2012}) as the set of functions $f:\mathcal{M}_Q\rightarrow\mathbb{R}$ given by the expectation value (EV) of linear operators over $\mathcal{M}_Q$, which are self-adjoint for the scalar product $\langle,\rangle_\xi$ employed in the EV. We denote the space of such operators as $\mathrm{Herm}_\xi(\mathcal{M}_Q)$. Thus:
\begin{equation}
\label{eq:AQ}
\mathcal{A}_Q:=\lbrace f_A(\Psi;\xi):=\langle\Psi\mid\hat A(\xi)\mid\Psi\rangle_\xi\,\vert \quad \forall \hat A\in \mathrm{Herm}_\xi(\mathcal{M}_Q)\rbrace
\end{equation}
Notice that the elements of $\mathcal{A}_Q$ inherit a threefold $\xi$-dependence from the complex structure $J_C$. The first one arises from the sectional nature of the states,  eq. \eqref{statesdependonxi}. The second one comes from the scalar product, as $\langle,\rangle_\xi$ is associated to the leaf dependent Gaussian measure $D\mu_\xi$. Lastly, the representation of operators is also $\xi$-dependent, because self-adjointness is referred to $\langle,\rangle_\xi$. We explicitly denote this as $\hat A=Q_\xi(A)$, where $Q_\xi(A)$ is a certain $\xi$-dependent quantization (see section \ref{example}) of a classical field function $A\in C^\infty(\mathcal{M}_F)$ yielding a self-adjoint operator for the $\xi$-dependent vacuum phase and measure. As the whole construction is tensorial and can be traced back to the Hermitian metric $\langle,\rangle_\xi$, all three dependencies must have a null covariant derivative w.r.t. the Hermitian connection, as we will see below.

There is a fourth leaf dependence of the elements of the algebra, but this one is natural of the classical field magnitudes before quantization, as they are assumed to have already been constructed in conjunction with geometrodynamics, so we may quantize (the matter degrees of freedom of) a function $A\in C^\infty(\mathcal{F}_C)$. For example, the classical field theoretical Hamiltonian in \eqref{classicalfieldhamiltonian} was lapse, shift and 3-metric dependent. Once quantized, it maintains such dependence and acquires the three former ones.

The set $\mathcal{A}_Q$ is a Poisson algebra for the Quantum PB (QPB) $\lbrace,\rbrace_Q$ that reproduces the Lie bracket of operators \cite{Kibble79,David2,Alonso2012}. For any two elements $f_A,f_B\in\mathcal{A}_Q$, defined as $f_A:=\langle\Psi\mid\hat A\mid\Psi\rangle$ and $f_B:=\langle\Psi\mid \hat B\mid\Psi\rangle$, its QPB is given by:
\begin{equation}
\label{QPB}
\lbrace f_A,f_B\rbrace_Q=\dfrac{-i}{\hbar}\langle\Psi\mid [A,B]\mid\Psi\rangle=-\hbar^{-1}f_{i[A,B]}\;.
\end{equation}
 For future use, we define the algebra resulting under the  completion of $\mathcal{A}_Q$ under the ordinary product of functions:
\begin{equation}
\label{completedalgebra}
\bar{\mathcal{A}}_Q=\lbrace \prod\limits_{i=1}^n f_{A_i}\,\vert\, f_{A_i}\in\mathcal{A}_Q \quad\forall i\in[1,n];\forall n\in\mathbb{N}\rbrace
\end{equation}

While the elements of $\mathcal{A}_Q$ are $\xi$-dependent, the PB itself must be $\xi-$independent:
\begin{equation}
\label{leafindependenceQPB}
\partial_\xi \lbrace f,g\rbrace_Q=\lbrace\partial_\xi f,g\rbrace_Q+\lbrace f,\partial_\xi g\rbrace_Q\ \forall \xi\in[h,\pi,N,N^i]
\end{equation}
This condition is physically justified. Firstly, to have the  compatibility condition for the hybrid Poisson bracket as for \eqref{totalPB}, which was discussed at the level of symplectic forms below eq. \eqref{LeafIndependence},  the QPB must be, at least, $h,\pi_h$-independent, so the sum  $\lbrace f,g\rbrace_G+\lbrace f,g\rbrace_Q$ fulfils Jacobi identity $\forall f,g\in C^\infty(\mathcal{F})$. This is necessary to find a simple Hamlitonian representation on $\mathcal{F}$ of hsf. deformations.  Besides, their Hamiltonian nature implies that hsf. deformations are symplectomorphisms. Naturally, hsf. deformations modify $h,\pi_h$, and so, if the PB (related with the Poisson bivector, weak inverse of the quantum symplectic structure $\omega_Q$) depended on them, $\omega_Q$ would change under hsf. deformations, breaking the symplectomorphic nature. On the other hand, the kinematical structure of the theory should be independent of $N$,$N^i$ in order to implement path independence, so that additional dependencies on the particular choice of foliation do not arise beyond the usual one for the Hamiltonian generating function, as in \eqref{totalHamiltonianClassical}. If done otherwise, the CR would depend on $N$ and $N^i$, and thus, such functions could not be regarded as Lagrange multipliers for the constraints. In fact, even if regarded as a design choice, in this construction we aim to build the kinematics \textit{before} the choice of dynamics, and the lapse and shift functions characterize the dynamics, defining the vector field spamming the foliation. All these arguments imply \eqref{leafindependenceQPB}.

In turn, \eqref{leafindependenceQPB} together with \eqref{QPB} imply a relationship between $\hat\Gamma$ and the $\xi$-dependence for operators and $\langle,\rangle_\xi$: $\hat\Gamma$ must indeed  be a Hermitian connection for $\langle,\rangle_\xi$. For simplicity, we will use the bra-ket notation. As the construction is tensorial, we can consider the covariant derivatives of different types of sections (of different bundles):
\begin{itemize}
\item For the quantum states, given by $(1,0)$ tensors (evaluation on {\color{black} $\xi\in\mathcal{B}$} of sections {\color{black} in $\mathcal{M}_s$}), their covariant definition implies \eqref{statesdependonxi}  with $\Psi=\vert\Psi\rangle$. We assume that the connection does not mix the bra-ket representation, as the one in section \ref{example}.
\item For their duals $\langle\Psi\mid$, we can derive its $\xi$-dependence. Now, horizontal directions correspond to tangents to sections of the bundle $\mathcal{F}^*$, the dual bundle of $\mathcal{F}$. From \eqref{adjointbasedependent} we know that the notion of adjoint is leaf dependent, as it depends on $\langle,\rangle_\xi$. We define the adjunction $C_\xi$ for a given $\xi\in\mathcal{B}$ as $C_\xi(\mid\Psi\rangle_\xi):=\langle\Psi\mid_\xi$, and analogously, $C_\xi(\hat A\mid\Psi\rangle)=\langle\Psi\mid{\hat A}^{\dagger_\xi}$. Notice that $C_\xi$ mixes maximally the bra-ket representation and thus we must proceed with care{\color{black}, as it is a bundle isomorphism $C:\mathcal{F}\to \mathcal{F^*}$.} With {\color{black}these} considerations:
\begin{equation}
\partial_\xi \langle\Psi\mid_\xi=\partial_\xi C_\xi(\mid\Psi\rangle_\xi)= C_\xi(\partial_\xi \mid\Psi\rangle_\xi)+\langle\Psi\mid_\xi\hat T(\xi)=-\langle\Psi\mid{\hat\Gamma}^{\dagger_\xi}_{\xi}+\langle\Psi\mid_\xi\hat T=-\langle\Psi\mid\hat\Gamma_{\xi}^{T}
\end{equation}
where we have defined $\hat T(\xi):=(C_\xi^{-1}\partial_\xi C_\xi)$ as the object of adjointness transport, and we have used eq. \eqref{statesdependonxi} in the second equality. Finally, 
we defined $\hat\Gamma_{\xi}^{T}:={\hat\Gamma}^{\dagger_\xi}_{\xi}-\hat T$. 
\item Linear operators, denoted by $\hat A$, are $(1,1)$ tensors whose construction through $Q_\xi$ is $\xi$-dependent. Therefore, we can write the derivative of the EVs of operators as:
\begin{equation}
\label{Aprime}
\partial_\xi f_A=\langle\Psi\mid \hat A^\prime \mid\Psi\rangle\quad\text{where}\quad \hat A^{\prime}:=\left(-\hat A\hat\Gamma_{\xi}-\tilde\Gamma_{\xi}^{T}\hat A+\partial_\xi(\hat A)\right)\;.
\end{equation}
\end{itemize}
Now we return to \eqref{leafindependenceQPB}, and see that its left hand side yields:
\begin{equation}
\label{leftsidebrackets}
-\hbar^{-1}\partial_\xi f_{i[\hat A,\hat B]}=-\hbar^{-1}\langle\Psi\mid (i[\hat A,\hat B])^\prime \mid\Psi\rangle\;,
\end{equation}
while the right hand side yields:
\begin{equation}
\label{rightsidebrackets}
\lbrace \partial_\xi f_A,f_B\rbrace_Q+\lbrace  f_A,\partial_\xi f_B\rbrace_Q= -\hbar^{-1}\langle\Psi\mid i([\hat A^\prime,\hat B]+[\hat A,\hat B^\prime])\mid\Psi\rangle\;.
\end{equation}
With these expressions to fulfill \eqref{leafindependenceQPB}, we must have:
\begin{equation}
\label{antiselfadjointconnection}
[\hat A^\prime,\hat B]+[\hat A,\hat B^\prime]-([\hat A,\hat B])^\prime=0\Leftrightarrow \tilde\Gamma^{T}_{\xi}=-\hat\Gamma_{\xi}\;.
\end{equation}
This means that this connection for the bra-ket notation of the sections representing the quantum states must be anti-selfadjoint (with the term $T$ correcting for the leaf dependent notion of adjointness) in order to have a leaf independent Quantum PB. Precisely, for any two sections (not necessarily covariant) $\sigma_1,\sigma_2\in \Gamma(\mathcal{F})$, this condition for the connection implies (locally on $\xi$):
\begin{equation}
\label{HermitianConnection}
\partial_\xi\langle\sigma_1,\sigma_2\rangle_\xi=\langle\nabla_\xi\sigma_1,\sigma_2\rangle_\xi+\langle\sigma_1,\nabla_\xi\sigma_2\rangle_\xi\;.
\end{equation} Thus, as in \cite{mostafazadeh2018energy}, we obtain that $\Gamma$ must be a Hermitian connection. For physical quantum states (given by covariant sections), \eqref{HermitianConnection} leads to the conservation of scalar products and norm: $\partial_\xi\langle\Psi\mid\Psi\rangle_\xi=0$. In the functional picture presented in section \ref{example}, this is seen as a change of the modulus of $\Psi(\phi)$ that compensates for the change of Gaussian measure. We think that this is a novelty of our framework, in contrast with the usual 3+1 picture of QFT in curved spacetime where norm loss is ubiquitous (see for example the phenomenon of \textit{quantum completeness} \cite{Schneider,Schneider2}).

Notice that \eqref{antiselfadjointconnection} implies  that $A^\prime$, defined in eq. \eqref{Aprime}, is the covariant derivative of the operator, i.e. $A^{\prime}:=\partial_\xi(\hat A)+[\hat\Gamma_\xi,\hat A]=\nabla_\xi\hat A$. Therefore:
\begin{equation}
\label{leafdependencefA}
\partial_\xi f_{\hat A}=\langle\Psi\mid\nabla_\xi Q(A)\mid\Psi\rangle=\langle\Psi\mid\left(\partial_\xi Q(A)+[\hat \Gamma_\xi,Q(A)]\right)\mid\Psi\rangle\;.
\end{equation}
This is natural, from the geometrical point of view, since we are considering the covariant derivative of a $(1,1)$--tensor for the fiber of $\mathcal{F}$. Our conclusion is that the $\xi$-dependent quantization is consistent if all the leaf dependencies of the quantum objects are given by the parallel transport w.r.t. a Hermitian connection.
Note also that instead of enforcing this Hermitian nature for $\hat\Gamma$, we have obtained it (in terms of condition  \eqref{antiselfadjointconnection}) from the leaf independence of the kinematics encoded in \eqref{leafindependenceQPB}. In section \ref{example}, for the case of holomorphic quantization of a scalar field, we show the existence of (and, in fact, derive) a quantum connection that fulfils this condition and further physical requirements from hybrid geometrodynamics derived in section \ref{sec:4}. In general, its existence (which may not be guaranteed for arbitrary field theories and must be derived case by case), must be considered in the context of the Gel'fand triple that allows us to embed the current Hilbert space in the space of Hida distributions, because, locally on $\xi$, $\hat\Gamma$ cannot be an inner operator to the current $L^2$ space. Similarly, the Radon-Nikodym derivatives that describe the change of Gaussian measure exist in this generalized setting, $\dfrac{D\mu_{\xi^1}}{D\mu_{\xi^2}}\in (\mathcal{N})^\prime$. This is where the consideration of Hida distributions comes in handy, to see the connection as an operator which is always inner to the fiber.
A more geometric approach can be found in the supp. material, where $\langle,\rangle_\xi$ is related to a quantum Kähler structure $(\omega_Q,J_Q,\mu_Q)$ and the connection must fulfil some relations with $\omega_Q$ and $J_Q$.

As a summary, the main difficulty with a Hamiltonian picture of QFT in generic dynamical spacetimes is that the quantization procedure associated to a choice of complex structure $J_C$ over $\mathcal{M}_F$ is, generally, constructed in a leaf dependent way (this seems also the most physically sensible choice of $J_C$, as the restriction to Cauchy data on a hsf. of the ones used in  \cite{ashtekar75,CCQ04,agullo2015unitarity}). This implies that, for each $\xi\in\mathcal{M}_G\times\mathcal{M}_N$:
\begin{itemize}
    \item  states acquire a different functional representation $\sigma_\Psi(\xi)=\Psi\in(\mathcal{N})^\prime$, 
    \item  the scalar product will be associated to a different Gaussian measure $D\mu_\xi$ over the space of fields and, in turn, the states are bound to be seen belonging to a different Hilbert space, $L^2(\mathcal{N}^\prime,D\mu_\xi)$,
    \item operators have a different representation to be self adjoint for $D\mu_\xi$ and a vacuum phase associated to $J_C$ \cite{CCQ04}.
\end{itemize}
Thus, the usual notions of Fock spaces, creation and destruction operators, number of particle operators, etc. are constructed in an \textit{a priori} $\xi$-dependent way, which does not come as a surprise to the reader familiar with QFT in curved spacetime. While this may spoil the particle interpretation of quantum states, this is natural and not relevant for the construction of consistent hybrid geometrodynamics. Therefore, the object of interest is the Poisson structure of QFT. Following the most sensible physical choice in our judgement, this structure has been made invariant in order to properly reproduce kinematical relations of symmetry generators and the foliation independence principle of geometrodynamics. This postulate is mathematically captured in \eqref{leafindependenceQPB}, which has been shown to be fulfilled if states are related to covariant sections for a Hermitian connection that fulfils \eqref{antiselfadjointconnection}. However, it is not necessary to impose \textit{a priori} this covariant nature for the quantum states. Instead, this can be rigorously derived from geometric quantization, because the symplectic potential defining the polarization is $\xi$-dependent (as so was $J_C$), as argued in section \ref{example}. In this case, \eqref{leafindependenceQPB} can be seen as consequence of this result.

%In order to separate clearly the leaf dependent (associated to physical interpretation) and leaf independent (associated to the symplectic) geometric structures, in the supplementary notes we abandon the usual functional picture of Hamiltonian QFT, in order to borrow the Poisson structure for QFT from the more abstract Geometric Formalism of Quantum Mechanics developed in \cite{Kibble79} where the symplectic structure is readily available, and the Poisson tensor will be associated to its (weak) inverse. This Kähler formalism was adapted in \cite{David2} to QFT for a time dependent quantization, and the resulting connection already fulfilled \eqref{antiselfadjointconnection}, but instead of claiming \eqref{leafindependenceQPB}, an equivalent condition is claimed for the time dependence of geometric structures.
Note that the construction so far provides with a rigourous measure over the space of fields (which happens to be $\xi$-dependent) to the usual Schrödinger wave functional picture of QFT \cite{Long1996TheSW} (in our case, in the canonical, not covariant, picture), and a way to consistently relate states belonging to Hilbert spaces with different measures $D\mu_\xi$.

\subsection{Generating functions and supermagnitudes.}
Usually, the link to physical magnitudes of quantum operators is given by the quantization procedure $Q(f)$ of the classical functions $f$ associated to such magnitudes. Thus,\textit{ in the absence of anomalies}, the classical subalgebra of generating functions $A_i\in C^\infty(\mathcal{M}_F)$ of some symmetries  $a_i:\mathcal{M}_F\rightarrow\mathcal{M}_F$, are mapped through $Q_\xi$ to linear operators $\hat A:=Q_\xi(A)\in\mathrm{Herm}_\xi(\mathcal{M}_q)$ that fulfil the same algebraic relations under the commutator $-i[\hat A_i,\hat A_j]$ as the classical PB, $\lbrace A_i,A_j\rbrace_M$. In turn, the operators $\hat A_i$ are mapped to quantum generating functions in the quantum Poisson algebra through the EV, $f_{\hat A_i}(\Psi)=\langle\Psi\mid\hat A_i\mid\Psi\rangle\in\mathcal{A}_Q$.
Therefore, on $\mathcal{M}_Q$ the infinitesimal generators of the symmetries are the quantum Hamiltonian fields $X_{f_{A_i}}:=\lbrace\cdot,f_{A_i}\rbrace_Q$, that, acting over any $f_B\in\mathcal{A}_Q$, is given by $
\mathcal{L}_{X_{f_A}}f_B=\omega_Q(X_{f_A},X_{f_B})=\lbrace f_B,f_A\rbrace_Q=\dfrac{-i}{\hbar}\langle\Psi\vert[\hat B,\hat A]\vert\Psi\rangle_\xi$.
In the presence of anomalies, the generating functions at the classical level and at the quantum level may not have the same CR. We will comment on this later.

%HERE-------------------------------------------------------------
%HERE-------------------------------------------------------------

Let us consider now the generating functions of hsf. deformations. Firstly, we must note that the C.R. (\ref{closeHH}, \ref{closeHM} and \ref{closeMM}) at the classical level,
 usually written for the generators of  \textit{local} deformations of the leaf, $\mathcal{H}^M_\mu(x)$, involve products of fields  with $D_i\delta(x-y)$. When fields are upgraded to distributions $\varphi^x,\pi^x\in\mathcal{N}^\prime$ (in order to define a measure on $\mathcal{M}_F$), this is a source of inconsistencies and possible anomalies \cite{tsamis1987factor}, so it is better to avoid local (on $x\in\Sigma$) representations. Instead, we define the generators of \textit{global} deformations of the leaf, associated to test functions $\tau$ on $\Sigma$:
 \begin{equation}
\mathcal{H}_\mu^M[\tau]:=\int\limits_\Sigma d^3x \tau(x)\mathcal{H}_\mu(x)\quad\forall\tau\in\mathcal{N}
\end{equation}
We will assume that $\mathcal{H}_\mu^M[\tau]$, as functionals on distributions $\varphi^x,\pi^x$, belong to a Hilbert space $\mathcal{O}_{cl}=L^2(\mathcal{N}'\times\mathcal{N}',  D\mathcal{W}_\xi)$ defined by
\begin{comment}
   \begin{equation}
\label{eq:integrability_condition}\mathcal{O}_{cl}=L^2(\mathcal{N}_{\mathbb{C}}',  D\nu_\xi)\textrm{ with }\int  D\nu_\xi(\phi^{\mathbf{x}})e^{i (\bar\rho_x\phi^x+\rho_x\bar \phi^x) }= e^{-\frac{\bar\rho_x\Delta^{xy}\rho_y}{2} }.
\end{equation} 
\end{comment}

{\color{black}
\begin{equation}
    \int  D\mathcal{W}_\xi(\varphi^{\mathbf{x}},\pi^{\mathbf{x}})e^{i (\alpha_x\varphi^x+\gamma_x \pi^x) }= \exp\Big({-\frac{h_C^{-1}[(\alpha_\mathbf{x},\gamma_\mathbf{x}),(\alpha_\mathbf{x},\gamma_\mathbf{x})]}{4} }\Big).
\end{equation}

That differs from \eqref{eq:measurednu} by a 2 factor dividing the covariance. This  integrability condition was shown in \cite{David2} to be necessary for Weyl's quantization (which can be traced back to a spatial point-splitting regularization for products of fields). The classical theory expressed in $\mathcal{M}_F=\mathcal{N}'\times\mathcal{N}'$, which is a space of distributions, is already cured of point evaluation ambiguities as the ones exposed in \cite{tsamis1987factor} because classical observables are functions of $\mathcal{O}_{cl}$ to which singular objects as $(\varphi^x)^2$ do not belong.  
We will also assume that the classical theory is built in accordance with the CR and, as such,  also fulfils the global analogue of the local algebraic relations (\ref{closeHH}, \ref{closeHM} and \ref{closeMM}):
\begin{equation}
\label{closeHHglobal}
\lbrace \mathcal{H}[\tau],\mathcal{H}[\tau^\prime]\rbrace=\mathcal{H}_i[\tau h^{ij}D_j\tau^\prime]-\mathcal{H}_i[\tau^\prime h^{ij} D_j\tau]\;,
\end{equation}
\begin{equation}
\label{closeHMglobal}
\lbrace \mathcal{H}_{i}[\tau],\mathcal{H}[\tau^\prime]\rbrace=\mathcal{H}[\tau D_i\tau^\prime]\;,
\end{equation}
\begin{equation}
\label{closeMMglobal}
\lbrace \mathcal{H}_{i}[\tau],\mathcal{H}_{j}[\tau^\prime]\rbrace=-\mathcal{H}_{i}[D_j\tau\tau^\prime]+\mathcal{H}_{j}[\tau D_i\tau^\prime]\;,
\end{equation}
where these equations, instead of $\forall x,x^\prime\in\Sigma$, hold $\forall \tau,\tau^\prime\in\mathcal{N}$, \textit{i.e.} for any pair of global deformations on the leaf. Being $\tau h^{ij}D_j\tau^\prime,\tau D_i\tau^\prime\in\mathcal{N}$ if $\tau,\tau^\prime\in\mathcal{N}$, the r.h.s. of these CR are just linear combinations of the global sH and sM. Thus, the CR are also regularized $\forall \tau,\tau^\prime\in\mathcal{N}$ given that $\mathcal{H}^M_\mu[\tau]\in\mathcal{O}_{cl}$. We have written \ref{closeHHglobal}, \ref{closeHMglobal} and \ref{closeMMglobal} for the total sH and sM and the total PB, but the global analogue to the matter CRs (\ref{sMG-sHM},\ref{sMM-sHM}) are also fulfilled. Analogously, the \textit{global} constraints must be:
\begin{equation}
    \mathcal{H}_\mu[\tau]\simeq 0 \quad \forall\tau\in\mathcal{N}\Longrightarrow\lbrace\mathcal{H}_\mu[\tau],\mathcal{H}_\nu[\tau^\prime]\rbrace\simeq 0
\end{equation}
Due to their linearity on $\mathcal{H}_\mu[\mathcal{N}]$, the global CR are null on shell. All these global considerations are necessary to have regularized versions of the generators $\mathcal{H}^M_\mu[\tau]$ at the  level of classical field theory written in terms of  distributions, which was necessary to define the measure $D\nu_\xi$ (Eq. \eqref{eq:measurednu}) on $\mathcal{M}_F$  which we  used to define the scalar product of the Hilbert space of quantum states. Nevertheless, these constraints and CR are physically equivalent to the usual local ones for classical theories defined over test functions, but they are better suited for quantization.

For the global matter sH or sM, $\mathcal{H}_{\mu}^M[\tau]$, associated to a $\tau$-weighted normal ($\mu=0$) or tangential deformation ($\mu\neq0$), we can define linear operators over the quantum states given by $\hat{\mathcal{H}}^M_\mu[\tau]:=Q_\xi(\mathcal{H}_\mu^M[\tau])$. In turn, the generating functions in $\mathcal{A}_Q$ for these deformations, \textit{i.e.} the quantum sH and sM, are given by:
\begin{equation}
\mathcal{H}^Q_\mu[\tau]:=\langle\Psi\mid \hat{\mathcal{H}}^M_\mu[\tau]\mid\Psi\rangle\in\mathcal{A}_Q,
\end{equation}
Note that if $\mathcal{H}_\mu^M[\tau]$ was linear on $\tau$, so should be $\mathcal{H}^Q[\tau]$, by the linearity of $Q_\xi$. We will expand on this later. With these ingredients we can define the quantum Hamiltonian operator $\hat H$ that usually governs the dynamics in Schr\"odinger wave functional picture as the quantization of the classical field matter Hamiltonian $\mathcal{H}^M[N]+\mathcal{H}_{i}^M[N^i]$:
\begin{equation}
\hat H:=\hat{\mathcal{H}}^M[N]+\hat{\mathcal{H}}^M_{i}[N^i],
\end{equation}
acquiring the extra dependence on $\xi\in\mathcal{B}$ through $Q_\xi$, as it must be self-adjoint for $\langle,\rangle_\xi$.

On the other hand, to be a faithful representation of Dirac's algebra at the quantum level, it  is necessary that the quantum sH and sM fulfil the same CR as their classical matter counterparts. 
However, it is well known from Groenewold's no-go theorem \cite{GROENEWOLD1946405} that Dirac's quantization relations:
\begin{equation}
\label{Groenewold}
Q(\lbrace f,g\rbrace_c)\overset{?}{=}\dfrac{-i}{\hbar}[Q(f),Q(g)]
\end{equation}
do not hold for arbitrary functions $f,g\in C^\infty(\mathcal{M}_F)$ over classical phase space, and, thus, the Poisson algebra structure of the functions $C^\infty(\mathcal{M}_F)$ is not homomorphic in general to the Lie algebra structure defined on the set of quantum operators by the commutator. Nevertheless, a crucial result from Groenewold's no-go theorem is that, if either $f$ or $g$ is a quadratic polynomial, then the equality \eqref{Groenewold} can be satisfied. This can easily be checked in QM by making use of Weyl's quantization $Q_W$ for operators and Wigner-Weyl equivalence. Making use of Moyal's star bracket $\lbrace,\rbrace_{\star_M}$ we have that
\begin{equation}
\label{eq:star_brackets}
    \dfrac{-i}{\hbar}[Q_W(f),Q_W(g)]=Q_W(\lbrace f,g\rbrace_{\star_M})\forall f,g\in C^\infty(\mathcal{M}_F)\;.
\end{equation}
Now, we see that the Moyal star bracket  is a deformation of the classical Poisson bracket $\lbrace,\rbrace_C$ such that $\lbrace,\rbrace_{\star_M}=\lbrace,\rbrace_C+\lbrace\lbrace,\rbrace\rbrace_{\star_{M-C}}$, where $\lbrace\lbrace,\rbrace\rbrace_{\star_{M-C}}$ is an anti-symmetric bi-differential operator acting on each entry with higher-than-second-order derivatives corresponding to a series of odd-powers of the classical Poisson bracket. Thus, it must vanish on quadratic functions.  Besides, if $\lbrace f,g\rbrace_c=0$, the equality is also trivially fulfilled, because $\lbrace f,g\rbrace_c=0\Rightarrow\lbrace\lbrace f,g\rbrace\rbrace_{\star_{M-C}}=0$. 

 The original Groenewold result, obtained for ordinary QM, missed an extension to QFT. In \cite{Dito1990} and \cite{Dito1992}, Dito made a definition of star brackets for quantized field theories in the covariant setting, and classified different star brackets and associated quantization procedures that were c-equivalent (thus, sharing the same algebraic relations). In \cite{David2}, we extend this construction to the canonical quantization of Cauchy data on a hypersurface, taking special care of the definition integral measure and derivatives over the space of fields treated as distributions. Therefore, we can claim the existence of a star bracket (of Moyal's shape if Weyl's quantization is considered) for the QFT under consideration, such that \eqref{eq:star_brackets} also holds in this case.  This is where the integrability condition of \cite{David2} comes into play: Weyl quantization is only well-defined on the functions (rather, functionals) in $C^\infty(\mathcal{M}_F)$ that fulfil it.  However, the star bracket of the set of functions fulfilling this condition is inner to them, so \eqref{eq:star_brackets} holds on this subset.
 
Note that, in particular, the classical field theoretical supermagnitudes are up to quadratic in momenta. Firstly, the matter sM $\mathcal{H}_{i}^M$ must be linear on matter momenta\footnote{This is a consequence of the representation postulate in \cite{HKT76}. It ensures that the transformation on the fields is properly associated to an stretching on their domain, which is  equivalent to a spatial diffeomorphism intrinsic to the leaf, and thus, if higher powers of the momenta appeared in its generating function, such deformation would acquire an incoherent extrinsic nature.} (quadratic in all field variables). On the other hand, in \cite{HKT76} the matter sH is restricted to be no more than quadratic on momenta  (sensibly for the usual Hamiltonian construction, even more so regarding  quantization), in order to have the usual canonical relations between fields and momenta in relation with the generated dynamics.
Therefore, (assuming $\mathcal{H}_\mu^M$ have been classically regularized) the global quantum sH and sM fulfil:
\begin{equation}
\label{eq:Groenewold_supermagntitudes_1}
\lbrace\mathcal{H}_{\mu}^Q[\tau],\mathcal{H}_{\nu}^Q[\tau^\prime]\rbrace_Q=\dfrac{-i}{\hbar}\langle\Psi\vert[Q(\mathcal{H}_{\mu}^M[\tau]),Q(\mathcal{H}_{\nu}^M[\tau^\prime])]\Psi\rangle=\langle\Psi\vert Q(\lbrace\mathcal{H}_{\mu}^M[\tau],\mathcal{H}_{\nu}^M[\tau^\prime]\rbrace_M)\Psi\rangle
\end{equation}
where $\lbrace,\rbrace_M$ is the classical matter Poisson bracket. Note that his cannot be done for the local version of the supermagnitudes, because $\lbrace\mathcal{H}_{\mu}^M(x),\mathcal{H}_{\nu}^M(x^\prime)\rbrace_M$ involves naked products of distributions $\varphi^x,\pi^x$ and $D_j\delta(x-y)$ on the same point, which cannot fulfil the integrability condition and, thus, cannot be consistently Weyl quantized. In this line, some further considerations regarding anomalies must  be made:
\begin{itemize}
    \item  Firstly, in the context of canonical quantization of gravity,  anomalies may appear in the quantization $Q_G$ of gravitational supermagnitudes (see e.g. section 5.3.5 of \cite{Kie1}). The CR \eqref{closeHH} have algebraic constants that depend on $h$, so the ordering prescription for the quantization of its r.h.s. mixes the algebraic constants with the supermagnitudes in a non-commuting way, spoiling the algebraic relations and Dirac's constraints,  as argued in \cite{Kie1}.  Besides, the gravitational sH is not quadratic, but (at least) quartic on $h,\pi^h$, so the exception for quadratic operators of Groenewold's theorem may not apply. None of these two obstructions apply to our case, as we quantize matter fields but gravity remains classical. Instead, in the CR of matter supermagnitudes, the constants of the algebra do not depend on matter fields and $h$ appears there as a parameter (as it is not quantized), so it does not produce ordering problems, even for the $\xi$-dependence of  $Q_\xi$, as we will see.

    \item Other obstructions come from the mechanisms used in the quantization of covariant field solutions, such as the Schwinger terms related to the time ordering procedure of the quantized stress-tensor commutators discussed in \cite{boulware1967stress}. Again, these kind of considerations do not apply to our work, since we are not considering the quantization of covariant field solutions, but of canonical Cauchy data of matter fields. Thus, our formalism is not explicitly covariant. Instead, we will see that, as happened in classical geometrodynamics, covariance will be recovered in hybrid geometrodynamics for the solution to the hybrid dynamics for hybrid Cauchy data subject to the Hamiltonian and momenta first-class constraints. 
    \item The necessity of regularization argued in \cite{tsamis1987factor} does apply to our case and it must be done prior to the quantization, and thus, to the implementation of an ordering procedure. When switching from test functions to distributions in order to define $D\nu_\xi$ and prepare the classical theory for quantization, one has to regularize the theory at the classical level as argued above, as otherwise the products of distributions are not well defined. This results in a regularized classical field theory based on distributions, that must still fulfil the matter CR  for the regularized sH and sM $\mathcal{H}^M_\mu[\tau]$, as otherwise it would be inconsistent with classical geometrodynamics.
\end{itemize}

%We have taken this into account and assume that the star bracket (and, in fact, even the Poisson bracket) only acts on smeared functionals of the classical fields, i.e., functions that fulfil an integrability condition under the Gaussian measure of choice. Therefore, we can consider that the delta functions appearing in the closing relations only make sense smeared, in terms of a spatial point-splitting regulator, and integrated over the spatial leaf. Only this kind of functions will be quantized. We have treated this problem in \cite{David2}, where the classical functions that can be quantized for Weyl's quantization fulfil an integrability condition that can be traced back to a spatial point-splitting regularization for classical distributions.

We have taken this into account when we considered for quantization only the functions fulfiled the integrabilty condition of \cite{David2}. Within this formalism, the quantization procedure does not introduce new ordering problems, and we have the QFT analogue of 
\eqref{eq:star_brackets} for the star bracket defined also in \cite{David2}. In this work we will not delve anymore into the regularization conditions, as our focus is on the interplay between matter QFT for Cauchy data and classical gravitational geometrodynamics. 

We remark that a key \textit{assumption} of our work is the existence of a classical field theory in terms of distributions whose supermagnitudes fulfil the appropriate integrability conditions and the (global) matter CR of geometrodynamics. The QFT we work with will be the quantization of such  classical matter theory. Based on this, \eqref{eq:Groenewold_supermagntitudes_1}) worked for regularized classical matter supermagnitudes, because they were quadratic. This result will be crucial to show in the following section that the quantum supermagnitudes fulfill the appropriate CR for a matter theory.

\section{Hybrid geometrodynamics}
\textbf{}\label{sec:4}
The objective of this section is to substitute the classical matter of section \ref{sec:2} by the QFT of section \ref{sec:3}. To do so, we combine the quantum and geometrical phase spaces ($\mathcal{M}_Q$ and $\mathcal{M}_G$), PB ($\lbrace,\rbrace_Q$ and $\lbrace,\rbrace_G$) and supermagnitudes ($\mathcal{H}_\mu^Q$ and $\mathcal{H}_\mu^G$), in such a way that the we can identify the generating functions of hsf. deformations inside a hybrid Poisson algebra that fulfil Dirac's CR and we can impose some constraints enforcing the path independence principle. Thus, in this section,  $(h_{ij}
(x),\pi^{ij}_h(x))$ do not represent a given background depending  parametrically on the hsf. label $s$, but kinematical variables whose dynamics  are not externally given, but coupled with quantum matter fields.
\subsection{Hybrid phase space}
Simple hybrid phase spaces $\mathcal{M}_H$ are usually constructed \cite{Alonso2012}, as the cartesian product of the quantum and classical submanifolds (which are constructed independently). In the case of hybrid geometrodynamics, instead, we must consider the manifold of hybrid states as $\mathcal{M}_H:=\mathcal{F}$, given by the bundle $\mathcal{F}$ of the previous section, because the quantum states are dependent on classical degrees of freedom as in eq. \eqref{statesdependonxi}. Any point of $\mathcal{M}_H$ defines a pair of objects: a classical state which is a point in the base of the bundle $\xi\in\mathcal{B}$, and a quantum state $\Psi$ (given by the evaluation of a covariant section on the given $\xi$. The tangent space  $T_{(\xi,\Psi)}\mathcal{F}$ can be decomposed as in \eqref{eq:decomposition}, and we introduce a connection that defines the corresponding  horizontal directions as tangent to the covariant sections.

 Note also that the field measure (or scalar product), states and operators were leaf-dependent, as so was the quantization procedure $Q_\xi$ due to the choice of $J_C$.  Therefore, the manifold of lapse and shift functions, $\mathcal{M}_N$, is present in the base manifold of $\mathcal{F}$ to capture the dependence of these quantum objects, providing a way to relate them (through the horizontal bundle) when $N$ and $N^i$ change throughout the foliation. This will not affect the kinematics, such as the definition of PB and CR for the generators of hsf. deformations of either $\mathcal{M}_G$ or $\mathcal{M}_Q$, and the dynamics of $N$ and $N^i$ can be seen as an externally given curve on $\mathcal{M}_N$. Each of such curves defines a \textit{path} among possible hsfs., and the physical content of the theory will be independent of it due to the implementation at the hybrid level of the \textit{path independence principle}, stated in \cite{HKT76}.

Therefore, a hybrid state provides the whole information on a Cauchy surface $\Sigma_0$ of initial data and, with any given curve for the dynamics of $N$ and $N^i$, we will be able to pose a Cauchy problem at the hybrid level. Their solutions will spawn the physical information for the whole spacetime, once the foliation of spacetime is undone. This provides an isomorphism between the whole quantum matter and geometrical content of the Universe and the initial conditions on the original hsf. $\Sigma_0$, given by the intrinsic 3-metric $h_{ij}(0)$, its conjugate momenta $\pi^{ij}_h(0)$, and the quantum state $\Psi=\sigma_0(h(0),\pi_h(0),N(0),N^i(0))$.
We remind that the connection will allow us to determine the change on the quantum state due to the change of the geometric variables, relating its representations inside non-unitarily equivalent Hilbert spaces.

\subsection{Hybrid Poisson algebra, supermagnitudes and closing relations}
In simple hybrid systems such as \cite{Alonso2012}, the  hybrid algebra of observables can be represented by the EV of self adjoint linear operators over the quantum manifold $\mathcal{M}_Q$ which had an infinitely differentiable dependence on the classical variables. In terms of abstract $C^\star$-algebras one can consider  as in \cite{Koopman} that the hybrid algebra of observables is given by  $\mathcal{A}_H:=\mathcal{A}_G\otimes\mathcal{A}_Q$, being $\mathcal{A}_G\equiv C^\infty(\mathcal{M}_G)$. In such framework, purely classical observables are given by a classical function multiplying the identity operator on the Hilbert space, while purely quantum ones are operators independent of the classical degrees of freedom. 

In the current geometrodynamical case, $\mathcal{A}_Q$ was already defined as the EV of $\xi$-dependent operators, with the additional $\xi$-dependence for Hermitian product and states, but the $\xi\in\mathcal{B}$ contain now the classical kinematical variables. \ref{sec:3}. Thus, in our framework the hybrid algebra of observables is simply:
 \begin{equation}
\mathcal{A}_H:=\bar{\mathcal{A}}_Q\;,
 \end{equation}
where $\mathcal{A}_Q$ is defined as in Equation \eqref{eq:AQ}, taking into account that now $\xi\in\mathcal{B}$ is not regarded as an external parameter as in section \ref{sec:3}. We are already considering the completion, $\bar{\mathcal{A}}_Q$,  as defined in \eqref{completedalgebra} for reasons that will become clear later. We will now illustrate a bit how this hybrid observables are constructed and how they represent physical magnitudes, either material or purely geometrical.

Consider first a function of the classical matter fields,  $A\in C^\infty(\mathcal{M}_F)$ that does not depend on geometric fields. It can be expressed as a polynomial on matter field distributions $\varphi^x$ and momenta $\pi^y$ with $\xi$-independent coefficients, $a_{ij\vec{x}\vec{y}}$, such that:
\begin{equation}
A=\sum\limits_{ij} a_{ij\vec{x},\vec{y}} \varphi^{i\vec{x}}\pi^{j\vec{y}}
\end{equation}
The element of $\mathcal{A}_Q$ given by $\langle\Psi\mid Q_\xi(A)\mid\Psi\rangle_\xi$ represents a purely material observable at the classical level, but, \textit{a priori}, has acquired a hybrid nature due to the leaf dependence of the quantization, states, and scalar product. On the other hand, if we consider a function $B$ over the classical matter fields whose coefficients were already $\xi$-dependent (for example, this is the case for the sH of a classical scalar field), such that
\begin{equation}
B:=\sum\limits_{ij} b_{ij\vec{x},\vec{y}}(\xi) \varphi^{i\vec{x}}\pi^{j\vec{y}},
\end{equation}
the EV $\langle\Psi\mid Q_\xi(B)\mid\Psi\rangle_\xi\in\mathcal{A}_Q$ is a mixed matter-geometrical observable both before and after quantization, with additional leaf dependence in the quantum case.

Lastly, if we consider a function $g$ solely of the geometric variables, $g(h,\pi_h)\in \mathcal{A}_G$, and perform a quantization procedure for the matter fields, given its independence on such fields we get $Q_\xi(g)=g(h,\pi_h)\hat{\mathbb{I}}$. Thus, the hybrid observable representing this purely geometrical observable  is given by $f_{g\mathbb{I}}=g(h,\pi)\langle\Psi\mid\mathbb{I}\mid\Psi\rangle_\xi\in\mathcal{A}_H$. Given the norm conservation under changes of leaf variables derived from \eqref{antiselfadjointconnection},  the total hybrid dynamics will preserve norm of the quantum state. Thus, under the initial constraint of norm 1, instead of $g(\xi)\langle\Psi\vert\hat{\mathbb{I}}\Psi\rangle_\xi\in\mathcal{A}_H$ we can simply write $g(\xi)\in\mathcal{A}_H$.
\footnote{
In the presence of norm loss for the quantum states one can define $\mathcal{A}_{H}:=\mathcal{A}_G\cup \bar{\mathcal{A}}_Q$, to be able to include the purely gravitational observables independently of the norm of the quantum state.}

Summarizing,  $\mathcal{A}_H$ is constituted by functions (and products of them) defined as the EV under the quantum state  $\Psi\in\mathcal{M}_Q$ and scalar product $\langle,\rangle_\xi$ of the self-adjoint (for  $\langle,\rangle_\xi$) operator $Q_\xi(f)$ for any function $f\in C^\infty(\mathcal{M}_G\times\mathcal{M}_F)$.

Given that the construction of  $\mathcal{A}_H$ is equivalent to $\mathcal{A}_G\otimes\mathcal{A}_Q$, we may endow $\mathcal{A}_H$ with a bilinear operator given by the sum of the geometrical and quantum PB:
\begin{equation}
\label{eq:hybridPB}
\pb{}{}{H}=\pb{}{}{G}+\pb{}{}{Q}\;,
\end{equation}
which can be shown to be a PB over the set of observables contained in $\mathcal{A}_H$.  Leibniz, antisymmetry and bilinearity immediately inherited from the properties of $\lbrace,\rbrace_G$ and $\lbrace,\rbrace_Q$  over their respective algebras. Nevertheless, to fulfill Jacobi identity the compatibility condition given by the leaf independence of the quantum PB, eq. \eqref{leafindependenceQPB}, must be enforced. It is trivial to check that $\mathcal{A}_H$ forms a Poisson algebra under $\lbrace,\rbrace_H$, while $\mathcal{A}_{Q}$ is not a proper subalgebra, given that 
\begin{equation}
  \lbrace f_A,f_B\rbrace_G=f_{\partial_h A+[\Gamma_h,\hat A]}f_{\partial_\pi B+[\Gamma_\pi,\hat A]}-(A\leftrightarrow B)\not\in\mathcal{A}_Q
\end{equation}  
where we have made use of \eqref{leafdependencefA}, compacted in the subindex notation. Hence, the necessity of $\mathcal{A}_H$ to be the completion of $\mathcal{A}_Q$ under the ordinary product, $\bar{\mathcal{A}}_Q$.

Over this hybrid Poisson algebra, the infinitesimal generators $X_A$ of any Hamiltonian transformation $a:\mathcal{M}_H\rightarrow\mathcal{M}_H$ acquire a Hamiltonian representation over $\mathcal{A}_H$ through its generating function, $f_A$, such that, for any hybrid observable $F$:
\begin{equation}
\mathcal{L}_{X_A} F=\lbrace F,f_A\rbrace_H \quad \forall F\in\mathcal{A}_H\;.
\end{equation}
In particular for the generating functions of hsf. deformations, the hybrid sH and sM are 
constructed invoking the equivalence principle as in \cite{HKT76}, where the hybrid supermagnitudes must be built as the sum of the pure geometrodynamical (classical) and the matter (quantum) supermagnitudes:
\begin{equation}
\label{hybridSH}
\mathcal{H}^H_\mu[\tau]:=\mathcal{H}^G_\mu[\tau]+\mathcal{H}^Q_\mu[\tau]\;.
\end{equation}
Note  that the gravitational term should appear multiplied by the norm of the quantum state for the supermagnitudes to strictly belong to $\mathcal{A}_H$, but in our framework $\langle\Psi\mid\Psi\rangle=1$ always holds, so we have used a simplified notation.

It is thoroughly argued in \cite{HKT76} that, on the classical field case the matter sM was independent of the gravitational variables, and the matter sH was ultralocal on $h$. Nevertheless, in this hybrid case the quantization procedure (and states and scalar product) adds non trivial dependences on $\xi\in\mathcal{B}$, which may even be of derivative nature. For example, the complex structure $J_C$ in \cite{ashtekar75} adapted to the foliation presents spatial derivatives of the metric and of $N$, $N^i$; and so will the quantum connection. Howbeit, we will restrict to the case where $J_C$ (and, thus, $Q_\xi$, $\langle,\rangle_\xi$ and states) does not depend on the geometrodynamical momenta. Therefore:
\begin{equation}
\label{pihindepencence}
\partial_{\pi_h}\sigma_\Psi(\xi)=0,\;\Gamma_{\pi_h}=0\text{ and } \partial_{\pi_h}\langle\Psi\mid Q_\xi(f)\mid\Psi\rangle_\xi=\langle\Psi\mid Q_\xi(\partial_{\pi_h}f)\mid\Psi\rangle_\xi
\end{equation}
where the last equality implies that we allow the observables to depend on $\pi_h$ as they did before the quantization, but they do not acquire further dependence on it from the quantization procedure, scalar product or states. The dependence on the 3-metric, lapse and shift remains fully general. We consider this particular case because, firstly, it includes Ashtekar and Magnon's complex structure\cite{ashtekar75}, and, secondly, because we judge it to be the most physically sensible choice, which can be argued as follows.  If the complex structure $J_C$ over the field manifold used to define the quantization is chosen following some physical criterium  (such as, positive energy flux along the foliation), then it must be chosen in reference to classical field theoretical information (such as the Hamiltonian field at the classical level, as in \cite{ashtekar75,CCQ04}). Thus, there is no source of $\pi_h$ dependence in the quantization procedure, given the geometrodynamical principle of equivalence for the classical matter theory. 

Returning to the hybrid supermagnitudes, we will now see how they fare regarding the CR appropriate for Dirac's generators, \eqref{closeHHglobal}, \eqref{closeHMglobal} and \eqref{closeMMglobal}.
Let us make two observations to simplify the problem. Firstly,  for $\mathcal{H}_\mu^G$ we know that: i) under the gravitational bracket they already fulfilled all CR, and ii) the quantum PB of them with any other function will be null, $\lbrace\mathcal{H}_\mu^G[\tau],f\rbrace_Q=0 \;\forall f\in C^\infty(\mathcal{M}_H)$, as $\mathcal{H}_\mu^G$ does not depend on $\Psi$. Secondly, the gravitational bracket  over any two quantum supermagnitudes is null $\lbrace\mathcal{H}_{\mu}^Q[\tau],\mathcal{H}_{\nu}^Q[\tau^\prime]\rbrace_G=0$ given \eqref{pihindepencence} and the $\pi_h$-independence of the classical field sH and sM, \cite{HKT76}. 
Taking this into account, for $\mathcal{H}_\mu^H$ (sH if $\mu=0$, and sM if $\mu\in\lbrace1,2,3\rbrace$),  we see that the hybrid PBs only have the following non-vanishing terms:
\begin{equation}
\label{crAB}
\lbrace\mathcal{H}_{\mu}^H[\tau],\mathcal{H}_{\nu}^H[\tau]\rbrace_H=\lbrace\mathcal{H}_{\mu}^G[\tau],\mathcal{H}_{\nu}^H[\tau]\rbrace_G+
\lbrace\mathcal{H}_{\mu}^Q[\tau],\mathcal{H}_{\nu}^Q[\tau]\rbrace_Q+\lbrace\mathcal{H}_{\mu}^Q[\tau],\mathcal{H}_{\nu}^G[\tau]\rbrace_G
\end{equation}
On the quantum side, we remark that we assume that we have a regularized classical field theory in terms of distributions that fulfils the matter CR. Now, we can choose our quantization procedure $Q_\xi$ to extend to polynomials through Weyl's ordering and apply the equivalence between Moyal's bracket and commutator of Weyl quantized observables. Given that the classical field supermagnitudes are up to quadratic on momenta (for the free scalar field, also on fields), their Moyal bracket is equivalent to the classical PB, and therefore, we can invoke eq. \eqref{eq:Groenewold_supermagntitudes_1}. As in quantum mechanics, no ordering problem arises in this case, because the functions are quadratic and the theory had been regularized. This implies that the quantum supermagnitudes $\mathcal{H}^Q[\tau]$ and $\mathcal{H}_i^Q[\tau]$ fulfill \eqref{closeHHglobal} and \eqref{closeMMglobal} for the quantum PB, as so did their classical field counterparts:
\begin{multline}
\lbrace \mathcal{H}^Q[\tau],\mathcal{H}^Q[\tau^\prime]\rbrace_Q=\dfrac{-i}{\hbar}\langle\Psi\vert \left[Q\left(\mathcal{H}^M[\tau]\right),Q\left(\mathcal{H}^M[\tau^\prime]\right)\right]\Psi\rangle=\langle\Psi\vert Q\left(\lbrace\mathcal{H}^M[\tau],\mathcal{H}^M[\tau^\prime]\rbrace_M\right)\Psi\rangle\\=\langle\Psi\vert Q\left(\mathcal{H}_i^M[\tau h^{ij}D_j\tau^\prime]-\mathcal{H}_i^M[\tau^\prime h^{ij} D_j\tau]\right)\Psi\rangle=\mathcal{H}_i^Q[\tau h^{ij}D_j\tau^\prime]-\mathcal{H}_i^Q[\tau^\prime h^{ij} D_j\tau]
\end{multline}
and 
\begin{multline}
\label{sMQ-sMQ}
\lbrace \mathcal{H}^Q_i[\tau],\mathcal{H}^Q_j[\tau^\prime]\rbrace_Q=\dfrac{-i}{\hbar}\langle\Psi\vert \left[Q\left(\mathcal{H}^M_i[\tau]\right),Q\left(\mathcal{H}^M_j[\tau^\prime]\right)\right]\Psi\rangle=\langle\Psi\vert Q\left(\lbrace\mathcal{H}^M_i[\tau],\mathcal{H}^M_j[\tau^\prime]\rbrace_M\right)\Psi\rangle\\=\langle\Psi\vert Q\left(-\mathcal{H}_{i}^M[D_j\tau\tau^\prime]+\mathcal{H}_{j}^M[\tau D_i\tau^\prime]\right)\Psi\rangle=-\mathcal{H}_{i}^Q[D_j\tau\tau^\prime]+\mathcal{H}_{j}^Q[\tau D_i\tau^\prime]\;.
\end{multline}
Note that, to check the PB of the total supermagnitudes,  we still lack the consideration of the terms $\lbrace\mathcal{H}^G[\tau],\mathcal{H}^Q[\tau^\prime]\rbrace_G+\lbrace\mathcal{H}^Q[\tau],\mathcal{H}^G[\tau^\prime]\rbrace_G$ and $\lbrace\mathcal{H}^G_i[\tau],\mathcal{H}^Q_j[\tau^\prime]\rbrace_G+\lbrace\mathcal{H}^Q_i[\tau],\mathcal{H}^G_j[\tau^\prime]\rbrace_G$. We will come back to them later.
On the other hand, the sM-sH PB is however a bit trickier. For such quantum PB, we obtain  the EV of the quantization of the global version of \eqref{sMM-sHM}:
\begin{multline}
\lbrace \mathcal{H}^Q_i[\tau],\mathcal{H}^Q[\tau^\prime]\rbrace_Q=\dfrac{-i}{\hbar}\langle\Psi\vert \left[Q\left(\mathcal{H}^M_i[\tau]\right),Q\left(\mathcal{H}^M[\tau^\prime]\right)\right]\Psi\rangle=\langle\Psi\vert Q\left(\lbrace\mathcal{H}^M_i[\tau],\mathcal{H}^M[\tau^\prime]\rbrace_M\right)\Psi\rangle\\=
\langle\Psi\mid Q\left((D_j\tau2h_{i\alpha})^x\dfrac{\delta\mathcal{H}^M[\tau^\prime]}{\delta h_{\alpha j}^x}\right)\mid\Psi\rangle+\mathcal{H}^{Q}[\tau D_i\tau^\prime ]\;,
\end{multline}
where, in this case, the contraction of $x$ indices is just a shortcut for integration on $\Sigma$. Now, using \eqref{SH} and \eqref{eq:classicalPB}, we obtain the analogue to \eqref{sMG-sHM}:
\begin{equation}
\label{eq:cond1}
\lbrace\mathcal{H}_{i}^G[\tau],\mathcal{H}^Q[\tau^\prime] \rbrace_G+\lbrace\mathcal{H}_{i}^Q[\tau],\mathcal{H}^G[\tau^\prime]\rbrace_G=-(D_j\tau 2h_{i\alpha})^x \dfrac{\delta\mathcal{H}^Q[\tau^\prime]}{\delta h_{\alpha j}^x}+(\tau^\prime G_{\alpha jkl}\pi^{\alpha j})^x\dfrac{\delta\mathcal{H}_i^Q[\tau]}{\delta h_{kl}^x}\;.
\end{equation}
Therefore, in order to fulfill \eqref{closeHM} for the total hybrid supermagnitudes, we must have:
\begin{equation}
\label{eq:cond3}(\tau^\prime G_{\alpha jkl}\pi^{\alpha j})^x\dfrac{\delta\mathcal{H}_i^Q[\tau]}{\delta h_{kl}^x}-(D_j\tau 2h_{i\alpha})^x \dfrac{\delta\mathcal{H}^Q[\tau^\prime]}{\delta h_{\alpha j}^x}
+
\langle\Psi\mid Q\left((D_j\tau2h_{i\alpha})^x\dfrac{\delta\mathcal{H}^M[\tau^\prime]}{\delta h_{\alpha j}^x}\right)\mid\Psi\rangle=0\;.
\end{equation}
Given that neither the quantization procedure nor the classical field supermagnitudes depend on $\pi_h$,  we obtain that the first term of equation \eqref{eq:cond3} must be null on its own, as the rest of the terms do not depend on $\pi_h$, so we must have:
\begin{equation}
\label{sMconnection}
   \dfrac{\delta\mathcal{H}_{i}^Q[\tau]}{\delta h_{\alpha j}(x^\prime)}=0 \quad \forall \Psi,h,N,N^i\Rightarrow
   \dfrac{\delta Q(\mathcal{H}_{i}^M[\tau])}{\delta h_{\alpha j}(x^\prime)}+[\hat \Gamma_h,Q(\mathcal{H}^M_i[\tau])]=0,
\end{equation}
obtaining a new constraint for $\hat\Gamma_h$. For the two remaining terms to cancel, we must have:
\begin{equation}
\dfrac{\delta\mathcal{H}^Q(\tau^\prime)}{\delta h_{\alpha j}(x)}-\langle\Psi\mid Q\left(\dfrac{\delta\mathcal{H}^M [\tau^\prime]}{\delta h_{\alpha j}(x)}\right)\mid\Psi\rangle=0\;,
\end{equation}
which, using eq. \eqref{leafdependencefA} implies one last constraint on the quantum connection:
\begin{equation}
\label{sHconnection}
\dfrac{\delta Q(\mathcal{H}^{M}[\tau^\prime])}{\delta h_{\alpha j}(x)}+[\hat \Gamma_{h(x)},Q(\mathcal{H}^{M}[\tau^\prime]]=Q\left(\dfrac{\delta\mathcal{H}^M [\tau^\prime]}{\delta h_{\alpha j}(x)}\right)\;.
\end{equation}
Now we reconsider the two terms we saved for later under \eqref{sMQ-sMQ}. We need to fulfil:
\begin{equation}
\lbrace\mathcal{H}_{\mu}^G[\tau],\mathcal{H}_{\nu}^Q[\tau^\prime]\rbrace_G+\lbrace\mathcal{H}_{\mu }^Q[\tau],\mathcal{H}_{\nu}[\tau^\prime]\rbrace_G=0\;\text{ for } \mu=\nu=0 \text{ and }\mu=i,\nu=j.
\end{equation}
For the sM-sM PB, this is immediate if equation \eqref{sMconnection} is fulfilled. For $\mu=\nu=0$, this PB was null in the classical case because of the ultralocality of $\mathcal{H}^M$, the local dependence on $\pi_h$ of $\mathcal{H}^G$ and the antisymmetry of the PB. In the quantum case, if eq. \eqref{sHconnection} is fulfilled, one directly inherits such property from classical theory.

At this point we can claim that we can reproduce the CR for the generators of Dirac's group of hsf. deformations on a hybrid geometrodynamical phase space for QFT and classical geometry, at the cost of assuming that no anomalies affect the validity of Wigner-Weyl/Moyal-Lie equivalence for a regularized classical field-distribution theory and the existence of a Hermitian connection for $\mathcal{F}$ which must fulfill \ref{antiselfadjointconnection}, \ref{sMconnection} and \ref{sHconnection}. Note that \ref{sMconnection} and \ref{sHconnection} imply that quantum supermagnitudes behave as their classical field counterparts regarding their dependence on $h$, i.e.:
\begin{equation}
\label{RequirementHM}
\langle\Psi\mid Q\left(\dfrac{\delta\mathcal{H}_{AC} [\tau]}{\delta h_{\alpha j}(x)}\right)\Psi\rangle=\frac{\delta}{\delta {h_{\alpha j}(x)}}\mathcal{H}_{AQ}[\tau]\;.
\end{equation}
The same result would have been obtained for $\pi_h$, if 
\eqref{pihindepencence} had not been imposed.
Generalizing this result, while it is not strictly necessary for hybrid geometrodynamics, it would appear desirable to extend this property to the generating functions of any symmetry of the system and, ideally, to the whole algebra of operators:
\begin{equation}
\label{RequirementHardcore}
\langle\Psi\vert \nabla_\xi \left(Q(f)\right)\Psi\rangle:=\langle\Psi\vert (\partial_\xi Q(f)+[\hat\Gamma_\xi,Q(f)])\Psi\rangle=\langle\Psi\vert Q(\partial_\xi f)\Psi\rangle\;\forall f\in C^\infty(\mathcal{M}_C)
\end{equation}
Notice that this expression represents a compatibility condition of the quantization mapping and the behavior of the covariant derivative $\nabla_\xi$. To be consistent with the fact that the covariant derivative of quantum states is null, the operators must have a covariant derivative that cancels the dependence of the quantization procedure on $\xi$, leaving solely the dependence on $\xi$ that the classical function had before quantization. We show in section \ref{example}, for the geometric quantization in the holomorphic picture of a real scalar field, the existence of a connection that fulfils this relation (and, thus, all the requirements for geometrodynamics). To what extent this can be achieved  in any field theory is beyond the scope of this paper and will be the subject of future investigations. 

Therefore, in this section we have found the hybrid generating functions of hsf. deformations given by \eqref{hybridSH}, that appropriately reproduce the CR of Dirac's group for the hybrid PB \eqref{eq:hybridPB}, under the consistency requirement given by \eqref{RequirementHM} (which implies constraints on the quantum connection) and in the absence of anomalies in the quantization of regularized quadratic functionals on classical field-distributions.
\begin{comment}
Note that all this construction depends on the generating function at the quantum level being given by eq. \eqref{quantumSM}.\\
This is the simplest case, a more general case is considered in the appendix, where the generating functions can be shifted by the EV of an operator proportional to the identity, but with \textit{a priori} non-trivial dependences on $h,\pi_h$. In such case, eq. \eqref{RequirementHM} can be softened in an appropriate manner to allow less restrictive quantizations or choices of $J_C$. This is possible because the quantum states would contain the same physical information if they belong to the same projective ray.
\end{comment}
\subsection{Constraints}
Lastly, we must enforce the path independence criterium, i.e. the physical Cauchy data $(h(0),\pi(0),\Psi(0))$ defined on an initial hsf. $\Sigma_0$ should yield under evolution the same Cauchy data $(h(t),\pi(t),\Psi(t))$ on a final hsf.  $\Sigma_t$, independently of the path chosen from one leaf to another, i.e. of the sheaf  of intermediate hsfs. $\Sigma_{\tau} \; \;\forall\tau\in(0,t)$ that conforms the foliation of that region of spacetime. Path independence implies at the infinitesimal level that any two arbitrary deformations should provide the same physical data independently of the order such deformations are applied. This  means that, when evaluated on  a physical hybrid state $(h,\pi_h,\Psi)$, the CR of the hybrid supermagnitudes should be null \textit{on shell} $\lbrace\mathcal{H}^H_\mu[\tau],\mathcal{H}^H_\nu[\tau^\prime]\rbrace\simeq0$, i.e. over the submanifold defined by the constraints. By linearity this means that, as in the classical case, the Hamiltonian and momenta first class constraints must be enforced on Hybrid Geometrodynamics:
\begin{equation}
\label{hybridconstraints}
\mathcal{H}^H_\mu[\tau]\simeq 0\quad\forall\tau\in\mathcal{N}
\end{equation}
Notice that the validity of the theory is subject to the existence of a non-trivial subspace for the quantum states that can fulfil these constraints. We can argue in favour of its existence. Firstly, these constraints are imposed in a Hamiltonian manner for the hybrid system (elements of $\mathcal{A}_H$ being null). Thus, for the leaf's geometric constant $\xi\in\mathcal{B}$, they appear as a constraint for some EV function on the set of admissible quantum states. This is a weaker condition than the usual $(\mathcal{H}^G[\tau]+\hat{\mathcal{H}}^M_\mu[\tau])\Psi=0$, as the state need not be an eigenstate of the corresponding operator, but any linear combination of eigenstates that fulfil this EV condition is valid.  This makes the condition much easier to satisfy and therefore defines a much larger set of admissible quantum states.
%n the particular case of Ashtekars choice\cite{ashtekar75} of $J_C$,
Besides, a non-trivial  subspace of the Hilbert space that  fulfils the constraints is easy to identify.  For the sake for  simplicity this example is described in the holomorphic picture, which is further explored in  section \ref{example} and  \cite{David2}. First, we define a coherent state (CS) as (see, e.g., \cite{David} or \cite{Hida}) $\Psi_f(\phi):=\dfrac{e^{\bar{f}_x\phi^y}}{e^{-f_x\Delta^{xy}\bar f_y}}$ (being $\Delta^{xy}$ the covariance matrix of $D\mu_\xi$, as in \eqref{eq:characteristiccomplex} of section \ref{example}).  
%The real and imaginary parts of the complex coefficients $f_x$ (relative to our choice of $J_C$ as indicated below in \eqref{eq:holomorphic_coordinates}) can be identified with some classical fields and momenta $\varphi_x,\delta_{xy}\pi^y$, as in the usual definitions of coherent states. 
As shown in the appendix, the EV in these CS of the quantization of matter supermagnitudes turn out to be the classical sH and sM evaluated in the holomorphic test functions  $f_x,\bar f_x$, and, in turn, in some classical fields and momenta $\varphi_x,\delta_{xy}\pi^y$ (see sec. \ref{example}). Thus, there exists a non-trivial subspace inside the Hilbert space  that fulfils the hybrid constraints, given by a subset of the CS whose $f_x$ fulfilled the constraints at the classical level (as in \cite{HKT76}).
%$\Psi_f(\phi):=\dfrac{e^{\bar{f}^xK_{xy}\phi^y}}{e^{-f^xK_{xy}\bar f^y/2}}$ 
%Note that this is physically consistent because Ashtekar's complex structure was proposed to preserve (upon quantization) the notion at each $\xi$ of positive energy particles moving to the future along the foliation in a way that was relative to the Hamiltonian structure of the classical field.
Nevertheless, this subspace may not be the largest one. The whole space can be characterized by taking the expectation value of the quantization of matter sH and sM under linear combinatons of CS, as they are dense in $L^2(\mathcal{N}^\prime,D\mu_\xi)$, and characterizing the constraints induced in the coefficients by eq. \eqref{hybridconstraints}. The characterization of this larger space can also be based on geometrodynamics. Consider any other geometric content $\xi^\prime\neq\xi$ which can be related with the current physical one, $\xi$, through some leaf deformations generated by $\mathcal{H}^{H}_\mu[\tau^\mu_{\xi^\prime\rightarrow\xi}(s)]$. As we will later prove, the hybrid constraints are preserved under hypersurface deformations. Then, we may consider $\forall\xi^\prime\in\mathcal{B}$ the set of CS associated to the classical $f_x$ that fulfilled the constraints. Once all of them are transported to $\xi$ through $\mathcal{H}^{H}_\mu[\tau^\mu_{\xi^\prime\rightarrow\xi}(s)]$, we  obtain a new set of quantum states that fulfil the constraints for $\xi$, which may not consist solely of CS.  This shows that the space of physical quantum states fulfilling \eqref{hybridconstraints} is non-empty  (\textit{assuming} that the constraints \eqref{constraints} can be solved for the classical fields).

%As a technical note, one may consider that the nulity required in these constraints for hybrid supermagnitudes that involve the EV of quantum local operators must be seen as the nulity of all EVs of hybrid operators constructed as contractions of any distribution with the supermagnitudes $f^x\mathcal{H}_{Hx}=0\ \forall\  f^x\in D'(\Sigma)$.\\

As argued in \cite{HKT76}, the Hamiltonian representation of Dirac's group of hsf. deformations together with these first class constraints ensure that we have successfully enforced the 3+1 equivalent to the general covariance in the hybrid theory.

%Besides, in relation with the dynamics that we will see in the following section, this implies that, on physical data, the total Hamiltonian function $f_H$ (Geometric and Quantum part, which contains the EV) is always null for the whole hybrid universe, similarly to the Wheeler-DeWitt equation, but in this case, this nulity applies at each leaf and the 3-metric and its momentum are still classical. 

\subsection{Dynamics for generic hybrid observables and preservation of first class constraints.}
For any  element of the hybrid algebra of observables the effect of the hsf. deformation given by a normal deformation of size of the lapse function, $N^x$, and tangential deformations given by  the shift vector, $N^{ix}$, must reproduce its dynamics throughout the foliation. The Hamiltonian representation of Dirac's algebra allows us to identify:
\begin{equation}
d_s F=\lbrace F,f_H\rbrace_H\;\forall F\in\mathcal{A}_H\quad\text{with}\quad 
f_H:=\mathcal{H}^H[N]+\mathcal{H}^H_{i}[N^i]\;.
\end{equation}
If $F$ is also lapse, shift and s-label dependent, its dynamics will be given by:
\begin{equation}
\label{functionalevolution}
d_s F=\lbrace F,f_H\rbrace_H+(\dot{N}^x\partial_{N^x}+\dot{N}^{ix}\partial_{N^{ix}}+\partial_{s})F\quad\forall F\in C^\infty(\mathcal{F},\mathbb{R})\;,
\end{equation}
where the lapse and shift functions contain their own external dynamics, given by  $\dot{N}^{x}:=\partial_s {N}^{x}$ and  $\dot{N}^{ix}:=\partial_s {N}^{ix}$, given by the choice of foliation.

Let us now examine the case of the Hamiltonian and momenta constraints and consider the effect of the Hamiltonian field over $\mathcal{H}_\mu^H[\tau]$. We have shown that $\mathcal{H}_\mu^H[\tau]$ fulfil (\ref{closeHHglobal}, \ref{closeHMglobal} and \ref{closeMMglobal}), which are linearly proportional to the constraints, and, being $f_H$ a linear combination of sH and sM, we have:
\begin{equation}
\lbrace \mathcal{H}^H_{\mu }[\tau],f_H\rbrace_H\simeq 0\quad \forall \mu=0,\cdots,3\quad\forall\tau\in\mathcal{N}
\end{equation}
Thus, given this result and according to \eqref{functionalevolution}, the only non-vanishing dynamics of the supermagnitudes on shell is given by its $N,N^i$ dependence. Such dependence was not present in the classical field theoretical case, but is acquired through $Q_\xi$, $\langle,\rangle_\xi$ and the covariant section nature of the quantum states. This dependence must be made null for the constraints to be preserved during the dynamics. Given that this must be true for any foliation or path along possible hsfs., it must be true for any choice of evolution for lapse and shift. Consequently, we must enforce:
\begin{equation}
\label{Lapseshiftindependence}
\frac{\delta }{\delta N^{\mu}}\mathcal{H}_\nu^H[\tau]=
\frac{\delta}{
\delta 
{N^\mu(x^\prime)}}\mathcal{H}_\nu^Q[\tau]= 0\ \forall\ \mu,\nu=0,\cdots ,3
\end{equation}
which implies, at the level of the quantum connection,
\begin{equation}
\label{lapseshiftconnection}
\langle\Psi\vert\left(\frac{\delta}{
\delta {N^{\mu}(x^\prime)}}Q_\xi(\mathcal{H}^M_\nu[\tau])+\left[\hat\Gamma_{N^\mu}(x^\prime),Q_\xi(\mathcal{H}^M_\nu[\tau])\right]\right)\Psi\rangle_\xi=0
\end{equation}
Note, therefore, that if eq. \eqref{Lapseshiftindependence} is not fulfilled, even though the constraints are conserved under the Hamiltonian dynamics given by  $\lbrace,f_H\rbrace_H$, they would not be conserved under the curve the lapse and shift functions follow. Therefore, at some  step in the evolution the hybrid states would abandon the submanifold of null supermagnitudes and the path independence would be lost from that point onward, falling into the Hamiltonian analogue of losing general covariance.
Note  also that eq. \eqref{lapseshiftconnection} is equivalent to extend equation \eqref{RequirementHM}  to apply for derivatives not only w.r.t. $h,\pi_h$ but also for lapse and shift (as the classical field sH and sM do not depend on them). This starts depicting a general trend: symmetry generating functions at the quantum level must present the same kinematical relations and leaf dependence as the classical field theory to be consistent with geometrodynamics, and the quantum connection is chosen to provide such compatibility  when the quantization procedure and scalar product are chosen in a leaf dependent way.
\begin{comment}
Furthermore, it could be argued that the generating functions of symmetries are to be considered as kinematical objects, and as such, even if built in terms of leaf dependent objects through this leaf dependent quantization, they must be intrinsically independent of lapse and shift and not acquire further dependences on $h$ due to the quantization procedure.\\
\end{comment}

\begin{comment}
Lastly, based on the representations over the hybrid phase space of hsf. deformations, we can evolve a general hybrid functional over hybrid phase space, $F(\Psi,\bar\Psi,h_{ij},\pi^{ij}):\mathcal{M}_H\rightarrow\mathbb{C}$, through the whole foliation. Infinitesimally, a general deformation of the hsf. acts on $F$ through its phase space representation as:
\begin{equation}
\dfrac{\delta F}{\delta s}=\pb{F}{\mathcal{H}_{Hx}}{H}N^x s+\pb{\Psi}{\mathcal{H}_{iHx}}{H}N^{ix}=\pb{F}{f_H}{H}\; .
\end{equation}
To write this in a more compact notation, we have defined the lapse-shift-dependent Hybrid Hamiltonian function as:
\begin{multline}
    f_H(N,N^i):\mathcal{M}_H\rightarrow\mathbb{R}\\
    f_H(h,\pi,\Psi,\bar\Psi;N,N^i):=\mathcal{H}_xN^x+\mathcal{H}_{ix}N^{ix}+\langle \Psi\mid\hat H_Q\mid\Psi\rangle .
\end{multline}
\end{comment}

\subsection{Hybrid equations of motion and properties of the dynamics.}
Having introduced above the evolution equations for any hybrid observable, we consider illustrative to write down explicitly the equations of motion governing the dynamics of the 3-metric, its associated momenta and the quantum states. For the 3-metric from a hsf. to a neighbouring one can only be by the extrinsic curvature and the spatial diffeomorphism, its differential equation is as in ordinary geometrodynamics:
\begin{equation}
\frac{d}{ds} h_{ij}(x)=\lbrace h_{ij},f_H\rbrace_G=2N(x)G_{ijkl}(x)\pi^{kl}(x)+2(D_iN^k(x))h_{kj}(x)\;.
\end{equation}
On the other hand, the diff. eq. for $\pi_h$ is coupled to quantum matter:
\begin{equation}
\frac{d}{ds}  \pi^{ij}(x)=\lbrace \pi^{ij},f_H\rbrace_G=-\partial_{h_{ij}(x)}{H_G}-\langle\Psi\mid Q(\partial_{h^{ij;x}}\mathcal{H}^M[N])\mid\Psi\rangle,
\end{equation}
 where the first term is the usual from matterless geometrodynamics and the second one is the coupling with quantum matter, which, by mercy of the constraints on the quantum connection, has no contribution from the quantum sM, given \eqref{sMconnection}, and the derivative w.r.t. $h$ of the quantum sH fulfills \eqref{sHconnection}.  We dub this last term \textit{backreaction} (of quantum matter into gravity). Lastly, the quantum state evolves as:
\begin{multline}
\label{quantumeveolution}
\frac{d}{ds}  \Psi=\lbrace \Psi,f_H\rbrace_Q+\lbrace \Psi,f_H\rbrace_G+\dot N^x\partial_{N^x}\Psi+\dot N^{ix}\partial_{N^{ix}}\Psi
=\\
\left(\dfrac{-i}{\hbar}\hat H-\lbrace h_{ij;x},f_H\rbrace_G\hat\Gamma_{h_{ij;x}}-\dot N^x\hat\Gamma_{N^x}-\dot N^{ix}\hat\Gamma_{N^{ix}}\right)\Psi
\end{multline}
where we remind that such quantum connections must fulfill  eqs. (\ref{antiselfadjointconnection},\ref{sHconnection},\ref{sMconnection},\ref{lapseshiftconnection}), and the notation for continuous indices implies the contraction through integration over $\Sigma$ of the local connections with the $s$-derivative of their associated local variables (e.g. $\lbrace h_{ij;x},f_H\rbrace_G\hat\Gamma_{h_{ij;x}}=\int\limits_\Sigma d^3x\partial_{\pi^{ij}(x)}f_H\hat\Gamma_{h_{ij}}(x)$, and equivalently $-\dot N^x\hat\Gamma_{N^x}-\dot N^{ix}\hat\Gamma_{N^{ix}}$). Besides, we remind that the quantum Hamiltonian operator $\hat H$ and the connections are $(h,N,N^i)$-dependent.

This system of equations defines the hybrid dynamics. Note that it cannot be unitary, given the non-linearity arising from the mutual backreaction. Nevertheless, norm conservation is assured by the quantum connection,  even if the definition of the scalar product (and thus, of the Hilbert space) is leaf dependent, because the quantum part of \eqref{quantumeveolution}  is given by the  antiselfadjoint operator $-i\hat H$, and, thus, it conserves norm. On the other hand, the connections appear acting as operators on $\Psi$ are not anti-self-adjoint  operators for $\langle,\rangle_\xi$. Thus, if the scalar product $\langle,\rangle_\xi$ associated to a fixed $\xi\in\mathcal{B}$, they would induce a change of norm. In fact, the main property of such connection is that it always maps the quantum state \textit{out} of the original Hilbert space to which it originally belonged (given by $\langle,\rangle_\xi$), in order to fit it in the infinitesimally neighbouring Hilbert space under the path along the section of scalar products $\langle,\rangle_{\xi+\delta\xi}$, given by the infinitesimal change $\delta\xi$ of $(h,\pi_h,N,N^i)$. Nevertheless, the quantum connection compensates for the change of the scalar product itself under changes of $\xi$, as seen in eq. \eqref{antiselfadjointconnection}, and for such leaf dependent definition of scalar product, we achieve norm conservation, as seen explained under \eqref{HermitianConnection}.  In other words, we cannot compute the norm of a state in the future (associated to $\xi^\prime$ with the scalar product of the past (associated to $\xi$). We may summarize this discussion by applying eq. \eqref{functionalevolution} to the norm:
\begin{equation}
\frac{d}{ds} \langle\Psi\mid\Psi\rangle_\xi=\lbrace\langle\Psi\mid\Psi\rangle_\xi,f_H\rbrace_G+\dot{N}^{\mu x}\partial_{N^{\mu x}} \langle\Psi\mid\Psi\rangle_\xi+\dfrac{i}{\hbar}\langle\Psi\mid[\hat H,\mathbb{I}]\Psi\rangle_\xi=0
\end{equation}
where the first two terms are null because the {\color{black} image} through horizontal sections of the tangent fields on $\mathcal{B}$ conserve the $\xi$-dependent scalar product, while the quantum PB is null because it involves the commutator of the self-adjoint (for the given $\langle,\rangle_\xi$) Hamiltonian operator with the identity.}
{\color{black}
Contrarily to the case of the horizontal lift of the change of $\xi$, the term with the quantum Hamiltonian is inner to the original Hilbert space, not accounting for changes of the base, but for the vertical evolution in the leaf.

Therefore, one may decompose the hybrid dynamics of the quantum state as the sum of a vertical vector field on the fibration and a horizontal one. We must remember that the quantum state as an element of $\mathcal{M}_Q$ is defined as $\sigma_\Psi\vert_\xi$, the evaluation of a covariant section $\sigma_\Psi\in\mathcal{M}_s$ on a given point $\xi\in\mathcal{B}$. In this context, the vertical vector, which is given by the quantum PB that provides the Schrödinger-like term,  provides the change within the space of $\mathcal{M}_s$, the covariant sections, but keeps evaluation point $\xi\in\mathcal{B}$ static.  On the other hand, the horizontal one, given by the  image through $\sigma_\Psi$ of the gravitational PB and the tangent vector to the curve defining the lapse and shift along the foliation, provides the change of the evaluation point in $\mathcal{B}$, without changing the covariant section $\sigma_\Psi$. Together they yield the total evolution of the quantum state, accounting both for the change of Hilbert space due to the change of $\xi$ and for the change of quantum state within the original Hilbert space, as illustrated in Figure \ref{fig:HorizontalVertical}.

}
\section{Example: quantum connection for holomorphic quantization of the real scalar field.}
\label{example}
Let us firstly fix notation. We will denote $\xi$ to any of $h_{ij},\pi^{ij},N,N^i$, as our results will be valid for all of them. We will formally work with $\partial_\xi$ as a derivative w.r.t. a parameter. Notice that for any tensor $\mathcal{T}$ that does not have $d\xi$ nor $\partial_\xi$ entries (just products of $d\varphi^x,d\pi^x,\partial_{\varphi^x},\partial_{\pi^x}$  with coefficients in  $C^\infty(\mathcal{F}_C)$), the Lie derivative  $\mathcal{L}_{X_\xi}\mathcal{F}\;\forall\, X_\xi\in T\mathcal{B}$, such that $X_\xi=v_\xi^x(\xi,\Psi)\partial_{\xi^x}$, is equivalent to $v_\xi\partial_{\xi^x}$ acting on the coefficients of $\mathcal{T}$. This is the kind of tensors (with parametric dependence on $\xi$) we need for geometric quantization on a Cauchy hypersurface for curved spacetime. Thus, by linearity, it is justified to use simply the  parametric derivative $\partial_\xi\mathcal{F}$ to alleviate the notation.

Consider now the real scalar fields and momenta $\varphi_x,\pi^x\in\mathcal{M}_F$ such that, as in \cite{HKT76}:
\begin{equation}
    \partial_\xi(\varphi_x)=\partial_\xi(\pi^x)=0 \qquad \forall \xi\in[h,\pi_h,N,N^i]\;.
\end{equation}
Therefore, the classical field symplectic structure, which is given by eq. \eqref{omegaC}, is leaf independent, as in \eqref{LeafIndependence}. % Formally $\mathcal{L}_{X_\xi}\pi^\star_F\omega_C=0$, being $\pi_C$ the projection  as in\eqref{LeafIndependence} from $\mathcal{F}_C$ to $\mathcal{M}_F$.
If we switch our description to distributions $\varphi_x\rightarrow\varphi^x$, we have:
\begin{equation}
    \omega_C=-\delta_{xy}d\varphi^x\wedge d\pi^y\;,
\end{equation}
where $\delta_{xy}$ is symmetric and appropriately contracts distributions (its inverse  $\delta^{xy}$ acts on test functions, see \cite{David2} for details on these $\delta$ operators). This symplectic structure is diffeomorphic to the previous one and is still leaf independent. Now, $\varphi^x=\delta^{xy}\varphi_y$ is leaf dependent, as so is $\delta^{xy}$, but $\partial_\xi(\delta_{xy}\varphi^y)=\partial_\xi(\delta_{xy}d\varphi^x)=0$, to be consistent with $\partial_\xi\varphi_x=0$.

Now, to prepare the space for quantization,  we add a complex structure $J_C$ to $\mathcal{M}_F$ that may depend on $\xi\in\mathcal{M}_C\times\mathcal{M}_N$. For example, the complex structure used in \cite{CCQ04} and proposed in \cite{ashtekar75} exhibits this kind of $\xi$-dependence (or $s$-dependence if spacetime is seen as a given background structure, as in \cite{agullo2015unitarity} and \cite{David2}). Generically, $J$ is written as:
\begin{equation}
-J_C= \partial_{\varphi^y}\otimes[A^{y}_xd\varphi^x+\Delta^{y}_xd\pi^x]+\partial_{\pi^y}\otimes[D^{y}_xd\varphi^x-(A^t)^y_xd\pi^x]
\end{equation}
with $A^2+\Delta D=-\mathbb{I}$, $\Delta^t=\Delta$, $D^t=D$, $A\Delta=\Delta A^t$ and $A^tD=DA$. We define $K^{x}_y$ as the inverse of $\Delta^{x}_y$. Note that $\delta^{xy}$ and $\delta_{yx}$ can be used to lower and raise indices, with function-distribution implications.
Thus, $\mathcal{M}_F$ has a single symplectic form, but a $\xi$-parametric family of K\"ahler structures (we can define $\mu_C=\omega_C(\cdot,J_C\cdot)$ as the  Riemannian tensor $ \forall \xi\in\mathcal{B}$) that yields different quantization mappings at each $\xi\in\mathcal{B}$. In this context, we  may define holomorphic and antiholomorphic coordinates w.r.t. $J$, $\phi^x$ and $\bar\phi^x$:
\begin{equation}
\label{eq:holomorphic_coordinates}
\phi^x=\dfrac{1}{\sqrt{2}}\left(\varphi^x+i\left(A^x_y\varphi^y+\Delta^x_y\pi^y\right)\right)\quad\text{and}\quad 
{\bar\phi}^x=\dfrac{1}{\sqrt{2}}\left(\varphi^x-i\left(A^x_y\varphi^y+\Delta^x_y\pi^y\right)\right)\;.
\end{equation}
In these variables, we can write $\omega_C=iK_{xy}d\phi^x\wedge d\bar\phi^y$, and the corresponding PB:
\begin{equation}
    \lbrace f,g\rbrace_M=\delta^{xy}\partial_{[\varphi^x}f\partial_{\pi^y]}g=-i\Delta^{xy}\partial_{[\phi^x}f\partial_{\bar\phi^y]}g\quad\forall f,g\in C^\infty(\mathcal{M}_F)
\end{equation}
The holomorphic coordinates depend on $\xi$, as so does $J_C$, so we can differentiate \eqref{eq:holomorphic_coordinates}:
\begin{equation}
\label{eq:dxiphi}
\partial_\xi\phi^x=\dfrac{1}{\sqrt{2}}\left(i(\dot{A}^{xu}\delta_{uy}\varphi ^y+\dot{\Delta}^x_y\pi^y)+\dot{\delta}^{xy}\delta_{yz}\varphi^z\right)=\dfrac{1}{2}\left(\mathrm{S}^x_y\phi^y+\mathrm{T}^x_y\bar\phi^y\right)
\end{equation}
where $\mathrm{T}^x_u:=i\dot{A}^{xy}\delta_{yu}-\dot{\Delta}^{xy}(K_{yz}+iK_{yz} A^z_u)+2\dot{\delta}^{xy}\delta_{yu}+i\dot{\delta}^{xy}\delta_{yz}A^{zv}\delta_{vu}$ and $\mathrm{S}^x_u:=i\dot{A}^{xy}\delta_{yu}+\dot{\Delta}^{xy}(K_{yz}-iK_{yz} A^z_u)+i\dot{\delta}^{xy}\delta_{yz}A^{zv}\delta_{vu}$.  Analogously, $\partial_\xi(\bar\phi^x)=\dfrac{1}{2}\left(\bar{\mathrm{T}}^x_y\phi^y+\bar{\mathrm{S}}^x_y\bar\phi^y\right)$. The  second equality in \eqref{eq:dxiphi} is reached through the change of variables:
\begin{equation}
\label{eq:varphipi}
\varphi_x=\delta_{xy}\dfrac{\phi^y+{\bar\phi}^y}{\sqrt{2}}\quad\text{and}\quad\pi^x=-i\delta^{xz} K_{zy}\dfrac{\phi^y-{\bar\phi}^y}{\sqrt{2}}-\delta^{xz} K_{zu}A^{uv}\delta_{vy}\dfrac{\phi^y+{\bar\phi}^y}{\sqrt{2}}
\end{equation}

Regarding 1-forms, we must note that the differential structure for matter fields does not depend on $\xi$, and, thus, $\partial_\xi d\phi^x=d(\partial_\xi\phi^x)$, while  $\partial_\xi d(\delta_{xy}\varphi^y)=\partial_\xi d\pi^x=0$. 

%The one forms $d\phi^x$, $d{\bar\phi}^x$ depend on $\xi$ just with the exterior derivative of this last expression and its conjugate, \textit{i.e.} $\partial_\xi d\phi^x=d(\partial_\xi\phi^x)$. This is because $d\varphi_x=\delta_{xy} \dfrac{d\phi^y+d{\bar{\phi}}^y}{\sqrt{2}}$ (and another linear combination for $d\pi^x$). In other words, this exterior derivative, as an operator for the classical matter fields, does not depend on $\xi$.

Let us now consider the holomorphic (relative to $J_\xi$) polarization through the choice of a $\xi$-dependent symplectic potential, $\theta$, of the form \cite{David2}:
\begin{equation}
\theta=-i\bar\phi^xK_{xy}d\phi^y\;,
\end{equation}
where the $\xi$-dependent terms are closed forms (easy to check in terms of $d\varphi_x$, $d\pi^x$, because $\phi^x,\bar\phi^x$ is a $\xi$-dependent chart for the same manifold as the $\xi$-independent chart $\varphi_x, \pi^x$), \textit{i.e.} $d\theta=\omega_C\;\forall\,\xi\in\mathcal{B}$ and $d(\partial_\xi\theta)=0\;\forall\,\xi\in\mathcal{B}$. We can verify that:
\begin{equation}
\partial_\xi\theta=-i\bar\phi^x\dot{K}_{xy}d\phi^y-i\partial_\xi(\bar\phi^x)K_{xy}d\phi^y
-i\bar\phi^x{K}_{xy}\partial_\xi(d\phi^y)
=\dfrac{-i}{2} \left(\bar\phi^v\mathcal{K}_{vu}d\bar\phi^u+\phi^v\bar{\mathcal{K}}_{vu}d\phi^u\right)
\end{equation}
where the dot notation implies $\partial_\xi$, \textit{i.e.} $\dot K_{xy}=\partial_\xi (K_{xy})$, and we have defined
\begin{equation}
\mathcal{K}_{vu}:=\delta_{vx}\left(\dot K^{xy}+i\partial_\xi(\delta^{xw}K_{wy}A^{yz})\right)\delta_{yu}=K_{vx} \mathrm{T^x_u}\;,
\end{equation}
with $\dot K^{xy}=\partial_\xi(\delta^{xu}K_{uv}\delta^{vy})$. Note that $\delta^{wx}K_{xy}A^{yz}=\left(\delta^{wx}K_{xy}A^{yz}\right)^t$, so $\mathcal{K}_{vu}$ is symmetric.
\subsection{Leaf dependence of prequantization of operators}We will consider the usual presctiption for geometric quantization for linear operators:
\begin{equation}
\label{eq:geometric_quantization}
Q(f)\Psi=(f-i\hbar\tilde{\nabla}_{X_f})\Psi=\big[f-i\hbar X_f-\hbar^{-1}\theta(X_f)\big]\Psi
\end{equation}
with $X_f:=\lbrace,f\rbrace_M$ and $\tilde{\nabla}_{X_f}$ being the covariant derivative for the connection $\dfrac{1}{i\hbar}\theta$. This is the usual holomorphic prequantization for linear functions, $Q(f_x\phi^x)=f_x\phi^x-f_x\Delta^{xy}\partial_{\bar\phi^y}$ and $Q(f_x\bar\phi^x)=f_x\Delta^{xy}\partial_{\phi^y}$. Note that when acting on holomorphic functions $\Psi(\phi)$, the $\partial_{\bar\phi^y}$ is always null.
Now we can see how this construction depends on $\xi$ as:
\begin{equation}
\label{eq:partial_xi_Q}
\partial_\xi Q(f)=\partial_\xi(f) -i\lbrace,(\partial_\xi f)\rbrace_{C}-\theta_\xi(\lbrace,f\rbrace_{M})-\theta(\lbrace,\partial_\xi(f)\rbrace_{M})=Q(\partial_\xi(f))-\theta_\xi(X_f)\;,
\end{equation}
where $\theta_\xi:=\partial_\xi(\theta)$. Note that, given that the symplectic form $\omega_C$ does not depend on $\xi$, neither does the PB structure. Note also that $\theta_\xi(X_f)$ is the term that the quantum connection must cancel with to obtain  \eqref{RequirementHardcore}.  These terms, for linear functions on the holomorphic field, $f[\phi]:=f_x\phi^x$ with $X_{f[\phi]}=\lbrace \cdot,f_w\phi^w\rbrace_M=-if_w\Delta^{wu}\partial_{\bar\phi^u}$, or on the antiholomorphic field, $f[\bar\phi]:=f_x\bar\phi^x$ with $X_{f[\bar\phi]}=if_x\Delta^{xy}\partial_{\phi^y}$, are given by:
    \begin{equation}
    \label{fallophi}
    \theta_\xi(X_{f[\phi]})=
    \dfrac{-1}{2}\bar\phi^v\mathcal{K}_{vw}\Delta^{uw}f_w\quad\text{and}\quad\theta_\xi(X_{f[\bar\phi]})=\dfrac{1}{2}\phi^v\bar{\mathcal{K}}_{vw}\Delta^{uw}f_w
    \end{equation}
We will come back to these terms to check if the quantum connection cancels with them.
\subsection{Leaf dependence of the Gaussian measure, connection and covariant sections.}
\label{sec:leafdependence}
On the other hand, as in \cite{David,David2}, we can define a Gaussian measure $D\mu$ over $\mathcal{M}_F$, understood as the set of complex distributions $\mathcal{N}^\prime_{\mathbb{C}}$. We invoke Minlos' theorem, such that we can characterize $D\mu$ in terms of the Hermitian form of the Kahler structure on $\mathcal{M}_F$: 
\begin{equation}
\label{eq:characteristiccomplex}
C(\rho,\bar\rho)=\int\limits_{\mathcal{M}_F\sim\mathcal{N}^\prime_{\mathbb{C}}} D\mu(\phi,\bar\phi)e^{i(\rho_x\bar{\phi}^x+\bar\rho_y\phi^y)}=e^{-\rho_x\Delta^{xy}\bar\rho_y}\quad\forall\rho,\bar\rho\in\mathcal{N}_\mathbb{C}
\end{equation}
where $\Delta^{xy}$ is the covariance of $D\mu$, identified with a $\xi$-dependent entry of $J_C$.  This Gaussian measure defines the scalar product (at each $\xi$) of our $\xi-$dependent Hilbert spaces of square integrable functionals over $\phi^x$.  To understand norm preservation, we must understand how a Gaussian measure changes at each $\xi$. Thus, we have to compute:
\begin{equation}
\label{partialxicaracteristic}
\partial_\xi C(\rho,\bar\rho)=-\rho_x\dot{\Delta}^{xy}\bar\rho_y C(\rho,\bar\rho)=\int\limits_{\mathcal{M}_F\sim\mathcal{N}^\prime_{\mathbb{C}}} D\mu(\phi,\bar\phi)\partial_{\bar\phi^x}\dot\Delta^{xy}\partial_{\phi^y}e^{i(\rho_x\bar{\phi}^x+\bar\rho_y\phi^y)}
\end{equation}
\begin{comment}
On the other hand, given that the integration variables $\phi,\bar\phi$ depends on $\xi$, we can undo the change of variables back to $\varphi^x,\pi^x$, obtaining a Gaussian measure  $D\mu(\varphi,\pi)$ (with non-null diagonal terms in the covariance matrix) and considering $\phi^x$, $\bar\phi^x$ as functions of $\varphi_x,\pi^x$. In this sense, the characteristic functional can also be written as:
\begin{equation}
C(\rho,\bar\rho)=\int\limits_{\mathcal{M}_F\sim\mathcal{N}^\prime_{\mathbb{C}}} D\mu(\phi,\bar\phi)e^{i(\rho_x\bar{\phi}^x(\varphi,\pi)+\bar\rho_y\phi^y(\varphi,\pi))}=e^{-\rho_x\Delta^{xy}\bar\rho_y}\quad\forall\rho,\bar\rho\in\mathcal{N}_\mathbb{C}
\end{equation}
\end{comment}
Now, consider a function given by a linear combination of plane waves, 
$f(\phi,\bar\phi)=\sum\limits_jf_je^{i(\rho_x^j\bar{\phi}^x+\bar\rho_y^j\phi^y)}$. Taking into account that this class of functions are dense on  any $L^2(\mathcal{N}^\prime_\mathbb{C},D\mu)$, the discussion that follows is general. Taking its $\xi$-derivative we see:
\begin{multline}
\label{dxif}
\partial_\xi f(\phi,\bar\phi)=\sum\limits_j\partial_\xi(f_j)e^{i(\rho_x^j\bar{\phi}^x+\bar\rho_y^j\phi^y)}+\sum\limits_jf_j\partial_\xi(e^{i(\rho_x^j\bar{\phi}^x+\bar\rho_y^j\phi^y)})=\\
=\sum\limits_j\partial_\xi(f_j)e^{i(\rho_x^j\bar{\phi}^x+\bar\rho_y^j\phi^y)}+\sum\limits_jf_j(\partial_\xi(\phi^x)\partial_{\phi^x}+\partial_\xi(\bar\phi^x)\partial_{\bar\phi^x})(e^{i(\rho_x^j\bar{\phi}^x+\bar\rho_y^j\phi^y)})
\end{multline}
Here, the $\rho$, $\bar\rho$ are just $\xi$-independent coefficients  for the plane waves, while $\phi^x$, $\bar\phi^x$ must be regarded as linear combinations on $\varphi_x,\pi^x$ with $\xi$-dependent coefficients. 
%In this sense, $f$ is still a function over the $\xi$-independent $\varphi_x$ and $\pi^x$, which justifies the third equality of \eqref{dxif}.\\

On the other hand, the $D\mu$ integration of $f(\phi,\bar\phi)$ depends on $\xi$ as:
\begin{equation}
\partial_\xi\int\limits_{\mathcal{M}_F\sim\mathcal{N}^\prime_{\mathbb{C}}} D\mu(\phi,\bar\phi) f(\phi,\bar\phi)=\sum\limits_j \partial_\xi(f_j)C(\rho^j,\bar\rho^j)+\sum\limits_j f_j\partial_\xi C(\rho^j,\bar\rho^j)\;.
\end{equation}

With some straightforward algebra, making use of (\ref{partialxicaracteristic}, \ref{dxif}) this is rewritten as:
\begin{equation}
\label{change_functional}
\partial_\xi\int\limits_{\mathcal{M}_F} D\mu(\phi,\bar\phi) f(\phi,\bar\phi)=\int\limits_{\mathcal{N}^\prime_{\mathbb{C}}} D\mu(\phi,\bar\phi)\left(\partial_{\bar\phi^x}\dot\Delta^{xy}\partial_{\phi^y}-\partial_\xi(\phi^x)\partial_{\phi^x}-\partial_\xi(\bar\phi^x)\partial_{\bar\phi^x}+\partial_\xi\right)f(\phi,\bar\phi)
\end{equation}
where $\partial_\xi\phi^x$ represents the linear combination of $\phi,\bar\phi$ given by \eqref{eq:dxiphi} and $\partial_\xi{\bar\phi}^x$, its complex conjugate. Besides, making use of Skorokhod integral as dual to Malliavin derivatives (see its applications to QFT in \cite{David}), we obtain that, in the former expression, the operator $(\partial_{\bar\phi^x}\dot\Delta^{xy}\partial_{\phi^y}-\partial_\xi(\phi^x)\partial_{\phi^x}-\partial_\xi(\bar\phi^x)\partial_{\bar\phi^x})$ can be substituted by:
\begin{equation}
\label{XI}
\Xi:=-\dfrac{1}{4}        \left(\partial_{\phi^x}\Delta^{xy}\mathcal{K}_{yv}\Delta^{vz}\partial_{\phi^z}+\phi^x\bar{\mathcal{K}}_{yv}\phi^y+\partial_{\bar\phi^x}\Delta^{xy}\bar{\mathcal{K}}_{yv}\Delta^{vz}\partial_{\bar\phi^z}+\bar\phi^x{\mathcal{K}}_{yv}\bar\phi^y\right)\;.
\end{equation}

If we compute the change of norm under $\xi$ of our wave functionals, we must consider simply that $f(\phi,\bar\phi)=\bar\Psi(\bar\phi)\Psi(\phi)$. In this context, the norm conservation condition
\begin{equation}
\partial_\xi\int\limits_{\mathcal{M}_F\sim\mathcal{N}^\prime_{\mathbb{C}}} D\mu(\phi,\bar\phi) \bar\Psi(\bar\phi)\Psi(\phi)=0\;,
\end{equation}
taking into account that $\partial_\xi\Psi(\phi)=-\Gamma_\xi\Psi(\phi)$ and $\partial_\xi\bar\Psi(\bar\phi)=-\bar\Gamma_\xi\bar\Psi(\bar\phi)\;$ (
as stated in \eqref{statesdependonxi}), implies $\Xi=\Gamma_\xi+\bar\Gamma_\xi$. This defines a candidate for the ($d\xi$-coordinate of) connection:
\begin{equation}
\label{eq:quantum_connection}
\Gamma_\xi:=-\dfrac{1}{4}        \left(\partial_{\phi^x}\Delta^{xy}\mathcal{K}_{yv}\Delta^{vz}\partial_{\phi^z}+\phi^x\bar{\mathcal{K}}_{yv}\phi^y\right)\,\text{and}\;\,\bar\Gamma_\xi:=-\dfrac{1}{4}        \left(\partial_{\bar\phi^x}\Delta^{xy}\bar{\mathcal{K}}_{yv}\Delta^{vz}\partial_{\bar\phi^z}+\bar\phi^x{\mathcal{K}}_{yv}\bar\phi^y\right)\quad
\end{equation}
This connection derived from norm conservation already fulfils \eqref{antiselfadjointconnection}. In fact, we can derive (not impose) the covariant section nature of the quantum states for this very same connection from the derivative of the polarization $\mathcal{P}$ of geometric quantization. The usual definition of polarization implies $\mathcal{P}:=\lbrace X\in T\mathcal{M}_F\, \vert \;\theta(X)=0\rbrace$, and the set of states adapted to the polarization, defining our quantum states, is given by $\mathcal{F}:=\lbrace\Psi(\phi)\in (\mathcal{N})^\prime \; \vert \;d\Psi(X)=0\;\forall \;X\in\mathcal{P}\rbrace$, being $\Psi(\varphi,\pi)$ functionals over $\mathcal{M}_F$. This is how the holomorphic nature of $\Psi(\phi)$ is obtained from $\theta$. However, being this  $\theta$  $\xi$-dependent (because ``holomorphic" is relative to $J_C$), we have a parametric family of polarizations $\mathcal{P}_\xi$ (one for each $\theta\vert_\xi$), related through the condition $\partial_\xi(\theta(X))=\theta_\xi(X)+\theta(\partial_\xi(X))=0$). In turn, this defines a parametric family of holomorphic functions, $\mathcal{F}_\xi$ (one for each $\mathcal{P}_\xi$), related through $\partial_\xi (d\Psi(X))=d(\partial_\xi\Psi)+d\Psi(\partial_\xi X)=0\;\forall X\in\mathcal{P}_\xi$), hence, the section nature of $\Psi$. After some calculations,  the same quantum connection as in \eqref{eq:quantum_connection} is derived, with the freedom to include  in it some extra terms. Of course, the correct mathematical tools for these computations are the parallel transported sections on vector bundles for each of these elements, but we have used an oversimplified notation.
\subsection{Covariant derivatives of linear operators.}
To check \eqref{RequirementHM} or \eqref{RequirementHardcore}, we must consider first that, for any quantum operator $\hat O$:
\begin{equation}
\partial_\xi\int\limits_{\mathcal{N}^\prime_C}D\mu \bar\Psi(\bar\phi)\hat O\Psi=\int\limits_{\mathcal{N}^\prime_C}D\mu\left( \Xi\bar\Psi(\bar\phi)\hat O\Psi-\bar\Gamma\bar\Psi(\bar\phi)\hat O\Psi-\bar\Psi(\bar\phi)\Gamma\hat O\Psi+\bar\Psi(\bar\phi)\partial_\xi\hat O\Psi\right)\;.
\end{equation}

Now, given that $\Xi=\Gamma_\xi+\bar\Gamma_\xi$, we can write:
\begin{equation}
\partial_\xi\int\limits_{\mathcal{N}^\prime_C}D\mu \bar\Psi(\bar\phi)\hat O\Psi=\int\limits_{\mathcal{N}^\prime_C}D\mu \bar\Psi(\bar\phi)\nabla_\xi(\hat O)\Psi=\int\limits_{\mathcal{N}^\prime_C}D\mu \bar\Psi(\bar\phi)(\partial_\xi\hat O+[\Gamma_\xi,\hat O])\Psi\;.
\end{equation}
Now, using \eqref{eq:partial_xi_Q}, it is immediate that, in order to reproduce \eqref{RequirementHardcore}, we must have:
\begin{equation}
\label{eq:integral_correction}
\int\limits_{\mathcal{N}^\prime_C}D\mu\bar\Psi(\bar\phi) \left(-(\theta_\xi)(X_f)+[\Gamma_\xi,Q(f)]\right)\Psi(\phi)=0
\end{equation}
This is immediately checked for linear functions. For  $f_x\phi^x$, using \eqref{fallophi}, we have:
\begin{equation}
\label{eq:correction_phi}
\zeta_{f_x\phi^x}:=\left(-(\theta_\xi)(X_{f_x\phi^x})+[\Gamma_\xi,Q({f_x\phi^x})]\right)=\dfrac{1}{2}\bar\phi^x\mathcal{K}_{xy}\Delta^{yz}f_z-\dfrac{1}{2}\partial_{\phi^x}\Delta^{xy}\mathcal{K}_{xy}\Delta^{yz}f_z\;,
\end{equation}
which becomes null once  inside the EV expression, as in \eqref{eq:integral_correction}. This is shown making use of the Malliavin derivative-Skorokhod integral duality (in this case,  equivalent to substituting $\partial_{\phi^x}\Delta^{xy}\rightarrow\bar\phi$). On the other hand, for $f_x\bar\phi^x$, making use \eqref{fallophi}, it is immediate that $-(\theta_\xi)(X_{f_x\bar\phi^x})+[\Gamma_\xi,Q({f_x\bar{\phi}^x})]=0$, even before integration.

With this, we have that $\langle\Psi\vert\nabla_\xi Q(\phi^x)\Psi\rangle=\langle\Psi\vert Q(\partial_\xi\phi^x)\Psi\rangle$ and $\nabla_\xi Q(\bar\phi^x)=Q(\partial_\xi\bar\phi^x)$, as desired. Thus, by the linearity of $\varphi^x$ and $\pi^x$ on $\phi^x$, $\bar\phi^x$ as in \eqref{eq:varphipi}, we find:
\begin{equation}
    \langle\Psi\vert\nabla_\xi Q(\delta_{xy}\varphi^y)=0\quad\text{and}\quad \langle\Psi\vert\nabla_\xi Q(\pi^y)\Psi\rangle=0\;.
\end{equation}
\subsection{Covariant derivative of polynomial operators, ordering and supermagnitudes.}

To begin with, it is well known that we cannot extend the geometric quantization procedure \eqref{eq:geometric_quantization} to arbitrary polynomials. Instead, we define the quantization procedure $Q$ for each monomial as the ordered product of the quantization of the fields, whose quantization is given as if they were linear, by \eqref{eq:geometric_quantization}. The choice of ordering defines different quantization procedures.   Notice that we can try to choose a fixed ordering (Weyl or Wick, for instance) for all $\xi$, even if the quantization mapping is $\xi$--dependent. 
%Nonetheless, as the quantization $Q(\partial_\xi\phi)$ and $Q(\partial_\xi\bar\phi)$ are both linear combinations of $Q(\phi)$,$Q(\bar\phi)$,  our choice of ordering may break down under changes of $\xi$. 
%In our work, Weyl's ordering, given by the complete symmetrization  of products of $\varphi^y$ and $\pi^x$. 
Thus, we can try to verify \eqref{RequirementHardcore}, but two obstacles must be overcome: 1) The cancellation through Skorokhod-Malliavin duality for $\nabla_\xi Q(\phi)$ for field products must be proven. 2) Given that, $Q(\partial_\xi\phi)$ and $Q(\partial_\xi\bar\phi)$ are both linear combinations of $Q(\phi)$,$Q(\bar\phi)$, our choice of ordering may break down under changes of $\xi$.  In this section, we solve both problems.

Let us start with the first obstacle we find in reproducing \eqref{RequirementHardcore} for polynomials. Consider a real quadratic function on the holomorphic fields such as $f_2:=\phi^xM_{xy}\phi^y+\bar\phi^x\bar M_{xy}\bar \phi^y$ to be quantized and let $M$ be symmetric. Its quantization is given by:
\begin{equation}
Q(f_2):=
Q(\phi^x)M_{xy}Q(\phi^y)+Q(\bar\phi^x)\bar M_{xy}Q(\bar \phi^y)
\end{equation}
because ordering does not affect products of solely holomorphic (or solely antiholomorphic) quantized fields, as they commute. Now, we compute $\nabla_\xi$ for this operator. Let us start with the term of the left, which is the most problematic one:
\begin{multline}
\nabla_\xi\left(Q(\phi^x)M_{xy}Q(\phi^y)\right)=Q(\partial_\xi\phi^x)M_{xy}Q(\phi^y)+Q(\phi^x)\partial_\xi M_{xy}Q(\phi^y)+Q(\phi^x)M_{xy}Q(\partial_\xi\phi^y)+\\
+\dfrac{1}{2}Q(\phi^x)M_{xy}\left(\Delta^{yz}\mathcal{K}_{zu}\bar\phi^u-\Delta^{yz}\mathcal{K}_{zu}\Delta^{uv}\partial_{\phi^v}\right)+
\dfrac{1}{2}\left(\bar\phi^v\mathcal{K}_{vu}\Delta^{ux}-\partial_{\phi^v}\Delta^{vz}\mathcal{K}_{zu}\Delta^{ux}\right)M_{xy}Q(\phi^y)
\end{multline}
where we have made use of Leibniz rule and of \eqref{eq:correction_phi} to derive the second line. We will denote this second line by $\zeta_{f_2}$,  which must be null for consistent quantization. Remembering that, on non-holomorphic functions, $Q(\phi^x)=\phi^x-\Delta^{xy}\partial_{\bar\phi^y}$, it can be shown that $\zeta_{f_2}$ acting on the holomorphic state $\Psi(\phi)$ is equivalent to:
$$\langle\Psi\vert
\zeta_{f_2}\Psi\rangle_\xi=\langle\Psi\vert\left(-\Delta_{xy}\mathcal{K}^{yz}\Delta_{zu} M^{ux}+\phi^xM_{xy}\Delta^{yz}\mathcal{K}_{zu}\bar\phi^u-\phi^xM_{xy}\Delta^{yz}\mathcal{K}_{zu}\Delta^{uv}\partial_{\phi^v}\right)\Psi\rangle_\xi\;.
$$
This expression is null inside the EV as in \eqref{eq:integral_correction}, through the Skorokhod-Malliavin duality, because $\Psi(\phi)$ is holomorphic.
This solves the first problem. On the other hand, for the products of $Q(\bar\phi)$, we have:
\begin{equation}
\nabla_\xi\left(Q(\bar\phi^x)M_{xy}Q(\bar\phi^y)\right)=Q(\partial_\xi\bar\phi^x)\bar M_{xy}Q(\bar\phi^y)+Q(\bar\phi^x)\partial_\xi \bar M_{xy}Q(\bar\phi^y)+Q(\bar\phi^x)\bar M_{xy}Q(\partial_\xi\bar\phi^y)
\end{equation}
because $\nabla_\xi Q(\bar\phi^x)=Q(\partial_\xi\bar\phi^x)$. Analogously, for terms such as $Q(\bar\phi^x)M_{xy}Q(\phi^y)$ or $Q(\phi^x)M_{xy}Q(\bar\phi^y)$ we obtain the same result once the EV is taken, \textit{i.e.} $\nabla_\xi$ enters the quantization procedure as $\partial_\xi$ for each quantized field.
\begin{comment}
we can show:
\begin{equation}
 \langle\Psi\vert\nabla_\xi Q(\phi^x)M_{xy}Q(\bar\phi^y)\Psi\rangle=\langle\Psi\vert\left(Q(\partial_\xi\phi^x)M_{xy}a^{y}+a^{+\;x}p\partial_\xi M_{xy}a^y+a^{+\;x}M_{xy}Q(\partial_\xi\bar\phi^y)\right)\Psi\rangle
\end{equation}
\end{comment}
These results can be generalized to arbitrary polynomials with analogous considerations, \textit{i.e.}, the terms containing $Q(\phi)$ only fulfill \eqref{RequirementHardcore} inside the EV, as in \eqref{eq:integral_correction}. This solves obstacle 1).

Obstacle 2) goes as follows. If the ordering procedure is prescribed in terms of creation and annihilation operators, $Q(\phi)$ and $Q(\bar\phi)$, then, given that $Q(\partial_\xi\phi)$ and $Q(\partial_\xi\bar\phi)$ are both non-trivial linear combinations of $Q(\phi)$ and $Q(\bar\phi)$, the covariant derivative $\nabla_\xi$ can spoil ordering. For example, $\nabla_\xi(Q(\phi^x)M_{xy}Q(\phi^y))$  provides terms such as $Q(\bar\phi^x)N_{xy}Q(\phi^y))$,  which are not normal ordered and, thus, $\nabla_\xi$ would not be inner to the space of normal ordered operators. \footnote{If a complex structure that allows to write the matter sH and sM as $\mathcal{H}^M_\mu[\tau]=\bar\phi^x M_{xy}^\mu[\tau]\phi^y\;\forall\tau\in\mathcal{N}$ exists, we can fulfill \eqref{RequirementHM} even for normal ordering. It seems that this choice of $J$ must be related with the generalization to generic spacetimes of the one proposed in \cite{ashtekar75}.  Nevertheless, these are our speculations and the case of normal ordering is still under study.} On the other hand, if Weyl's quantization is chosen, ordering can be implemented through the symmetrization of products of $\varphi^x$ and $\pi^x$. We remind the reader that $\nabla_\xi Q(\delta_{xy}\varphi^y)=\nabla_\xi Q(\pi^x)=0$ (at least, for their EVs), which is also true for products of them because obstacle 1) is solved. Therefore, for Weyl's quantization, $Q_W$, the covariant derivative becomes a derivative $\partial_\xi$ on the coefficients of each monomial, preserving Weyl's ordering, and we can conclude that:
\begin{equation}
\langle\Psi\vert\nabla_\xi Q_W(f(\varphi,\pi))\vert\Psi\rangle=\langle\Psi\vert Q_W(\partial_\xi f(\varphi,\pi))\vert\Psi\rangle\quad \forall f\in C^\infty(\mathcal{M}_F)
\end{equation}
This is precisely  the EV version of \eqref{RequirementHardcore}, and a sufficient condition  for the quantum connection to fulfill all the requirements imposed by geometrodynamics. In fact, for our purposes, it would be enough to fulfil its reduced version, \eqref{RequirementHM}. With this, we conclude our example for the $\xi$-dependent holomorphic quantization of the scalar field, having shown the existence (and proposed a candidate) of a connection fulfilling all the necessary requirements for geometrodynamics. \footnote{Note that this connection may not be the only one fulfilling \eqref{RequirementHM} and \eqref{antiselfadjointconnection}. For example, one can always add a term of the type $i f(\xi)d\xi$ to the connection 1-form $\Gamma$, which would be the equivalent of including global (regarding the Hilbert space) $\xi$-dependent Berry connection.} 
\subsection{Connection 1-form, horizontal lift for geometric vector fields and curvature.}
Note that we have been working for simplicity with the entries $\xi$ of the connection. Besides, we have been working with $\partial_\xi$ as a derivative with respect to a parameter for simplicity. Nevertheless, the well-defined connection form can be written as 
\begin{equation}
\Gamma=\sum\limits_{\xi\in\lbrace h_{ij},\pi^h_{ij},N,N^i\rbrace}\Gamma_{\xi^x} d{\xi^x}\;\;\text{where}\;\; \Gamma_{\xi^x}:=\dfrac{-1}{4} \left(\partial_{\phi^a}\Delta^{av}\mathcal{K}_{vu;\xi^x}\Delta^{ub}\partial_{\phi^b}+\phi^v\bar{\mathcal{K}}_{vu;\xi^x}\phi^u\right)
\end{equation}
with $\mathcal{K}_{vu;\xi^x}:=\delta_{vx}\left(\dfrac{\delta}{\delta\xi(x)}( \delta^{xz}K_{zu}\delta^{uy})+i\dfrac{\delta}{\delta\xi(x)}(\delta^{xw}K_{wy}A^{yz})\right)\delta_{yu}$.

In this sense, if we wanted to make the covariant derivative of a section $\sigma\in\Gamma(\mathcal{M}_F)$ with respect to $X_\xi=\sum\limits_{\xi\in\lbrace h_{ij},\pi^h_{ij},N,N^i\rbrace}v_\xi^x(\xi,\Psi)\partial_{\xi^x}$, we would simply have:
\begin{equation}
\nabla_{X_\xi}\sigma=X_\xi(\sigma)+\Gamma(X_\xi)\sigma\;,\quad\text{where, in coord.,}\quad\Gamma(X_\xi)=\sum\limits_{\xi\in\lbrace h_{ij},\pi^h_{ij},N,N^i\rbrace}v_{\xi}^x\Gamma_{\xi^x}\;.
\end{equation}
We can now find a compatibility condition between  $\Gamma$ and $\theta$ that characterizes our success in relating non unitarily-equivalent Hilbert spaces and quantized operators on them, which is given in terms of a curvature tensor. For example, for the linear classical observables that did not depend on $\xi$ before quantization ($\partial_\xi f(\varphi,\pi)=0$), such as $\varphi_x$ and $\pi^x$, the condition of consistent quantization \eqref{eq:integral_correction} is equivalent to:
\begin{equation}
\label{curvature}
\int\limits_{\mathcal{N}^\prime_C}D\mu\bar\Psi(\bar\phi)\left(\nabla_{X_G}\tilde{\nabla}_{X_f}-\tilde\nabla_{X_f}\nabla_{X_G}\right)\Psi(\phi)=0
\end{equation}
where $X_f=\lbrace,f\rbrace_M$ and $\tilde\nabla_{X_f}=X_f+\theta(X_f)$, being $\theta$ the $\xi$-dependent 1-form for geometric quantization, and $\nabla_{X_\xi}f= X_\xi f+\Gamma(X_\xi)f$ (being $f$ a section of the bundle). We remind also that $\nabla_{X_\xi}\Psi(\phi)=0$. Then, the term $\left(\nabla_{X_G}\tilde{\nabla}_{X_f}-\tilde\nabla_{X_f}\nabla_{X_G}\right)$ reminds us of the definition of curvature tensor for a connection 1-form given by some combination of (the pullbacks to $\mathcal{F}_C$ of)  $\theta$ and $\Gamma$, and \eqref{curvature} claims that its EV must be null. This avenue is also mentioned in \cite{David2}.

In this example, we have chosen the holomorphic picture for simplicity, but the procedure is analogous in other pictures. Besides, the holomorphic connection can be related with the one for the Schrödinger case (real polarization, the domain of the states is given by $\varphi^x$) through the Segal-Bargman map between pictures developed in \cite{David,David2}.
%\subsection{Hybrid dynamics and reduction to QFT in CS: FLRW example.}
%Aqui podriamos  poner tambien las constraints y el espacio que las cumple, asi como la ecuacion de movimiento.
\begin{comment}
\footnote{For easier reproducibility, while this is not rigourous, such duality in this case is operationally equivalent to compute total derivatives for $\partial_{\phi^x}$ and $\partial_{\bar{\phi}^x}$ and consider the usual exponent for the Gaussian, $-\phi^x K_{xy}{\bar\phi}^y$. Equivalently, one can formally use that $\phi^x$ can be obtained from $-\Delta^{xy}\partial_{\bar\phi^y}$ acting on $D\mu$.} 
\end{comment}
\section{Discussion.}
\label{sec:6}
We have constructed a hybrid geometrodynamical framework with classical 3-metric, its momenta  and quantum field theoretical matter described with the Schr\"odinger wave functional picture in a Hamiltonian framework.  In order to succesfully represent the infinitesimal generators of hypersurface deformations in a Hamiltonian way, the characterization of the hybrid manifold as a fibration provided with a Hermitian quantum connection has shown to be crucial. Such connection is related to a notion which is widespread in the literature: the time dependence of the vacuum state in QFT in curved spacetime, which leads to time dependent Fock spaces \cite{Long1996TheSW,CCQ04}. In our work, instead of vacuum states, we worked with field measures defining a Hilbert space and a quantization procedure for operators. These non-unitarily-equivalent Hilbert spaces usually lead to norm loss of quantum states \cite{Schneider}, of difficult interpretation from the probabilistic point of view. In our formalism, norm loss is entirely avoided, given that the Hermitian connection provides a parallel transport that preserves the scalar product (of the non-unitarily-equivalent Hilbert spaces) and quantum states are associated to covariant sections.  These results, as well as the necessity of the quantum connection, are valid beyond hybrid geometrodynamics, also for QFT in a given curved spacetime, as the (densely) Hermitian connection is a requisite for quantum states when the scalar product of the Hilbert space and the quantization procedure depend on the geometric content of each leaf of the foliation. Such connection had to fulfil eqs. (\ref{antiselfadjointconnection}, \ref{RequirementHM} and \ref{lapseshiftconnection}) for geometrodynamics to be consistent.  In section \ref{example}, for the case of the holomorphic quantization of the scalar field, we have derived a connection fulfilling all these constraints, and, in fact, the more restrictive condition \eqref{RequirementHardcore}.

While the  nature of quantum states as covariant sections for the Hermitian connection was deduced as a requirement for the construction of hybrid geometrodynamics, one does not need to enforce it by hand. Instead, as argued at the end of section \ref{sec:leafdependence}, this is naturally obtained from the leaf-dependence of the quantization procedure. Arguing beyond geometric quantization used in section \ref{example}, an abstract quantum state $\omega$ is a dual to a $C^\ast$ algebra of observables, which yields its EV as $\omega(a)\in\mathbb{R}\;\forall a\in C^\ast_Q$. Quantization is a representation of such algebra as (bounded) operators on a certain Hilbert space $\mathcal{H}$, $Q(a)=\hat A\in\mathcal{B}(\mathcal{H})\;\forall a\in C^\star_Q$.  In this context, $\omega$ is represented by $\Psi\in\mathcal{H}$, and $\omega(a)=\langle\Psi\vert \hat A\vert\Psi\rangle$. If the quantization  $Q_\xi$ depends smoothly on $\xi$, so does the representation $\hat A_\xi$ (of the abstract observable $a$) and the Hilbert space, $\mathcal{H}_\xi$. Therefore, the same observable $a\in C^\ast$ is represented differently at each $\xi$. To reproduce the same correct value of $\omega(a)=\langle\Psi\vert\hat A_\xi\vert\Psi\rangle_\xi$ at each $\xi$ (at least for a suitable subset of observables) the state $\Psi\in \mathcal{H}_\xi$ must acquire different representations at each $\xi$, which are related precisely through parallel transport for a Hermitian connection.

In particular, based on the parametric family of complex structures of \cite{ashtekar75,agullo2015unitarity}, we have introduced the concept of a $\xi$-parametric family of quantizations, each of them characterized by a measure, a representation of operators and states. Such quantizations are not unitarily equivalent, contrarily to the finite dimensional case. Therefore, we introduce a non-unitary operator to relate them, which does not preserve the Hilbert state structure it acts on, but instead transports the state to the new Hilbert space resulting from the change of $\xi$. Such operator represents functionally the quantum connection 1-form. From the requirements for consistent hybrid geometrodynamics, we obtain that it must fulfill certain constraints, which in the auxiliary mathematical notes are related to the covariance of Kibble's Kähler structure (\cite{Kibble79}). In section \ref{example} we have shown the existence and provided a candidate of the quantum connection for the holomorphic quantization of the scalar field that fulfils all the necessary conditions. 
This construction provides norm conservation of quantum states, ensures that the hybrid generating functions of hsf. deformations reproduce Dirac's algebra and helps preserving the hybrid geometrodynamical constraints throughout any foliation. 

\begin{comment}
On the other hand, the reader  might have notice that the process that has lead to the identification of the quantum supermagnitudes is slightly misaligned with the spirit of geometrodynamics \cite{HKT76}. Instead of invoking the\textit{ representation postulate} to find the sM as purely spatial diffeomorphisms on leaf, $\mathcal{L}^\star_{N^i}\Psi=\lbrace\Psi,\mathcal{H}_i^Q\rbrace_Q$, we have resorted to the quantization of the appropriate generating functions over the classical field theory to identify such generating function. In the end, the procedure is analogous, for a consistent quantization procedure of the observables yields the appropriate operators (specially if they are no more than quadratic on the fields) representing the transformation for the functions adapted to a certain $L^2$, taking into acount that momentum operator is represented as a self-adjoint  derivative w.r.t. the measure and vacuum phase, in order to be self adjoint.
\end{comment}

The geometrodynamical principle of equivalence is a key notion in \cite{HKT76} for the inclusion of classical matter sources in geometrodynamics. We keep this notion, but apply it only to the classical field matter considered for quantization (so, its supermagnitudes do not depend on $\pi_h$ and only ultralocally on $h$) and, given that the choice of quantization is made referent to the classical matter field structures as in \cite{ashtekar75}, the quantized version does not acquire extra dependencies on $\pi_h$. Nevertheless, we allow the quantization procedure to add any non-ultralocal dependence on the 3-metric, which shall be compensated by the connection.
%On the other hand, one may speculate that, to some extent, some sort of non-minimally coupled field theories could still emerge as effective theories for matter test fields obtained from the current theory  were matter fields are sources. The idea would be that the effect of the backreaction of matter on gravity, and its subsequent effect on the propagation of the fields due to the change of metric, makes up for a gravity-mediated self-interaction. Therefore, if the perturbation on the geometry is small, it could be ``integrated out", i.e. absorbed into the matter theory as a new dynamical pole on the propagator of the fields coupled to a background gravity, considering for this new effective theory the geometry unaffected by matter ``test fields".
On the other hand, $N,N^i$ have a dynamical origin, not  a kinematical one. Thus, the fact that the quantization procedure is dependent on them makes some quantum kinematical structures (scalar product and states) $N,N^i$-dependent, apparently mixing kinematics and dynamics. This is solved thanks to eq. \eqref{Lapseshiftindependence} and the role of the connection (which yields \eqref{leafindependenceQPB}), which renders the generating functions of symmetries and the quantum PB $N,N^i$-independent.
\begin{comment}
{\color{black} In order to heal this unnatural dependence a possible path to follow would be to consider the whole Kähler structure dependent on both $h$ and $\pi_h$ and try to recover \cite{ashtekar75} {\color{black}choosing a submanifold whose constraints allowed us to rewrite} $\pi_h$ in terms of $h$, lapse and shift. This procedure is, nonetheless, very difficult to attain in practice because of the non polynomial nature of the geometrodynamical constraints. The reader might be familiar with this kind of problems from the study of the Wheeler-DeWitt equation. There is another possibility that we can borrow from the history of quantum gravity. In terms of the Hamiltonian description of gravity in terms of the Ashtekar-Barbero variables, the ones used in Loop Quantum Gravity for canonical quantization,  the constraint becomes much more tractable and thus the dependence  of the Kähler structure might be easier to elucidate. We refer to \cite{Kie1} for further details. This programme would be called Hybrid connection dynamics and will be the subject of future work.}
{\color{black} No sé si me convence mucho. Yo no daría indicaciones sobre soluciones alternativas que queremos seguir, porque nos faltan todavía algunos meses para poder tener resultados. Otra cosa es de resultados ya casi listos. ¿Qué gana el paper con esta información que damos? }
\end{comment}
In our construction the enforcing of the equivalence principle for Quantum Field Theory is not required at the quantum level. However, one of the implications of hybrid geometrodynamics yields some light on it. In the same sense that Ehrenfest's theorem \textit{almost} reproduces Newton's equations for the EVs, we can argue that the EVs of the operators resulting from quantizing classical matter supermagnitudes \textit{almost} obey this hybrid geometrodynamical equivalence principle.  In this way we find an interpretation under the light of the equivalence principle for the leaf dependence of hybrid observables\;,
\begin{equation}
\partial_\xi \langle\Psi\mid Q(A)\mid\Psi\rangle=\langle\Psi\mid Q(\partial_\xi A)\mid\Psi\rangle\;,
\end{equation}
for  any suitable classical function $A$. Note that, in the case of leaf dependent quantization, this is only possible thanks to the inclusion of the quantum connection. Therefore, physical observables with matter content, have the appropriate dependence on the geometry of spacetime for their EVs, as $A$ fulfilled the classical geometrodynamical principle of equivalence.

Another important principle in the construction of geometrodynamics is path independence, achieved under the hybrid constraints $\mathcal{H}_{\mu}^H\simeq 0$ (ensured their preservation through \eqref{Lapseshiftindependence} and their first class nature for the Hamiltonian $\lbrace,f_H\rbrace_H$).  These constraints might have important phenomenological consequences, as we have hybrid conserved magnitudes for the whole universe.  For example, starting from non-divergent initial data for matter and geometry, if at any point of the evolution the pure geometrodynamical part becomes divergent (formation of a singularity, for example in a hybrid dust collapse model), it must be accompanied by a same-sized divergent EV of the quantum operator, as we know that the Hamiltonian constraint is preserved. To what extent this might prevent the dynamical formation of  singularities when matter sources are quantum will be the subject of future work.

In this line, one realizes that, if the phenomenon of quantum completeness as stated in \cite{Schneider},\cite{Schneider2} had a backreaction of quantum matter on gravity as in this work, the quantum matter fields would not be able to act as sources of divergent geometry. Such phenomenon is based on the loss of norm for the quantum states while approaching geometric singularities. In that sense, singularity formation starting with non-singular initial data would be to some extent protected in hybrid geometrodynamics if norm loss was allowed, as, in the process of forming the singularity, the material sources would become smaller the nearer to the formation of the singularity. Nevertheless, as argued above, in our framework norm loss does not take place thanks to the Hermitian connection. To what extent the phenomenology of quantum completeness could find its way into our formalism, given that norm loss is healed by the quantum connection being Hermitian for the $\xi$-parametric family of scalar products, is still a matter of discussion.

{\color{black}
Another interesting application of the formalism is the study of the phenomenon of particle creation (for a decomposition of wave functionals into the usual particle interpretation of Fock spaces, see \cite{David,David2}). In particular, if we considered a closed path on the base of the hybrid fibration,  starting from an Euclidean measure (and a suitable associated geometromomenta), going through arbitrary Riemannian 3-metrics, back into the Euclidean case, we are reproducing the usual ``past Minkowski", intermediate curved spacetime, ``future Minkowski" set up. The holonomy of the quantum connection along this curve allows for the precise mathematical characterization of the particle creation phenomenon in our framework, while leaving room for more general phenomena, such as the physical interpretation of a possible torsion or leaf-dependent Berry phase for such quantum connection. This will be the subject of study of future works.
}

We conclude this work by quoting Robert M. Wald, who has guided the intertwining of QFT and curved spacetime for almost half a century. The quote is an extract from \cite{wald1977} on the consistency of quantum fields as source for classical gravity:
\begin{center}
\textit{``But if the five axioms} [of QFT in Curved Spacetime] \textit{are inconsistent in a nontrivial manner, then unless one can somehow evade the arguments of Section II} [leading to the five axioms]\textit{ one would be forced to conclude that ``back reaction" effects cannot be treated within the context of the semiclassical approximation."}
\end{center} 
In this context, a natural question that we will need to answer in future investigations is whether or not something has been gained regarding this issue in our depiction of the backreaction in terms of Hamiltonian framework for Dirac's spatial hsf. deformations.


\section*{Acknowledgements.}

We express our sincere gratitude to Prof. J.L. Cortés for his insightful comments. We also extend our appreciation to Jan Głowacki for engaging discussions and sharing his extensive knowledge on geometrodynamics. Furthermore, we would like to thank M. Schneider for stimulating conversations that sparked our curiosity in the phenomenology of quantum completeness in relation with our framework.

The authors acknowledge partial finantial support of Grant PID2021-123251NB-I00
funded by MCIN/AEI/10.13039/501100011033 and by the European Union, and of
Grant E48-23R funded by Gobierno de Aragón. C.B-M
and D.M-C acknowledge financial support by Gobierno de Aragón through the grants
defined in ORDEN IIU/1408/2018 and ORDEN CUS/581/2020 respectively.
\appendix
\section{Proofs related with Skorkhov-Malliavin duality}
\subsection{Quantum supermagnitudes for coherent states and constraints.}
The classical matter supermagnitudes are quadratic on $\varphi_x,\pi^x$. Once they are written in classical holomorphic and antiholomorphic coordinates $\phi$ and $\bar\phi$ (see section \ref{example} for these coordinates) they are quadratic operators of the type 
\begin{equation}
\label{holomorphic_supermagnitudes}
\mathcal{H}^M_\mu [\tau;\phi^x,\bar\phi^x]=\phi^xM_{xy}\phi^y+\bar\phi^x\bar{M}_{xy}\bar\phi^y+\phi^xN_{xy}\bar\phi^y\;,
\end{equation}
for some complex symmetric matrix $M_{xy}$ and real symmetric matrix $N_{xy}$ that depend on $\tau$. Its quantization, therefore, acting on an holomorphic state $\Psi(\phi)$ will be given by (see section \ref{example} for holomorphic quantization):
\begin{equation}
\label{quantized_holomorphic_supermagnitudes}
Q_\xi(\mathcal{H}^M_\mu)\Psi=\left(\phi^xM_{xy}\phi^y+\partial_{\phi^u}\Delta^{ux}\bar{M}_{xy}\Delta^{yv}\partial_{\phi^v}+\bar\phi^y+\phi^xN_{xy}\Delta^{yv}\partial_{\phi^v}\right)\Psi\;.
\end{equation}
Now,  as in section \ref{sec:4}, we consider a coherent state given by $\Psi_f=\dfrac{e^{\bar f_x\phi^x}}{e^{f_x\Delta^{xy}\bar f_y/2}}$. These kind of coherent states are normalized, $\int D\mu\bar\Psi_f\Psi_f=1$. The EV of $Q_\xi(\mathcal{H}^M_\mu)$ defines the quantum supermagnitudes. We proceed to compute now, term by term, the EV of \eqref{quantized_holomorphic_supermagnitudes} under this coherent state. Firstly, it is easy to check that:
\begin{equation}
\int D\mu\bar\Psi_f\phi^x M_{xy}\phi^y\Psi_f=\int D\mu \partial_{\bar\phi^u}\Delta^{ux}M_{xy}\Delta^{yv}\partial_{\bar\phi^v}\bar\Psi_f\Psi_f=f_u\Delta^{ux}M_{xy}\Delta^{yv}f_v\;,
\end{equation}
where in the first equality we have made use of Skorokhod-Malliavin duality (in this case, $\phi^x$ acting on $\Psi$ becomes $\Delta^{xy}\partial_{\bar\phi^y}$acting on $\bar\Psi$). Besides
\begin{equation}
\int D\mu\bar\Psi_f\partial_{\phi^u}\Delta^{ux}M_{xy}\Delta^{yv}\partial_{\phi^v}\Psi_f=\bar f_u\Delta^{ux}M_{xy}\Delta^{yv}\bar f_v
\end{equation}
and
\begin{equation}
\int D\mu\bar\Psi_f\phi^x M_{xy}\Delta^{yv}\partial_{\phi^v}\Psi_f=\int D\mu \partial_{\bar\phi^u}\Delta^{ux}\bar\Psi_fM_{xy}\Delta^{yv}\partial_{\phi^v}\Psi_f=f_u\Delta^{ux}M_{xy}\Delta^{yv}\bar f_v\;,
\end{equation}
where we have also made use of Skorokhod-Malliavin duality.
With these results, we have shown, term by term, that
\begin{equation}
\langle\Psi_f\vert Q_\xi(\mathcal{H}_\mu^M)\Psi_f\rangle_\xi=\mathcal{H}_\mu^M[\tau;\Delta^{yv}f_v,\Delta^{xz}\bar f_z]
\end{equation}
for the classical definition of $\mathcal{H}_\mu^M[\tau;\phi^x,\bar\phi^x]$ given in \eqref{holomorphic_supermagnitudes}, but now, the role of the classical holomorphic field is played by $\Delta^{yv}f_v$, associated to the coherent state.

Thus, the hybrid constraints \eqref{hybridconstraints} for coherent states reduce to the classical constraints (the global analogue to  \eqref{constraints}) written in holomorphic coordinates. In turn, as in section \ref{example}, the real and imaginary parts of the complex coefficients $f_x\Delta^{xy}$ (relative to our choice of $J_C$ as indicated below in \eqref{eq:holomorphic_coordinates}) can be identified with some classical fields and momenta $\varphi_x,\pi^x$  Given that the classical constraints have solutions for a certain $\varphi_x,\pi^x$ at each $\xi$, the hybrid constraints also have at least a solution for the coherent state $\Psi_f$ that has its $f_y$ associated to those classical fields fulfilling such constraints.

\subsection{Characterization of the change of measure as $\Xi$}
Let as firstly showthat the operator inside \eqref{change_functional}, $\left(\partial_{\bar\phi^x}\dot\Delta^{xy}\partial_{\phi^y}-\partial_\xi(\phi^x)\partial_{\phi^x}-\partial_\xi(\bar\phi^x)\partial_{\bar\phi^x}\right)$, is indeed equivalent to \eqref{XI}. We will do so term by term. Let us start with the term $\Xi_1:=\int D\mu\left(\partial_\xi(\phi^x)\partial_{\phi^x}+\partial_\xi(\bar\phi^x)\partial_{\bar\phi^x}\right)f(\phi,\bar\phi)$, which, making use of \eqref{eq:dxiphi},  becomes:
\begin{multline}
\Xi_1=\dfrac{1}{2}\int D\mu (\phi^y \mathrm{S}^x_y\partial_{\phi^x}+\bar\phi^y\mathrm{T}^x_y\partial_{\phi^x}+\bar\phi^y \bar{\mathrm{S}}^x_y\partial_{\bar\phi^x}+\phi^y\bar{\mathrm{T}}^x_y\partial_{\bar\phi^x})f(\phi,\bar\phi)=\\
\dfrac{1}{2}\int D\mu (\partial_{\bar\phi^z}\Delta^{zy} \mathrm{S}^x_y\partial_{\phi^x}+\partial_{\phi^z}\Delta^{zy} \mathrm{T}^x_y\partial_{\phi^x}+\partial_{\phi^z}\Delta^{zy} \bar{\mathrm{S}}^x_y\partial_{\bar\phi^x}+\partial_{\bar\phi^z}\Delta^{zy}\bar{\mathrm{T}}^x_y\partial_{\bar\phi^x})f(\phi,\bar\phi))=\\
\dfrac{1}{2}\int D\mu \left( \partial_{\bar\phi^z}\left(\Delta^{zy} \mathrm{S}^x_y+(\Delta^{zy} \bar{\mathrm{S}}^x_y)^t\right)\partial_{\phi^x}+\partial_{\bar\phi^z}\Delta^{zy}\bar{\mathrm{T}}^x_y\partial_{\bar\phi^x}+\partial_{\phi^z}\Delta^{zy} \mathrm{T}^x_y\partial_{\phi^x}\right)f(\phi,\bar\phi))
\end{multline}
where, in the second line, we have made use of Skorkof-Malliavin duality, which, in this case, it is simply substituting $\phi^y\rightarrow\partial_{\bar\phi^z}\Delta^{zy}$ and $\bar{\phi}^y\rightarrow\partial_{\phi^z}\Delta^{zy}$.  It is immediate to check that
\begin{equation}
\mathrm{T}_x^y=\mathcal{K}_{xz}\Delta^{zy}
\end{equation}
Let us have a closer look to the terms with $\mathrm{S}^x_y$, which we remind it was given by $\mathrm{S}^x_y:=i\dot{A}^{xu}\delta_{uy}+\dot{\Delta}^{xu}(K_{uy}-iK_{uz} A^z_y)+i\dot{\delta}^{xu}\delta_{uz}A^{zv}\delta_{vy}$ which can be rewritten as $\mathrm{S}^x_y=\dot{\Delta}^{xu}K_{uy}+i\Delta^x_n\partial_\xi(K^{n}_mA^m_v\delta^{vu})\delta_{uy}$.
\begin{equation}
    \dfrac{1}{2}\left(\Delta^{zy} \mathrm{S}^x_y+(\Delta^{zy} \bar{\mathrm{S}}^x_y)^t\right)=\dot{\Delta}^{xz}+\dfrac{i}{2}\Delta^x_n\partial_\xi(K^{n}_mA^m_v\delta^{vu})
    \Delta_u^z-\dfrac{i}{2}\Delta_u^z\partial_\xi(K^{n}_mA^m_v\delta^{vu})^t\Delta^x_n=\dot{\Delta}^{xz}
\end{equation}
where the two imaginary terms cancel because $KA=A^tK$ and $A\Delta=\Delta A^t$ and both $\delta^{xy}$ and $\partial_\xi\delta^{xy}$ commute with $\Delta$.
%+\dfrac{i}{2}\left(\dot{A}^{xu}\Delta^z_u+\Delta^{xu}\dot K_{uv}A^{v}_y\Delta^{yz}+\dot{\delta}^{xu}A_u^v\Delta_v^{z}\right)\\
%-\dfrac{i}{2}\left(\dot{A}^{xu}\Delta^z_u+\Delta^{xu}\dot K_{uv}A^{v}_y\Delta^{yz}+\dot{\delta}^{xu}A_u^v\Delta_v^{z}\right)=\Delta\partial_\xi(KA\delta)\Delta
With these results, we see that:
\begin{equation}
\Xi_1=\dfrac{1}{2}
\int\limits_{\mathcal{N}^\prime_{\mathbb{C}}} D\mu\left(\partial_{\bar\phi^z}\Delta^{zy}\bar{\mathcal{K}}_{yu}\Delta^{ux}\partial_{\bar\phi^x}+\partial_{\phi^z}\Delta^{zy} \mathcal{K}_{yu}\Delta^{ux}\partial_{\phi^x}+2\partial_{\bar\phi^z}\dot\Delta^{zx}\partial_{\phi^x}\right)f(\phi,\bar\phi))
\end{equation}
and, therefore,
\begin{multline}
\int\limits_{\mathcal{N}^\prime_{\mathbb{C}}} D\mu \left(\partial_{\bar\phi^x}\dot\Delta^{xy}\partial_{\phi^y}-\partial_\xi(\phi^x)\partial_{\phi^x}-\partial_\xi(\bar\phi^x)\partial_{\bar\phi^x}\right)f(\phi,\bar\phi)=\int\limits_{\mathcal{N}^\prime_{\mathbb{C}}} D\mu \partial_{\bar\phi^x}\dot\Delta^{xy}\partial_{\phi^y}f(\phi,\bar\phi)-\Xi_1=\\-\dfrac{1}{2}
\int\limits_{\mathcal{N}^\prime_{\mathbb{C}}} D\mu\left(\partial_{\bar\phi^z}\Delta^{zy}\bar{\mathcal{K}}_{yu}\Delta^{ux}\partial_{\bar\phi^x}+\partial_{\phi^z}\Delta^{zy} \mathcal{K}_{yu}\Delta^{ux}\partial_{\phi^x}\right)f(\phi,\bar\phi))\;.
\end{multline}
In turn, using Skorkhof-Malliavin duality, it is immediate to show that:
\begin{equation}
    \int\limits_{\mathcal{N}^\prime_{\mathbb{C}}} D\mu \partial_{\bar\phi^z}\Delta^{zy}\bar{\mathcal{K}}_{yu}\Delta^{ux}\partial_{\bar\phi^x}f(\phi,\bar\phi)=\int\limits_{\mathcal{N}^\prime_{\mathbb{C}}} D\mu \left(c_1\partial_{\bar\phi^z}\Delta^{zy}\bar{\mathcal{K}}_{yu}\Delta^{ux}\partial_{\bar\phi^x}+(1-c_1)\phi^y\bar{\mathcal{K}}_{yu}\phi^u\right)f(\phi,\bar\phi)
\end{equation}
and
\begin{equation}
    \int\limits_{\mathcal{N}^\prime_{\mathbb{C}}} D\mu \partial_{\phi^z}\Delta^{zy}{\mathcal{K}}_{yu}\Delta^{ux}\partial_{\phi^x}f(\phi,\bar\phi)=\int\limits_{\mathcal{N}^\prime_{\mathbb{C}}} D\mu \left(c_2\partial_{\phi^z}\Delta^{zy}{\mathcal{K}}_{yu}\Delta^{ux}\partial_{\phi^x}+(1-c_2)\bar\phi^y{\mathcal{K}}_{yu}\bar\phi^u\right)f(\phi,\bar\phi)
\end{equation}
for any constants $c_1,c_2\in\mathbb{C}$. Any choice of these constants would provide a different notion of Hermitian quantum connection, all of them providing norm conservation. In order to preserve quantization of operators, we must choose $c_1=c_2=\dfrac{1}{2}$. With this choice, we are finally able to proof that, 
as defined in \eqref{XI},
\begin{equation}
\int\limits_{\mathcal{N}^\prime_{\mathbb{C}}} D\mu \left(\partial_{\bar\phi^x}\dot\Delta^{xy}\partial_{\phi^y}-\partial_\xi(\phi^x)\partial_{\phi^x}-\partial_\xi(\bar\phi^x)\partial_{\bar\phi^x}\right)f(\phi,\bar\phi)=\int\limits_{\mathcal{N}^\prime_{\mathbb{C}}} D\mu\Xi f(\phi,\bar\phi)
\end{equation}
\subsection{Covariant derivatives of higher order operators and ordering.}
Let us consider, as in section \ref{example}, the quadratic function on the holomorphic fields given by $f_2:=\phi^xM_{xy}\phi^y+\bar\phi^x\bar M_{xy}\bar \phi^y$ with $M$ symmetric, and the covariant derivative of its quantization must be:
\begin{multline}
\nabla_\xi Q(f_2)=
\nabla_\xi Q(\phi^x)M_{xy}Q(\phi^y)+Q(\phi^x)M_{xy}\nabla_\xi Q(\phi^y)+Q(\phi^x)\partial_\xi(M)_{xy} Q(\phi^y)+\\+\nabla_\xi Q(\bar\phi^x)\bar M_{xy}Q(\bar \phi^y)+ Q(\bar\phi^x)\bar M_{xy}\nabla_\xi Q(\bar \phi^y)+ Q(\bar\phi^x)\partial_\xi(\bar M)_{xy} Q(\bar \phi^y)=Q(\bar \phi^y)
\end{multline}
Notice that the $M$ behaves as a function in what regards the dependence on $\xi$, and hence the expression $\partial_\xi(\bar M)_{xy} $. Now, taking into account that $\nabla_\xi Q(\phi^x)=\left(Q(\partial_\xi\phi^x)-\theta_\xi(X_{\phi^x})+[\hat\Gamma_\xi,Q(\phi^x)]\right)$ (with $X_{\phi^x}=\lbrace,\phi^x\rbrace_M$) and $\nabla_\xi Q(\bar\phi^x)=Q(\partial_\xi\bar\phi^x)$, because  $-\theta_\xi(X_{\bar\phi^x})+[\hat\Gamma_\xi,Q(\bar\phi^x)]=0$,we  find:
\begin{multline}
    \nabla_\xi Q(f_2)= Q(\partial_\xi\phi^x)M_{xy}Q(\phi^y)+ Q(\phi^x)\partial_\xi M_{xy}Q(\phi^y)+ Q(\phi^x)M_{xy}Q(\partial_\xi\phi^y)+\\
    Q(\partial_\xi\bar\phi^x)\bar M_{xy}Q(\bar \phi^y)+Q(\bar\phi^x)\partial_\xi\bar M_{xy}Q(\bar \phi^y)+Q(\bar\phi^x)\bar M_{xy}Q(\partial_\xi\bar \phi^y)+
    \zeta_{\phi}^xM_{xy}Q(\phi^y)+Q(\phi^x)M_{xy}\zeta_{\phi}^x
\end{multline}
where the sum of the last two terms must be null for a consistent quantization as in \eqref{RequirementHardcore}, for which we have defined $\zeta_{\phi}^x:=-\theta_\xi(X_{\phi^x})+[\hat\Gamma_\xi,Q(\phi^x)]$, which has been shown to be $\zeta_{\phi}^x=\dfrac{1}{2}\left(\bar\phi^v\mathcal{K}_{vu}\Delta^{ux}-\partial_{\phi^v}\Delta^{vz}\mathcal{K}_{zu}\Delta^{ux}\right)$ in section \ref{example}. Thus, the last two terms, acting on the holomorphic wave functional $\Psi(\phi)$, yield:
\begin{multline}
\label{phibarphi}
\left[\zeta_{\phi}^xM_{xy}Q(\phi^y)+Q(\phi^x)M_{xy}\zeta_{\phi}^x\right]\Psi(\phi)=\dfrac{1}{2}\big[\left(\bar\phi^v\mathcal{K}_{vu}\Delta^{ux}-\partial_{\phi^v}\Delta^{vz}\mathcal{K}_{zu}\Delta^{ux}\right)M_{xy}\phi^y+\\+(\phi^x-\partial_{\bar\phi^z}\Delta^{zx})M_{xy} \left(\Delta^{yz}\mathcal{K}_{zu}\bar\phi^u-\Delta^{yz}\mathcal{K}_{zu}\Delta^{uv}\partial_{\phi^v}\right)\big]\Psi(\phi)
\end{multline}
where, in the last term, we have made use of the fact that $\zeta_{\phi}^x$ contains antiholomorphic fields (because $\theta_\xi$ mixes holomorphic and antiholomorphic polarizations), and therefore, the quantization of $\phi^x$ acting on them must retain all the terms of its prequantization, because the function on which it acts is not adapted to the holomorphic polarization, \textit{i.e.} $Q(\phi)\zeta_{\phi}^x\Psi=(\phi^x-\Delta^{xz}\partial_{\bar\phi^z})\zeta_{\phi}^x\Psi$. Keep in mind, however, that we still have $\partial_{\bar\phi^x}\Psi(\phi)=0$. We can see that the first term in the r.h.s. of the latter equation is null under the EV:
\begin{multline}
    \int D\mu\bar\Psi\left(\bar\phi^v\mathcal{K}_{vu}\Delta^{ux}-\partial_{\phi^v}\Delta^{vz}\mathcal{K}_{zu}\Delta^{ux}\right)M_{xy}\phi^y\Psi= \\
    \int D\mu\bar\Psi\left(\partial_{\phi^v}\Delta^{vz}\mathcal{K}_{zu}\Delta^{ux}-\partial_{\phi^v}\Delta^{vz}\mathcal{K}_{zu}\Delta^{ux}\right)M_{xy}\phi^y\Psi=0
\end{multline}
where  we have made use of Skorkhof-Malliavin duality and  substituted  $\bar\phi^x$  by $\partial_{\phi^z}\Delta^{zx}$, because it was at the left of any function of the holomorphic variables ($\bar\Psi$ is function of $\bar\phi$). Now, the second term of the r.h.s. of eq. \ref{phibarphi} can be shown to be:
\begin{multline}
    \int D\mu\bar\Psi\left(\phi^x-\partial_{\bar\phi^v}\Delta^{vx}\right)M_{xy} \left(\Delta^{yz}\mathcal{K}_{zu}\bar\phi^u-\Delta^{yz}\mathcal{K}_{zu}\Delta^{uv}\partial_{\phi^v}\right)\Psi=\\
    \int D\mu\bar\Psi\left(\phi^xM_{xy}\Delta^{yz}\mathcal{K}_{zu}\bar\phi^u-\Delta^{ux}M_{xy}\Delta^{yz}\mathcal{K}_{zu}-\phi^xM_{xy}\Delta^{yz}\mathcal{K}_{zu}\Delta^{uv}\partial_{\phi^v}\right)\Psi=\\
    \int D\mu\bar\Psi\left(\partial_{\phi^v}\Delta^{vu}\phi^xM_{xy}\Delta^{yz}\mathcal{K}_{zu}-\Delta^{ux}M_{xy}\Delta^{yz}\mathcal{K}_{zu}-\phi^xM_{xy}\Delta^{yz}\mathcal{K}_{zu}\Delta^{uv}\partial_{\phi^v}\right)\Psi=0
\end{multline}
where, again, $\bar\phi^u$ has been mapped to $\partial_{\phi^v}\Delta^{vu}$ in the last equality. Note that, in order to ensure this nulity, it is crucial to keep the trace term appearing in the second line from the antiholomorphic derivative acting on $\zeta_{\phi}^x$, \textit{i.e.} $\partial_{\bar\phi^v}\Delta^{vx}M_{xy}\Delta^{yz}\mathcal{K}_{zu}\bar\phi^u$.  With this, we have shown that $\zeta_{f_2}$ has null EV, as explained in section \ref{example}.

Considering products of holomorphic and antiholomorphic fields, we may define $\dot f_{H-A}:=\nabla_\xi \left(Q(\phi^x)M_{xy}Q(\bar\phi^y)\right)$, and, uwith analogous considerations as before, we see: 
\begin{equation}
\dot f_{H-A}=Q(\partial_\xi\phi^x)M_{xy}Q(\bar\phi^y)+Q(\phi^x)\partial_\xi M_{xy}Q(\bar\phi^y)+Q(\phi^x)M_{xy}Q(\partial_\xi\bar\phi^y)+\zeta_{\phi}^xM_{xy}\Delta^{yz}\partial_{\phi^z}
\end{equation}
the EV of the last term must be null for the requirement \eqref{RequirementHardcore} to hold. Indeed, we can check that:
\begin{equation}
\int D\mu\bar\Psi\zeta_{\phi}^xM_{xy}\Delta^{yz}\partial_{\phi^z}\Psi=\int D\mu\bar\Psi(\bar\phi^x-\partial_\phi^u\Delta^{ux})M_{xy}\Delta^{yz}\partial_{\phi^z}\Psi=0\;,
\end{equation}
where, again, this holds because, when $\bar\phi^x$ is at the left of any function of holomorphic fields, it is equivalent to $\partial_\phi^u\Delta^{ux}$ acting to the right.  If we inverted the ordering of $\phi$ and $\bar\phi$ and defined $\dot f_{A-H}:=\nabla_\xi \left(Q(\bar\phi^x)M_{xy}Q(\phi^y)\right)$, we would then see:
\begin{equation}
\dot f_{A-H}=Q(\partial_\xi\bar\phi^x)M_{xy}Q(\phi^y)+Q(\bar\phi^x)\partial_\xi M_{xy}Q(\phi^y)+Q(\bar\phi^x)M_{xy}Q(\partial_\xi\phi^y)+Q(\bar\phi^x)M_{xy}\zeta_{\phi}^y\;,
\end{equation}
because $Q(\bar\phi^x)$ commutes with $\zeta_{\phi}^y$ and therefore the computations are analogous to the former case, being the last term of the latter equation null under the EV. 

With this we have covered all the possible quadratic monomials, which is what we need in geometrodynamics in order to proof \eqref{RequirementHM}. The generalization to higher order monomials, needed to proof the more demanding requirement \eqref{RequirementHardcore}, is straightforward.
\newpage

\bibliography{Bibliography_source}

\begin{thebibliography}{10}

\bibitem{agullo2015unitarity}
I.~Agullo and A.~Ashtekar.
\newblock Unitarity and ultraviolet regularity in cosmology.
\newblock {\em Physical Review D}, 91(12):124010, 2015.

\bibitem{Alonso2012}
J.~Alonso, A.~Castro, J.~Clemente-Gallardo, J.~Cuch\'i', P.~Echenique, and F.~Falceto.
\newblock Statistics and {N}os\'e formalism for {E}hrenfest dynamics.
\newblock {\em J. Phys.A.-Math. Theor.}, 44, 04 2011.

\bibitem{David}
J.~L. Alonso, C.~{Bouthelier-Madre}, J.~{Clemente-Gallardo}, and D.~{Mart{\'i}nez-Crespo}.
\newblock Geometric flavours of {{Quantum Field}} theory on a {{Cauchy}} hypersurface. {{Part I}}: {{Geometric}} quantization and star products.
\newblock {\em ArXiV}, (arXiv:2306.148442), June 2023.

\bibitem{David2}
J.~L. Alonso, C.~{Bouthelier-Madre}, J.~{Clemente-Gallardo}, and D.~{Mart{\'i}nez-Crespo}.
\newblock Geometric flavours of {{Quantum Field}} theory on a {{Cauchy}} hypersurface. {{Part II}}: {{Canonical}} and {{Geometrical QFT}}.
\newblock {\em ArXiV}, (arXiv:2402.07953), 2024.

\bibitem{Schneider2}
A.~Ashtekar, T.~De Lorenzo, and M.~Schneider.
\newblock Probing the big bang with quantum fields.
\newblock {\em arXiv}, (arXiv:2107.08506), 2021.

\bibitem{ashtekar75}
A.~Ashtekar and A.~Magnon.
\newblock Quantum fields in curved space-times.
\newblock {\em Proceedings of the Royal Society of London. A. Mathematical and Physical Sciences}, 346(1646):375--394, 1975.

\bibitem{ashtekar1999geometrical}
A.~Ashtekar and T.~A Schilling.
\newblock Geometrical formulation of quantum mechanics.
\newblock In {\em On Einstein’s Path: Essays in Honor of Engelbert Schucking}, pages 23--65. Springer, 1999.

\bibitem{BERNARD1977201}
C.~Bernard and A.~Duncan.
\newblock Regularization and renormalization of quantum field theory in curved space-time.
\newblock {\em Annals of Physics}, 107(1):201--221, 1977.

\bibitem{boulware1967stress}
D.~G Boulware and S~Deser.
\newblock Stress-tensor commutators and schwinger terms.
\newblock {\em Journal of Mathematical Physics}, 8(7):1468--1477, 1967.

\bibitem{Koopman}
C.~{Bouthelier-Madre}, L.~{Gonz{\'a}lez-Bravo}, J.~{Clemente-Gallardo}, and D.~{Mart{\'i}nez-Crespo}.
\newblock Hybrid {K}oopman ${C}^\star$–formalism and the hybrid quantum-classical master equation.
\newblock Number arXiv:2306.15601, June 2023.

\bibitem{Canarutto2004QuantumCA}
D.~Canarutto.
\newblock Quantum connections and quantum fields.
\newblock {\em Rend. Istit. Mat. Univ. Trieste}, 36:27--47, 2004.

\bibitem{CCQ04}
A.~Corichi, J.~Cortez, and H.~Quevedo.
\newblock {S}chrödinger and {F}ock representation for a field theory on curved spacetime.
\newblock {\em Annals of Physics}, 313(2):446--478, 2004.

\bibitem{Dirac50}
P.~A.~M. Dirac.
\newblock Generalized hamiltonian dynamics.
\newblock {\em Canadian Journal of Mathematics}, 2:129–148, 1950.

\bibitem{Dirac51}
P.~A.~M. Dirac.
\newblock The hamiltonian form of field dynamics.
\newblock {\em Canadian Journal of Mathematics}, 3:1–23, 1951.

\bibitem{Dito1990}
J.~Dito.
\newblock Star-product approach to quantum field theory: the free scalar field.
\newblock {\em letters in mathematical physics}, 20(2):125--134, 1990.

\bibitem{Dito1992}
J.~Dito.
\newblock Star-products and nonstandard quantization for klein--gordon equation.
\newblock {\em Journal of mathematical physics}, 33(2):791--801, 1992.

\bibitem{Ebo2}
O~Eboli, So-Young Pi, and M~Samiullah.
\newblock Renormalizability of the functional {S}chr{\"o}dinger picture in {R}obertson-{W}alker space-time.
\newblock {\em Annals of Physics}, 193(1):102--141, 1989.

\bibitem{Kie}
D.~Giulini and C.~Kiefer.
\newblock The canonical approach to quantum gravity: General ideas and geometrodynamics.
\newblock {\em Lecture Notes in Physics}, 721:131--150, 09 2007.

\bibitem{GROENEWOLD1946405}
H.~J. Groenewold.
\newblock On the principles of elementary quantum mechanics.
\newblock {\em Physica}, 12(7):405--460, 1946.

\bibitem{Glow20}
J.~Głowacki.
\newblock Inevitability of the {P}oisson bracket structure of the relativistic constraints.
\newblock 12 2020.

\bibitem{Heslot}
A.~Heslot.
\newblock Quantum mechanics as a classical theory.
\newblock {\em Phys. Rev. D}, 31:1341--1348, Mar 1985.

\bibitem{Hida}
T.~Hida, H.~Kuo, J.~Potthoff, and L.~Streit.
\newblock {\em White noise: an infinite dimensional calculus}, volume 253.
\newblock Springer Science \& Business Media, 2013.

\bibitem{Schneider}
S.~Hofmann and M.~Schneider.
\newblock Classical versus quantum completeness.
\newblock {\em Phys. Rev. D}, 91:125028, Jun 2015.

\bibitem{HKT76}
S.~A Hojman, K.~Kuchař, and C.~Teitelboim.
\newblock Geometrodynamics regained.
\newblock {\em Annals of Physics}, 96(1):88--135, 1976.

\bibitem{Holl}
S.~Hollands and R.M. Wald.
\newblock Quantum fields in curved spacetime.
\newblock {\em Physics Reports}, 574:1--35, 2015.

\bibitem{Husa}
V.~Husain and S.~Singh.
\newblock Semiclassical cosmology with backreaction: The {F}riedmann-{S}chr{\"o}dinger equation and inflation.
\newblock {\em Physical Review D}, 99(8):086018, 2019.

\bibitem{husain2021quantum}
V.~Husain and S.~Singh.
\newblock Quantum backreaction on a classical universe.
\newblock {\em Physical Review D}, 104(12):124048, 2021.

\bibitem{Kibble79}
T.~Kibble.
\newblock Geometrization of quantum mechanics.
\newblock {\em Communications in Mathematical Physics}, 65:189--201, 01 1979.

\bibitem{Kie1}
C.~Kiefer.
\newblock {\em Quantum {G}ravity}.
\newblock Oxford University Press UK, 2004.

\bibitem{Kie2}
Claus Kiefer.
\newblock {The Semiclassical approximation to quantum gravity}.
\newblock {\em Lect. Notes Phys.}, 434:170--212, 1994.

\bibitem{kobayashi2014differential}
S.~Kobayashi.
\newblock {\em Differential geometry of complex vector bundles}, volume 793.
\newblock Princeton University Press, 2014.

\bibitem{kriegl1997}
A.~Kriegl and P.W. Michor.
\newblock {\em The convenient setting of global analysis}, volume~53.
\newblock American Mathematical Soc., 1997.

\bibitem{Long1996TheSW}
Do~Viet Long and G.~M. Shore.
\newblock The {S}chr{\"o}dinger wave functional and vacuum states in curved spacetime.
\newblock {\em Nuclear Physics}, 530:247--278, 1996.

\bibitem{maniccia2023qft}
G.~Maniccia, G.~Montani, and S.~Antonini.
\newblock Qft in curved spacetime from quantum gravity: Proper wkb decomposition of the gravitational component.
\newblock {\em Physical Review D}, 107(6):L061901, 2023.

\bibitem{minlos1959generalized}
R.~A. Minlos.
\newblock Generalized random processes and their extension in measure.
\newblock {\em Trudy Moskovskogo Matematicheskogo Obshchestva}, 8:497--518, 1959.

\bibitem{mostafazadeh2018energy}
A.~Mostafazadeh.
\newblock Energy observable for a quantum system with a dynamical hilbert space and a global geometric extension of quantum theory.
\newblock {\em Physical Review D}, 98(4):046022, 2018.

\bibitem{Oeckl}
R.~Oeckl.
\newblock The schr{\"o}dinger representation and its relation to the holomorphic representation in linear and affine field theory.
\newblock {\em Journal of mathematical physics}, 53(7):072301, 2012.

\bibitem{teitelboim1973hamiltonian}
C.~Teitelboim.
\newblock {\em The Hamiltonian structure of spacetime}.
\newblock PhD thesis, Princeton University, 1973.

\bibitem{Til2}
A.~Tilloy.
\newblock Binding quantum matter and space-time, without romanticism.
\newblock {\em Foundations of Physics}, 48(12):1753--1769, 2018.

\bibitem{Til1}
A.~Tilloy.
\newblock Does gravity have to be quantized? lessons from non-relativistic toy models.
\newblock In {\em Journal of Physics: Conference Series}, volume 1275, page 012006. IOP Publishing, 2019.

\bibitem{torre1999functional}
C.~G Torre and M.~Varadarajan.
\newblock Functional evolution of free quantum fields.
\newblock {\em Classical and quantum gravity}, 16(8):2651, 1999.

\bibitem{tsamis1987factor}
NC~Tsamis and RP~Woodard.
\newblock The factor-ordering problem must be regulated.
\newblock {\em Physical Review D}, 36(12):3641, 1987.

\bibitem{wald1977}
R.~M. Wald.
\newblock The back reaction effect in particle creation in curved spacetime.
\newblock {\em Communications in Mathematical Physics}, 54(1):1--19, 1977.

\bibitem{Wald}
R.~M. Wald.
\newblock {\em Quantum field theory in curved spacetime and black hole thermodynamics}.
\newblock University of Chicago press, 1994.

\end{thebibliography}
\bibliographystyle{plain}

\end{document}